\documentclass[twocolumn]{aastex631}
\usepackage{amsmath}

\newcommand\kms{km$\,$s$^{-1}$}
\newcommand\Msol{M$_{\odot}$}

\newcommand{\hi}{H\,{\sc i}}

\submitjournal{AAS Journals}

\shorttitle{Gas and star formation in satellites}
\shortauthors{Jones et al.}
\graphicspath{{./}{figures/}}

\begin{document}

\title{Gas and star formation in satellites of Milky Way analogs}

\correspondingauthor{Michael G. Jones}
\email{jonesmg@arizona.edu}

\author[0000-0002-5434-4904]{Michael G. Jones}
\affiliation{Steward Observatory, University of Arizona, 933 North Cherry Avenue, Rm. N204, Tucson, AZ 85721-0065, USA}

\author[0000-0003-4102-380X]{David J. Sand}
\affiliation{Steward Observatory, University of Arizona, 933 North Cherry Avenue, Rm. N204, Tucson, AZ 85721-0065, USA}

\author[0000-0001-8855-3635]{Ananthan Karunakaran}
\affiliation{Department of Astronomy \& Astrophysics, University of Toronto, Toronto, ON M5S 3H4, Canada}
\affiliation{Dunlap Institute for Astronomy and Astrophysics, University of Toronto, Toronto ON, M5S 3H4, Canada}
\affiliation{Instituto de Astrof\'{i}sica de Andaluc\'{i}a (CSIC), Glorieta de la Astronom\'{i}a, 18008 Granada, Spain}

\author[0000-0002-0956-7949]{Kristine Spekkens}
\affiliation{Department of Physics and Space Science, Royal Military College of Canada P.O. Box 17000, Station Forces Kingston, ON K7K 7B4, Canada}
\affiliation{Department of Physics, Engineering Physics and Astronomy, Queen’s University, Kingston, ON K7L 3N6, Canada}

\author[0000-0001-9857-7788]{Kyle A. Oman}
\affiliation{Institute for Computational Cosmology, Durham University, South Road, Durham DH1 3LE, UK}
\affiliation{Department of Physics, Durham University, South Road, Durham DH1 3LE, UK}

\author[0000-0001-8354-7279]{Paul Bennet}
\affiliation{Space Telescope Science Institute, 3700 San Martin Drive, Baltimore, MD 21218, USA}

\author[0000-0003-0715-2173]{Gurtina Besla}
\affiliation{Steward Observatory, University of Arizona, 933 North Cherry Avenue, Rm. N204, Tucson, AZ 85721-0065, USA}

\author[0000-0002-1763-4128]{Denija Crnojevi\'{c}}
\affil{Department of Physics \& Astronomy, University of Tampa, 401 West Kennedy Boulevard, Tampa, FL 33606, USA}

\author[0000-0002-3263-8645]{Jean-Charles Cuillandre}
\affiliation{AIM, CEA, CNRS, Universit\'{e} Paris-Saclay, Universit\'{e} Paris-Diderot, Sorbonne Paris-Cit\'{e}, Observatoire de Paris, PSL University, 91191 Gif-sur-Yvette Cedex, France}

\author[0000-0001-8245-779X]{Catherine E. Fielder}
\affiliation{Steward Observatory, University of Arizona, 933 North Cherry Avenue, Rm. N204, Tucson, AZ 85721-0065, USA}

\author[0000-0001-8221-8406]{Stephen Gwyn}
\affiliation{National Research Council of Canada, Herzberg Astronomy and Astrophysics, 5071 West Saanich Road, Victoria, BC V9E 2E7, Canada}

\author[0000-0001-9649-4815]{Bur\c{c}in Mutlu-Pakdil}
\affil{Department of Physics and Astronomy, Dartmouth College, Hanover, NH 03755, USA}



\begin{abstract}

We have imaged the entirety of eight (plus one partial) Milky Way-like satellite systems, a total of 42 (45) satellites, from the Satellites Around Galactic Analogs (SAGA) II catalog in both H$\alpha$ and \hi \ with the Canada-France-Hawaii Telescope and the Jansky Very Large Array. In these eight systems we have identified four cases where a satellite appears to be currently undergoing ram pressure stripping (RPS) as its \hi \ gas collides with the circumgalactic medium (CGM) of its host. We also see a clear suppression of gas fraction ($M_\mathrm{HI}/M_\ast$) with decreasing (projected) satellite--host  separation; to our knowledge, the first time this has been observed in a sample of Milky Way-like systems. 
Comparisons to the Auriga, APOSTLE, and TNG50 cosmological zoom-in simulations show consistent global behavior, but they systematically under-predict gas fractions across all satellites by roughly 0.5~dex.
Using a simplistic RPS model we estimate the average peak CGM density that satellites in these systems have encountered to be $\log \rho_\mathrm{cgm}/\mathrm{g\,cm^{-3}} \approx -27.3$. 
Furthermore, we see tentative evidence that these satellites are following a specific star formation rate-to-gas fraction relation that is distinct from field galaxies. 
Finally, we detect one new gas-rich satellite in the UGC~903 system with an optical size and surface brightness meeting the standard criteria to be considered an ultra-diffuse galaxy.

\end{abstract}



\section{Introduction} 
\label{sec:intro}

In recent years two major efforts, the Satellites Around Galactic Analogs \citep[SAGA,][]{Geha+2017,Mao+2021} and the Exploration of Local VolumE Satellites \citep[ELVES,][]{Carlsten+2022} projects, have built up consistently defined, large samples of satellite galaxies around tens of hosts analogous to the Milky Way (MW) for the first time, significantly extending earlier works \citep[e.g.][]{Zaritsky+1993,Zaritsky+1997,Crnojevic+2016,Bennet+2019,Carlsten+2020}. These new samples have opened the door to statistical studies of satellites in MW-mass systems other than those in the Local Group (LG), which until now have been almost the sole point of detailed comparison with simulated systems in this mass regime. However, in the case of satellite quenching, the removal of gas and the permanent cessation of star formation (SF), SAGA and ELVES do not necessarily align with findings from the LG, or perhaps even with each other. 

In the LG virtually all satellites within the virial radius of either the MW or M~31 ($\sim$300~kpc) are devoid of gas and quenched \citep{Spekkens+2014,Putman+2021}, with the most notable exceptions being the LMC and SMC. Initial analyses of the ELVES population \citep{Carlsten+2022} suggests that it follows a similar trend. However, in the case of SAGA, the vast majority of the satellites are still star-forming. \citet{Karunakaran+2021} found that over 85\% of the SAGA satellites are star-forming, in comparison to less than 40\% of equivalently selected satellites from the A Project Of Simulating The Local Environment \citep[APOSTLE,][]{Fattahi+2016,Sawala+2016} and Auriga \citep{Grand+2017} simulations. 

The satellite mass ranges sampled by SAGA and ELVES are overlapping but different, complicating attempts at direct comparisons. \citet{Greene+2023} find that the quenched fraction in ELVES systems is comparable to the LG, and \citet{Font+2022} and \citet{Carlsten+2022} argue that incompleteness in SAGA for low surface brightness (LSB) dwarfs could cause the apparent discrepancy between SAGA and simulations and the LG. However, \citet{Karunakaran+2023} explored a number of SF metrics and selection biases in the SAGA and ELVES samples in an attempt to make like-for-like comparisons, concluding that the quenched fraction of satellites is similar for both (after accounting for differences in sample selection) and in tension with the high quenched fraction seen in the LG and simulations.

Quenching is a pivotal process that galaxies undergo when they fall into a group or cluster. Quenching transforms a galaxy from a member of the star-forming population typically found in the field, to a quiescent member of a larger structure \citep{Dressler1980,Grebel+2003}. Theory and simulations indicate that ram pressure stripping (RPS) likely places a significant role in driving satellite quenching all the way from Milky Way-mass groups to the largest clusters \citep[e.g.][]{Gunn+Gott1972,Vollmer+2001,Tonnesen+2009,Simpson+2018,Oman+2021,Boselli+2022,Wright+2022}. There are even some claims of relatively isolated, low-mass galaxies undergoing RPS in the field, presumably due to collisions with denser-than-average regions of the intergalactic medium \citep[e.g.][]{Benitez-Llambay+2013,Benitez-Llambay+2017,Beale+2020,Yang+2022}.

RPS can occur when a gas-bearing galaxy moves at high speed through the intracluster medium (ICM) in a galaxy cluster or the circumgalactic medium (CGM) of the central galaxy in a galaxy group. If the ram pressure exerted on the gas content of the infalling galaxy exceeds the gravitational attraction between the gas and the other components of the disk, then gas will be stripped \citep{Gunn+Gott1972}. If a galaxy loses all or most of its gas reservoir through RPS, then it will no longer be capable of sustained SF and will quench.

There are numerous examples of RPS in progress in galaxy clusters \citep[e.g.][]{Kenney+1999,Kenney+2004,Koopmann+2004,Chung+2009,Poggianti+2017,Boselli+2018,Ramatsoku+2019,Cortese+2021,Jones+2022,Boselli+2023}, which are often called ``jellyfish'' galaxies because of the tendrils of gas \citep[and, frequently, in situ SF, e.g.][]{Jachym+2014,Boselli+2018} that extend from their disks. The phenomenon is also common in large groups \citep[e.g.][]{Roberts+2021}. However, for MW-mass systems there are only a handful of clear examples of ongoing RPS, such as, the LMC in the Local Group (LG) itself \citep[e.g.][]{Luks+Rholfs1992,deBoer+1998}, Holmberg~II in the M~81 system \citep{Bureau+2002,Bernard+2012}, and ESO~324-G024 in the Cen~A system \citep{Fritz+1997,Johnson+2015}. It has also been suggested that the LG Phoenix and Pegasus dwarfs might be experiencing RPS \citep{Young+2007,McConnachie+2007}.

In this work we approach the issue from a different perspective by trying to understand the details of the process of quenching in the SAGA systems, where most satellites have either yet to quench or are currently in the process of quenching. We have conducted observations of eight (plus one partial) SAGA systems, 42 (45) confirmed satellites, with both the Jansky Very Large Array (VLA) and the Canada-France-Hawaii Telescope (CFHT). The VLA \hi \ observations are to map the gas reservoirs of the satellites and to look for signs that gas is being lost. We identify four cases of likely RPS in progress, effectively doubling the number of known examples in MW-like systems. The CFHT H$\alpha$ observations trace recent SF, providing a uniform census of star formation rates (SFRs) and a robust means to separate quenched and star-forming satellites. By assessing both the SF, as well as the gas that fuels it, we aim to build a clearer picture of how quenching is proceeding in these systems.

This paper is organized as follows. In the following section we discuss the SAGA sample from which our targets were selected, as well as the VLA and CFHT observations. In \S\ref{sec:results} we present our results, including \hi \ masses (and limits) and SFR estimates for all targeted satellites. We discuss these results in \S\ref{sec:discussion} and present our conclusions in \S\ref{sec:conclusions}. 

Throughout this work we assume that all satellites are at the same distances as their SAGA hosts (unless stated otherwise) and we use host distance estimates from \citet{Mao+2021}, which are reproduced in Table~\ref{tab:hosts}.

\section{Sample \& Observations}
\label{sec:obs}

\begin{table*}
\centering
\caption{SAGA systems targeted}
\begin{tabular}{lcccccccccc}
\hline \hline
Host Name & RA  & Dec & $cz_\odot$            & Dist & $\log M_\ast$        & $\sigma_\mathrm{rms}$ & No. of VLA & Beam                     & PA & $\log M_\mathrm{HI}$ \\
     & deg & deg & $\mathrm{km\,s^{-1}}$ & Mpc  & $[\mathrm{M_\odot}]$ & mJy/beam              & pointings & \arcsec $\times$ \arcsec & deg & $[\mathrm{M_\odot}]$ \\ \hline
UGC~903   & 20.449  & 17.592  & 2516 & 38.4 & 10.8 & 1.2 & 2 & $76\times58$ & 76 & $9.69\pm0.04$ \\
NGC~1309  & 50.527  & -15.400 & 2136 & 34.3 & 10.8 & 1.4 & 2 & $90\times50$ & 20 & $9.92\pm0.04$ \\
PGC~13646 & 55.734  & -12.916 & 2163 & 31.9 & 10.6 & 1.2 & 1 & $85\times54$ & 12 &  \\
UGC~4906  & 139.416 & 52.993  & 2273 & 36.1 & 10.7 & 1.5 & 2 & $65\times47$ & -81 & $9.39\pm0.05$ \\
NGC~4158  & 182.792 & 20.176  & 2450 & 36.2 & 10.6 & 1.3 & 3 & $80\times56$ & 51 & $9.34\pm0.04$ \\
NGC~4454  & 187.211 & -1.939  & 2329 & 35.3 & 10.8 & 1.3 & 1$^\dagger$ & $83\times53$ & 40 & $<8.73$ \\
NGC~6278  & 255.210 & 23.011  & 2795 & 39.3 & 10.9 & 1.2 & 3 & $86\times53$ & -60 & $<8.79$ \\
PGC~68743 & 335.913 & -3.432  & 2865 & 39.1 & 10.9 & 1.4 & 1 & $70\times55$ & -8 & $<8.67$ \\
NGC~7541  & 348.683 & 4.534   & 2680 & 38.1 & 11.2 & 1.0 & 2 & $68\times57$ & 41 & $10.07\pm0.04$ \\
\hline
\end{tabular}\\
Stellar masses were estimated based on the K-band absolute magnitudes in \citet{Mao+2021} and assuming a mass-to-light ratio of 0.96 and $M_\mathrm{\odot,K} = 3.41$, as in \citep{Longhetti+2009}. Distance estimates are from \citet{Mao+2021}. The RMS noise and beam size values quoted here are (at the pointing center) for the robust=2 \hi \ cubes. There is no \hi \ mass measurement for PGC~13646 as it was outside of the VLA primary beam. $\dagger$ Note that two pointings are required to image all of the NGC~4454 satellites in SAGA, but only one pointing was observed.
\label{tab:hosts}
\end{table*}

Our target sample was selected from the SAGA-II release \citep{Mao+2021}; here we briefly describe the SAGA selection process. The SAGA project \citep{Geha+2017} used the HyperLEDA database \citep{Paturel+2003,Makarov+2014} to select all MW-like host galaxies in the distance range $25 \lesssim D/\mathrm{Mpc} \lesssim 40$, based on the distance estimates of the NASA-Sloan Atlas \citep{Blanton+2011}. SAGA defines ``MW-like'' as $-23 > M_\mathrm{K} > -24.6$, intended to approximately correspond to the MW's stellar mass, with the K-band magnitudes taken from the 2MASS Redshift Survey catalog \citep{Huchra+2012}. The final list of hosts was trimmed to reject those with bright foreground stars and those that are themselves within the virial radius of a much larger system \citep[][Section 2.1.2]{Mao+2021}.

SAGA then selected satellites around each host within 300~kpc (projected), taken as the approximate virial radius of a MW-like galaxy. Potential satellites were first selected from the photometric catalogs of SDSS DR14 \citep{SDSSDR14}, the Dark Energy Survey DR1 \citep{DESDR1}, and the DESI legacy imaging surveys DR6 and DR7 \citep{Dey+2019}. These catalogs were combined and cleaned of duplicates, Galactic cirrus, and other spurious detections \citep{Mao+2021}. The satellite candidates without prior redshift measurements were then targeted with follow-up spectroscopic observations. As not all candidates could be observed, SAGA target primarily those within a specific region of surface brightness--color space \citep[][equation 3]{Mao+2021}, designed to capture all galaxies with $z < 0.015$. Finally, only the candidates within 275~\kms \ of the host galaxy redshift are considered as satellites and retained in the final SAGA-II catalog.

Starting with the SAGA-II catalog, for our \hi \ and H$\alpha$ observations we selected all systems with three or more satellites that are North of Dec = -25, and therefore straightforward to observe with the VLA and CFHT, a total of 17 systems and 85 satellites. We then excluded systems in the range $13.6\,\mathrm{h} < \mathrm{RA} < 16.8\,\mathrm{h}$ as these were not observable in the CFHT 2021B semester during which our project began. This resulted in an initial sample of 10 hosts and 49 satellites.

\subsection{VLA observations}
\label{sec:VLA_obs}

These 10 SAGA systems were observed between July and September 2022 as part of the VLA project 22A-023 (PI: M.~Jones). Despite the fact that the SAGA systems nominally extend over approximately 1~sq~deg, in practice the satellites are usually arranged such that all can be imaged with the VLA in just a few 30\arcmin-wide L-band primary beams. Of the 10 systems observed, the maximum number of separate pointings required was three, with the average being two. In general these observations were lightly impacted by radio frequency interference (RFI) and antenna outages, however, NGC~2906, NGC~4454, and UGC~4906 were all more severely affected. For NGC~2906 (a single pointing) no usable data were collected, while for NGC~4454 and UGC~4906 about half of the data were lost to RFI. In addition, due to an error in the scheduling block files, the second pointing for NGC~4454 was not observed, meaning that 3/4 satellites in that system have either poor or no coverage. This reduced the sample to 8 systems (42 satellites) with complete \hi \ coverage and one additional system with partial coverage.

The data were reduced using \texttt{CASA} \citep{CASA} and the \hi \ pipeline\footnote{\url{https://github.com/AMIGA-IAA/hcg_hi_pipeline}} of \citet{Jones+2023}. Each observation was reduced separately following standard automated and manual flagging and calibration proceedures. Multiple pointings were combined into mosaics for imaging. All fields were imaged with Briggs robust values of 0.5 and 2. The former represents a compromise between sensitivity and spatial resolution, while the latter optimizes sensitivity for sources that are comparable in size to the beam (or larger). Multi-scale CLEANing and auto-masking were used within \texttt{CASA}'s \texttt{tclean} task. The images were CLEANed down to approximately 2.5$\sigma$ within the mask. Spectral resolution was also re-gridded to 5~\kms \ during imaging. The final RMS noise values and synthesized beam sizes for each system are shown in Table \ref{tab:hosts}. 

Source masking in the image cubes was performed using \texttt{SoFiA} v1.3.2 \citep{Serra+2015}. The standard smooth and clip algorithm was implemented with no spatial smoothing, as well as smoothing at roughly half and equal to the synthesized beam diameter. In addition, the data were smoothed to a spectral resolution of $\sim$15~\kms. A threshold of 4$\sigma$ was used to clip the smoothed data cube and the candidate detections were then merged. A reliability threshold of 95\% was enforced to remove spurious detections. These source masks were then used to generate moment maps and spectra of each detected target using the primary beam-corrected \hi \ data cube. 

In the case of non-detections, we returned to the (primary beam-corrected) data cube and extracted spectra over an area equal in size to one synthesized beam (from the robust=2 cube), centered at the optical center of each target. The RMS noise in each of the spectra was measured within $\pm100$~\kms \ of the known velocity of each target. These noise values were then used to estimate 3$\sigma$ upper limits on the \hi \ masses of each non-detection, assuming a nominal velocity width of 30~\kms \ and a roughly top-hat shaped spectral profile. There is one satellite (LS-429812-2469 of UGC~903) that we marginally detected. However, we still use the upper limit as the peak S/N is only $\sim$2.

For the detected satellites \hi\ masses were calculated simply by integrating all the flux within their source mask generated by \texttt{SoFiA}. As the \hi\ morphology of a number of the satellites is quite disturbed, and because they mostly have relatively narrow spectral profiles (not exhibiting double-horn features), this simplistic approach was preferred to profile fitting. However, in three cases manual adjustments were made to the \texttt{SoFiA} masks. The \hi \ emission from two satellites were lightly blended with their hosts, while NSA-542307 was blended with a low-level RFI feature that we were unable to remove via flagging. These were all manually separated using the 3-dimensional visualization tool \texttt{SlicerAstro} \citep{Punzo+2016,Punzo+2017} in order to measure their individual \hi \ masses.

We use the same approaches as above to measure the host \hi \ masses and limits in Table~\ref{tab:hosts}. However, for the undetected hosts we assumed a fiducial velocity width of 100~km/s. Three of the hosts (UGC~903, NGC~4158, and NGC~7541) are detected in the Arecibo Legacy Fast ALFA (Arecibo L-band Feed Array) survey, or ALFALFA \citep{Giovanelli+2005,Haynes+2018}, and after correcting for the different distance estimates all three of our \hi \ mass measurements agree within 1$\sigma$ with those from ALFALFA.

\subsection{CFHT observations}
\label{sec:CFHT_obs}

All SAGA systems observed with the VLA, except NGC~4454, were also observed in H$\alpha$ and $r$-band with the MegaCam instrument on CFHT during programs 21BC05 and 23AC01 (PI: K.~Spekkens). The MegaCam imager consists of 40 9.5 megapixel CCDs, covering a full square degree field of view (FoV). This meant that every target system could be observed in its entirety in a single pointing, although in some cases, where the satellite distribution was lopsided, the host was not placed at the center of the FoV. Each target was observed with either a five- or seven-point large dither pattern to fill in gaps between the CCDs. The nominal exposure times for each system were 2000~s in H$\alpha$ and 300~s in $r$-band, however, in some cases usable exposures were taken during partial executions of observing blocks, leading to slightly longer total integrations. These images were generally taken during grey time with seeing below 1.2\arcsec. The images were processed with the \texttt{Elixir-LSB} pipeline \citep{Ferrarese+2012} and mosaicked with \texttt{MegaPipe} \citep{Gwyn2008}. Continuum-subtracted H$\alpha$ images were produced by subtracting the $r$-band images from the H$\alpha$ images. 

H$\alpha$ fluxes were measured using the Aperture Photometry Tool \citep{Laher+2012}. The H$\alpha$ morphology of most of the detected satellites is highly irregular and clumpy, thus elliptical apertures (with annuli for sky subtraction) were manually adjusted for each source. In cases of H$\alpha$ non-detections, a circular aperture of 5.61\arcsec \ (30 pixels) in radius, and centered on the satellite coordinates from \citet{Mao+2021}, was used to estimate a 3$\sigma$ upper limit for the H$\alpha$ flux. To convert the observed magnitudes, $m_{\mathrm{H}\alpha}$, to H$\alpha$ fluxes we used the expression: $F_{\mathrm{H}\alpha} = 3.621\times10^{-20} \Delta\nu 10^{-0.4 m_{\mathrm{H}\alpha}} \; \mathrm{erg\,s^{-1}\,cm^{-2}}$, where $\Delta\nu$ is the filter bandwidth in Hz ($7.18\times10^{12}$~Hz). Fluxes were corrected for Galactic extinction \citep{Fitzpatrick+1999,Schlafly+2011}, but not internal extinction, and then converted to luminosities (in $\mathrm{erg\,s^{-1}}$) based on the host distance estimates (Table~\ref{tab:hosts}) and finally to SFRs following \citet{Kennicutt1998}, i.e. $\mathrm{SFR}_{\mathrm{H}\alpha} = 7.9\times10^{-42} (L_{\mathrm{H}\alpha}/\mathrm{erg\,s^{-1}}) \; \mathrm{M_\odot\,yr^{-1}}$. We assumed an uncertainty in this conversion of 0.2~dex, as in \citet{Kennicutt1998}, which was added in quadrature with the flux measurement uncertainties.

\section{Results}
\label{sec:results}

\subsection{Gas fraction of satellites}

\begin{figure*}
    \centering
    \includegraphics[width=1.2\columnwidth]{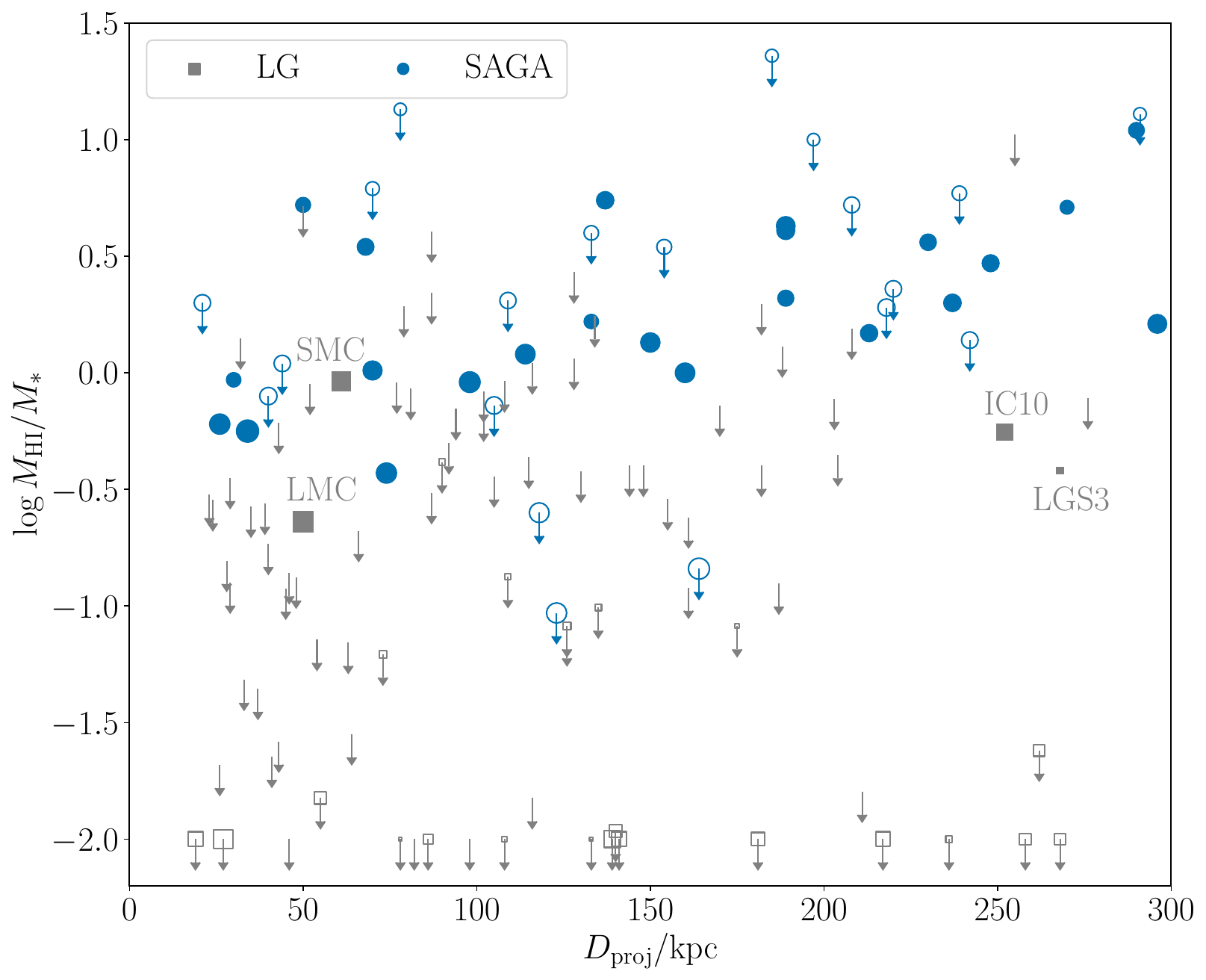}
    \caption{Gas fraction of SAGA satellites as a function of projected distance from their hosts (blue points and limits). Values for Local Group satellites are shown in grey for comparison \citep{Putman+2021}. In this case the separation is not the projected separation, but the smaller of the physical separation between the satellite and either the MW or M~31. Filled symbols indicate \hi\ detections and unfilled symbols are upper limits. The symbol diameters are scaled by $\log{M_\ast}$. A floor in gas fraction upper limits is set at $\log M_\mathrm{HI}/M_\ast = -2.0$. Error bars are omitted for clarity. We remind the reader that many Local Group satellites would be too faint/low-mass to be included in SAGA, hence many LG satellites have symbols that are too small to see ($\log{M_\ast/\mathrm{M_\odot}} \lesssim 6$).}
    \label{fig:gas_frac}
\end{figure*}

The \hi\ mass measurements and limits from \S\ref{sec:VLA_obs} (Table~\ref{tab:sats}) were combined with the stellar mass estimates of \citet{Mao+2021}, to calculate gas fractions, $M_\mathrm{HI}/M_\ast$, for every satellite in the nine systems targeted (with the exception of LS-321038-4238 in the NGC~4454 system, which was not observed with the VLA). These gas fractions are plotted as a function of projected separation between host and satellite in Figure~\ref{fig:gas_frac}. Detections are plotted as filled blue circles, while upper limits are unfilled circles. The LG satellites from \citet{Putman+2021} are shown as filled and unfilled grey squares for comparison \citep[stellar mass estimates from][]{Carlsten+2022}. In both cases the symbol sizes are scaled with $\log M_\ast$. A floor for gas fraction upper limits is set at $\log M_\mathrm{HI}/M_\ast = -2.0$ in order to prevent the vertical axis becoming too stretched.

Focusing first on the detections, we find a pronounced trend of decreasing gas fraction with decreasing host--satellite projected separation, declining approximately 1~dex over 300~kpc. This is the qualitative expectation if the satellites are being stripped of their gas by interactions with the central galaxy and its hot gas halo, however, the situation appears notably distinct from that of the LG where virtually all satellites within 300~kpc are devoid of gas. This is discussed further in \S\ref{sec:discuss_gas_frac}. Related trends in the \hi\ content of satellite galaxies in groups have been found in various works \citep[e.g.][]{Hess+2013,Odekon+2016,Brown+2017,Jones+2020,Zhu+2023}, but to our knowledge this is the first such measurement for a sample focused on satellites around MW-like hosts.

The upper limits are mostly mixed in with the detections and are not ruling out the possibility that the non-detections could be following the same trend as the detections. The exceptions are the three limits that are significantly below all of the detections. These are the mostly likely cases where satellites have lost their \hi \ reservoirs. These are all fairly massive ($\log M_\ast/\mathrm{M_\odot} \sim 9$) satellites with low specific star formation rates, that have likely been members of their current systems for some time, and as a result have lost the majority of their initial gas reservoir. The other non-detections are likely a mixture of genuinely gas-poor satellites and those with \hi\ masses too low to be detected simply because these satellite are low mass (both stellar and \hi).

\subsection{Star formation rates}

\begin{figure}
    \centering
    \includegraphics[width=\columnwidth]{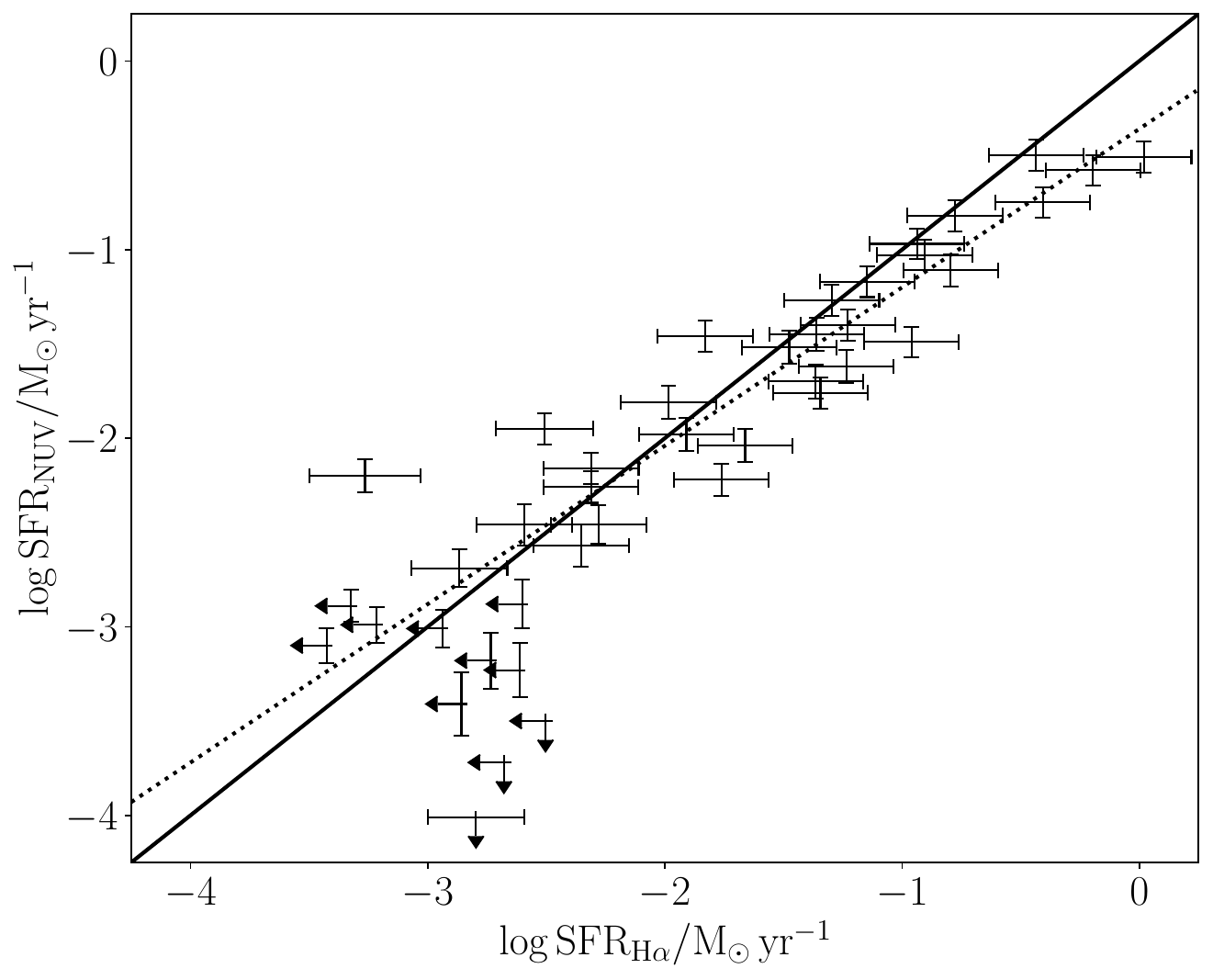}
    \caption{Comparison between NUV- and H$\alpha$-derived SFR estimates for the SAGA satellites in common between this work and \citet{Karunakaran+2021}. The solid black line is the line of equality and the dotted black line is a fit to the satellites detected in both NUV and H$\alpha$.}
    \label{fig:SFR_comp}
\end{figure}

\begin{figure}
    \centering
    \includegraphics[width=\columnwidth]{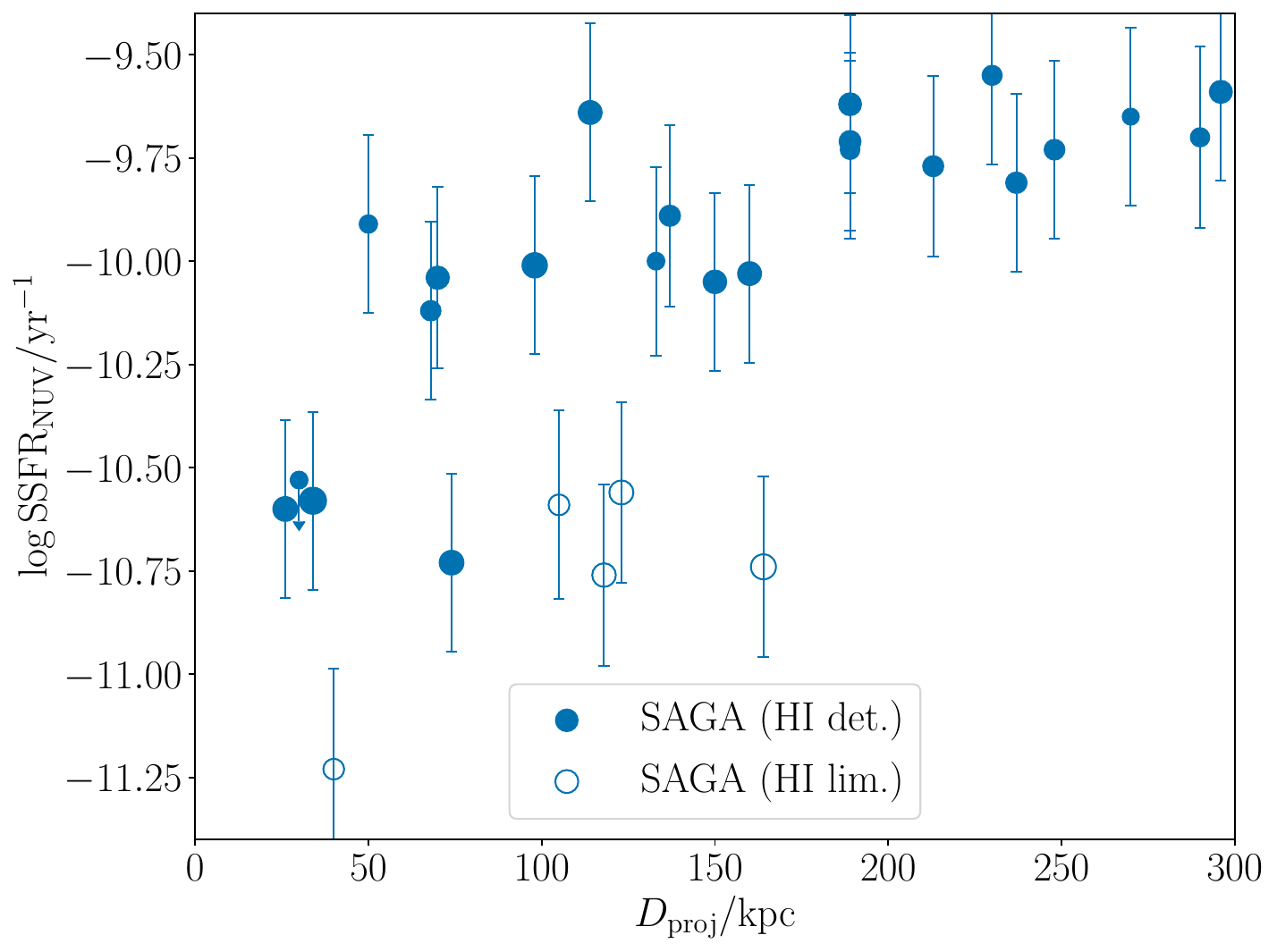}
    \caption{Specific star formation rate of SAGA satellites as a function of projected distance from their hosts. Filled blue points indicate \hi \ detections, while unfilled points show non-detections (only for objects where the \hi \ mass limits are less than the stellar mass estimates, i.e. only those guaranteed to be gas-poor). Symbols are scaled by $\log{M_\ast}$, as in Figure~\ref{fig:gas_frac}.}
    \label{fig:SSFR}
\end{figure}

SFRs for most SAGA-II satellites have been estimated previously based on GALEX NUV (and some FUV) fluxes \citep{Karunakaran+2021}. H$\alpha$ traces SFR over the past $\sim$10~Myr, as opposed to SFR over a timescale on the order of $\sim$100~Myr with NUV \citep[e.g.][]{Lee+2011}, thus the two tracers are highly complementary. Here we present H$\alpha$-based SFR estimates from the CFHT imaging and compare to the previous NUV-based estimates.

Figure~\ref{fig:SFR_comp} shows the comparison between our SFR estimates and those based on NUV from \citet{Karunakaran+2021}.\footnote{To estimate the NUV SFR uncertainties we add in quadrature the NUV flux uncertainties and 20\% scatter, as found by \citet{Iglesias-Paramo+2006}.} In general we see excellent agreement between the two measurements. There is a slight tendency for points at high SFRs to be below the one-to-one line (i.e. $\mathrm{SFR_{H\alpha} < SFR_{NUV}}$) and those at low SFRs to be above it. 
There is one satellite detected in H$\alpha$, but undetected in NUV, and with the H$\alpha$ SFR measurement in tension with the SFR limit from NUV. This is probably a result of the very shallow GALEX observations for this particular object causing the previous upper limit to be in error. We also note that optical spectra from \citet{Mao+2021} detect H$\alpha$ line emission in this object. There is another outlier (LS-595052-1940) approximately 1~dex above of the main locus of points. In this case we suspect that a background UV source in close proximity to the dwarf led to an overestimate of $\mathrm{SFR_{NUV}}$.

Using \texttt{HyperFit} \citep{hyperfit} we fit a straight line to all the satellites with detections in both H$\alpha$ and NUV (excluding the outlier LS-595052-1940). This fit is shown by the dashed black line in Figure~\ref{fig:SFR_comp}. As expected the best fit slope is slightly below unity and there is a non-zero intercept. The gradient and intercept values are $0.84 \pm 0.06$ and $-0.36 \pm 0.09$, respectively.

Figure~\ref{fig:SSFR} shows the specific star formation rates (SSFRs) of the SAGA satellites as a function of (projected) host--satellite separation. There is a drop-off in SSFR for \hi-detected satellites that are particularly close to their host ($D_\mathrm{proj} \lesssim 100$~kpc). For the \hi \ non-detections we only include those for which the upper limit provides a tight constraint on the gas fraction ($\log M_\mathrm{HI}/M_\ast < 0$), as the other non-detections could be just as gas-rich as the detections (cf. Figure~\ref{fig:gas_frac}).
These non-detections all have low SSFRs, indicating that proximity to the host is not the only relevant factor here. We will return to this point in \S\ref{sec:discuss_SFR}.

\subsection{Ram pressure stripping in progress}

\begin{figure*}
    \centering
    \includegraphics[width=\textwidth]{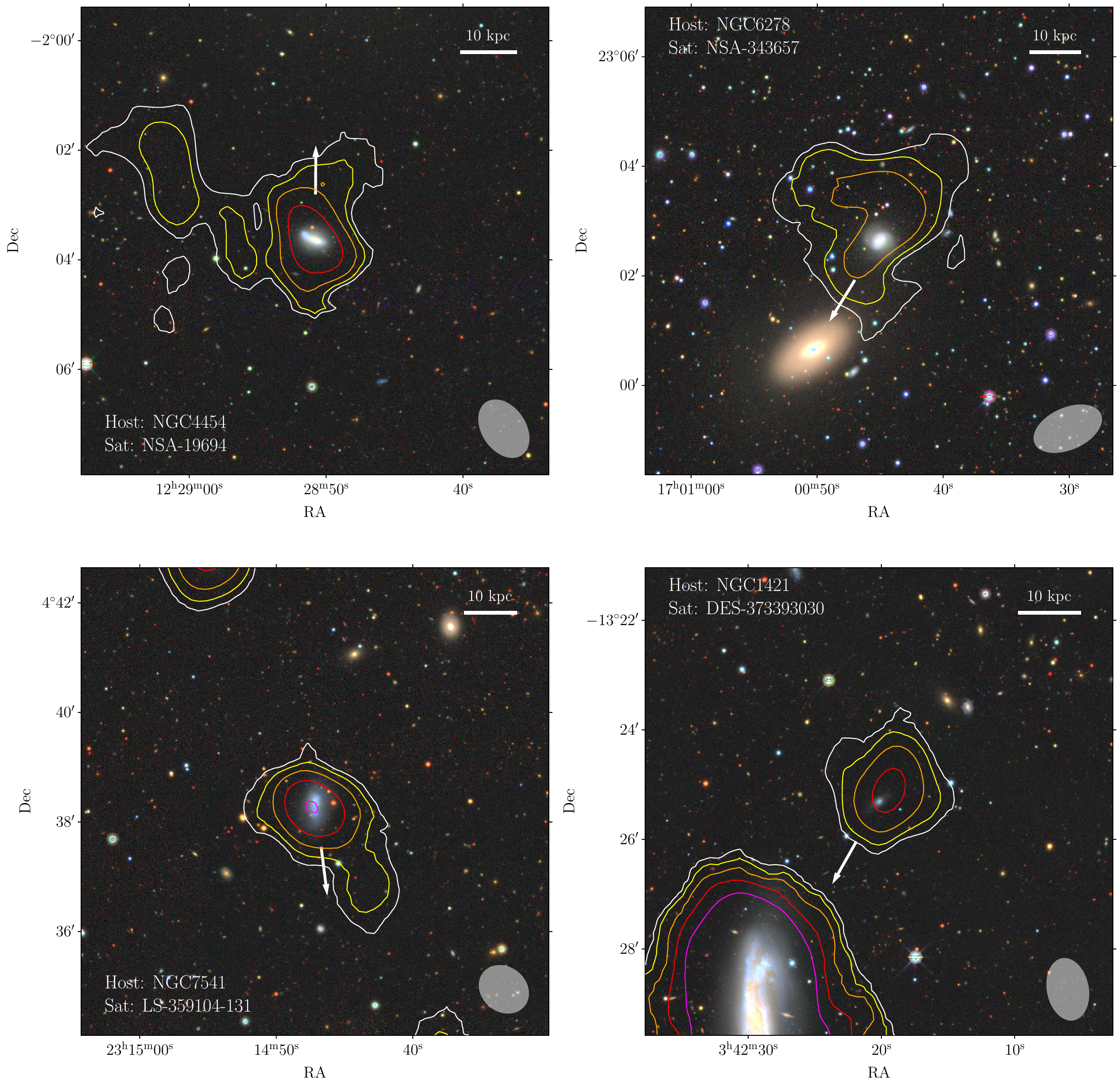}
    \caption{VLA \hi \ moment zero (robust=0.5) contours overlaid on DECaLS $grz$ images for the four examples of likely ram pressure stripping in progress. From top-left to bottom-right the satellites (hosts) are NSA-19694 (NGC~4454), NSA-343657 (NGC~6278), LS-359104-131 (NGC~7541), and DES-373393030 (NGC~1421). In each case a white arrow points in the direction of the host galaxy. The ellipses in the lower-right corner of each panel indicates the VLA synthesized beam size. The contours begin at 3$\sigma$ and each subsequent contour is double the previous. For top-left to bottom-right these lowest contour values are $8.4 \times 10^{18}$, $4.1 \times 10^{18}$, $5.7 \times 10^{18}$, and $6.9 \times 10^{18}$~$\mathrm{cm}^{-2}$ over 10~\kms.}
    \label{fig:rps_examples}
\end{figure*}

\begin{figure}
    \centering
    \includegraphics[width=\columnwidth]{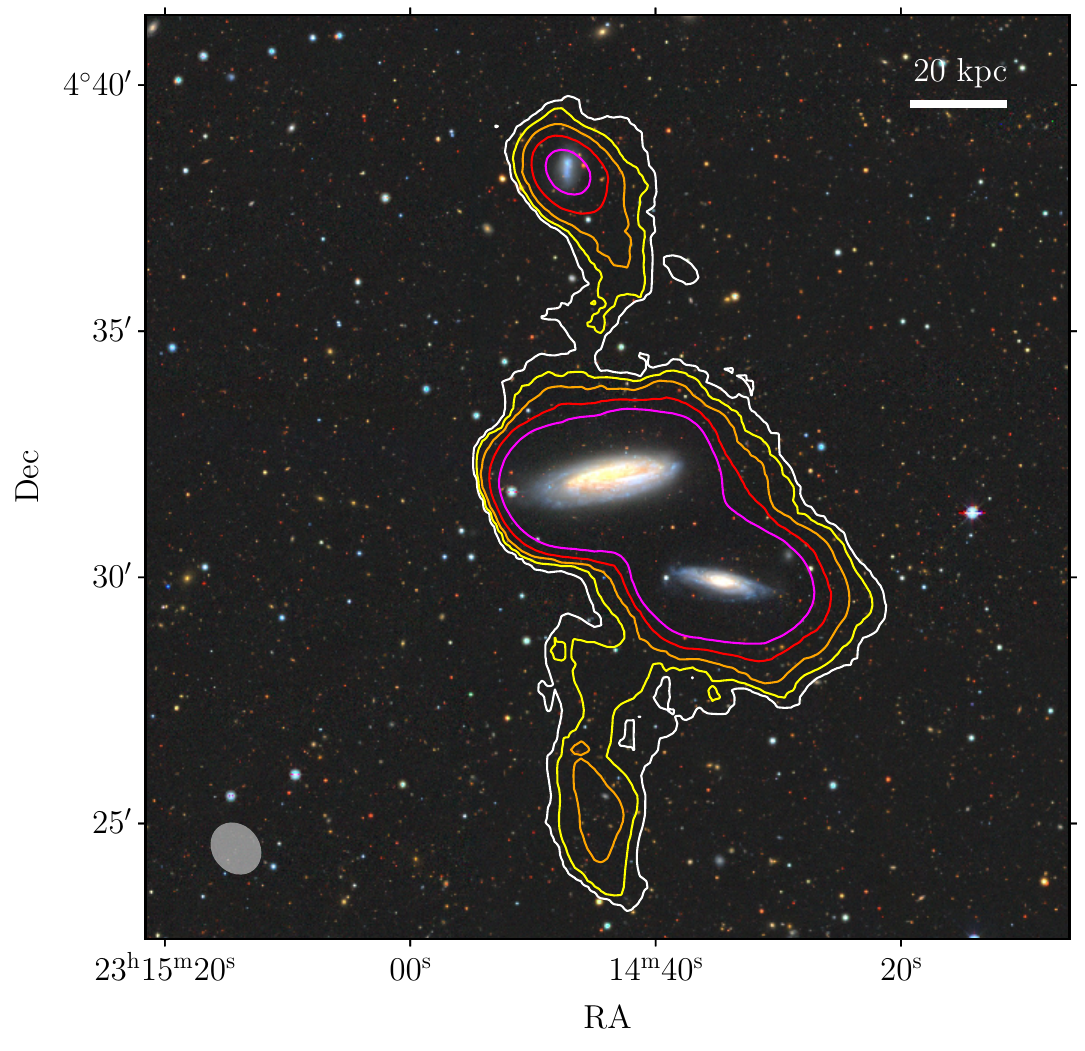}
    \caption{A wider view of the \hi \ moment zero contours of LS-359104-13 for the robust=2 \hi \ data overlaid on a DECaLS $grz$ image (cf. Figure~\ref{fig:rps_examples}, lower-left panel). The gas without a stellar counterpart in the south is likely an extension of the ram pressure tail still attached to LS-359104-13 in the north. The lowest contour is $4.5\times10^{18} \; \mathrm{cm}^{-2}$ over 10 \kms \ and each subsequent contour is double the previous.}
    \label{fig:rps_example_wide}
\end{figure}

Upon visual inspection of the \hi \ maps of the detected SAGA satellites it became apparent that four objects display signs of possible ongoing RPS in their \hi \ morphology (Figure~\ref{fig:rps_examples}). As ram pressure exerts a force on the gas in a galaxy, but not its stars, a telltale (necessary, but not sufficient) sign of RPS is a gas tail containing no stars (or only stars that likely formed recently in situ). In addition, RPS tails should occur in only one direction, as a galaxy's gas is swept back into its wake as it moves through the CGM or ICM. In contrast, tidal tails typically form both leading and trailing arms as the tidal field is roughly symmetric about the satellite. Tides are also gravitational and therefore impact all matter equally. Thus, if stars are coincident with gas at the stripping sites, then stellar tails should accompany the gas tails. However, in any real scenario gas stripping is unlikely to be driven purely by tides or ram pressure and they can be difficult to reliably separate in practice. Below we will argue that there are a number of factors that indicate that these four cases (Figure~\ref{fig:rps_examples}) are very likely dominated by RPS, but ultimately the confirmation of this will require further, more detailed observations.

In the cases of NSA-19694 and LS-359104-131 (Figure~\ref{fig:rps_examples}, left panels) we see clear \hi \ tails with no apparent accompanying stellar component. For LS-359104-131, the tail extends only to the SW, with no convincing signs of extensions in any other direction (given the sensitivity and resolution of the data). For NSA-19694 there is a main tail that extends to the NE as well as a small nub on the NW side of the galaxy. Although we do not know the 3-dimensional velocity of the galaxy, and therefore cannot know the direction of any ram pressure exerted on it (we return to this point below), this configuration would make sense if ram pressure is forcing gas to the NW and the two tail components come from opposite sides of the galactic disk. The gas distribution of galaxies undergoing RPS can often form a bow shape as the outskirts of the disk (on either side) are stripped before the center (where the gravitational restoring force is stronger). This bow shape can be slightly asymmetric either because the original \hi \ distribution is asymmetric \citep[which is exceedingly common, e.g.][]{Espada+2011} and/or because the orientation of the galaxy results in one side of the disk taking the brunt of the ram pressure interaction.

Returning to LS-359104-131, in Figure~\ref{fig:rps_example_wide} we show a wider field moment zero map made with the robust=2 \hi \ data cube (\S\ref{sec:VLA_obs}), which has better column density sensitivity for extended emission (but poorer angular resolution). To produce this map we also repeated the masking process with \texttt{SoFiA} using a lower threshold (3.5$\sigma$), accompanied by an even higher reliability criterion ($>$99\%), in an attempt to include low significance \hi \ emission surrounding brighter emission without adding spurious detections.

In Figure~\ref{fig:rps_example_wide} we see an \hi \ cloud with no apparent optical counterpart approximately 13\arcmin \ ($\sim$140~kpc in projection) to the south, on the other side of the host (NGC~7541) and its largest satellite (NSA-637123). Although it first appears that this \hi \ is likely from an ongoing interaction between NGC~7541 and NSA-637123, inspection of their \hi \ morphologies and kinematics indicates that neither of these galaxies are significantly disturbed, and have likely not yet begun strongly interacting, despite their small angular and velocity separations. Thus, it is unlikely that the orphaned gas cloud originated from either of these galaxies and we instead speculate that it is an extension of the RPS tail of LS-359104-131. Indeed, we note that its velocity is only $\sim$60~\kms \ offset from LS-359104-131 and there appears to be low significance indications that LS-359104-131 and the southern \hi \ cloud may connect via a spatially (and kinematically) continuous stream of gas. While a detailed investigation into the nature of this potential connection is beyond the scope of this work, we share the results of our preliminary investigation via an interactive \texttt{Jupyter} notebook hosted on \texttt{Binder}\footnote{\url{https://mybinder.org/v2/gh/ananthankarunakaran/NGC7541_3D/HEAD}}. We leverage the 3D rendering capability of \texttt{IPyvolume} and a masked sub-cube of the NGC7541 field to highlight the low S/N connection between LS-359104-131 and the southern cloud.
Deeper \hi \ imaging might reveal a continuous structure at higher S/N or perhaps the stripped gas does not have a high enough density to self-shield and there may be gaps, especially near NGC~7541, where a portion could already have been photoionized. We note that with a full end-to-end length of $\sim$15\arcmin \ ($\sim$170~kpc, in projection) and an assumed maximum velocity of 275~\kms, if this is one continuous structure then the southern portion must be at least $\sim$600~Myr old.

NSA-343657 and DES-373393030 (Figure~\ref{fig:rps_examples}, right panels) show a slightly different morphology, but are still consistent with RPS. In these cases the VLA data do not have sufficient resolution to clearly resolve tails, but the \hi \ distributions are offset in one direction from the optical centers of these galaxies. DES-373393030 is almost entirely unresolved, but its \hi \ distribution is clearly centered to the NW. If we assume that it has not yet reached pericenter and is still falling towards its host, NGC~1421\footnote{We note here that this galaxy was assigned the incorrect host in SAGA-II. It is just within 300~kpc (projected) of PGC~13646, but is clearly associated with NGC~1421 immediately to the SE.}, then this would mean that the bulk of its \hi \ gas is trailing in its wake. 
Finally, for NSA-343657 the \hi \ distribution forms almost a ``V'' shape, which we suspect is another example of the bow shape typical of RPS, but smeared out by the large and elongated beam shape. However, it should also be noted that there is a slight suggestion of a leading tail, which could make the distribution more of a ``Z'' shape, which would instead suggest a tidal interaction as the cause. The resolution of the current data notwithstanding, the fact that most of the galaxy's gas appears to be residing in its wake (again assuming it is falling towards its host), still means this object is a strong candidate for RPS in progress.

The white arrows in Figure~\ref{fig:rps_examples} point in the direction of the host galaxy of each satellite. All four RPS candidates are significantly less than 100~kpc (projected) from their hosts, comparable to the separation between the MW and LMC. However, we do not know if they are pre- or post-pericenter passage in their orbits. Before pericenter passage, a satellite's RPS tail should point roughly away from the host, whereas shortly after pericenter passage it would point towards the host. In all four cases we see that the tails are roughly (anti-)aligned with the direction of the host, strengthening the notion that they are indeed the result of RPS.

\subsection{A new satellite of UGC~903} \label{sec:new_sat}

\begin{figure*}
    \centering
    \includegraphics[width=\columnwidth]{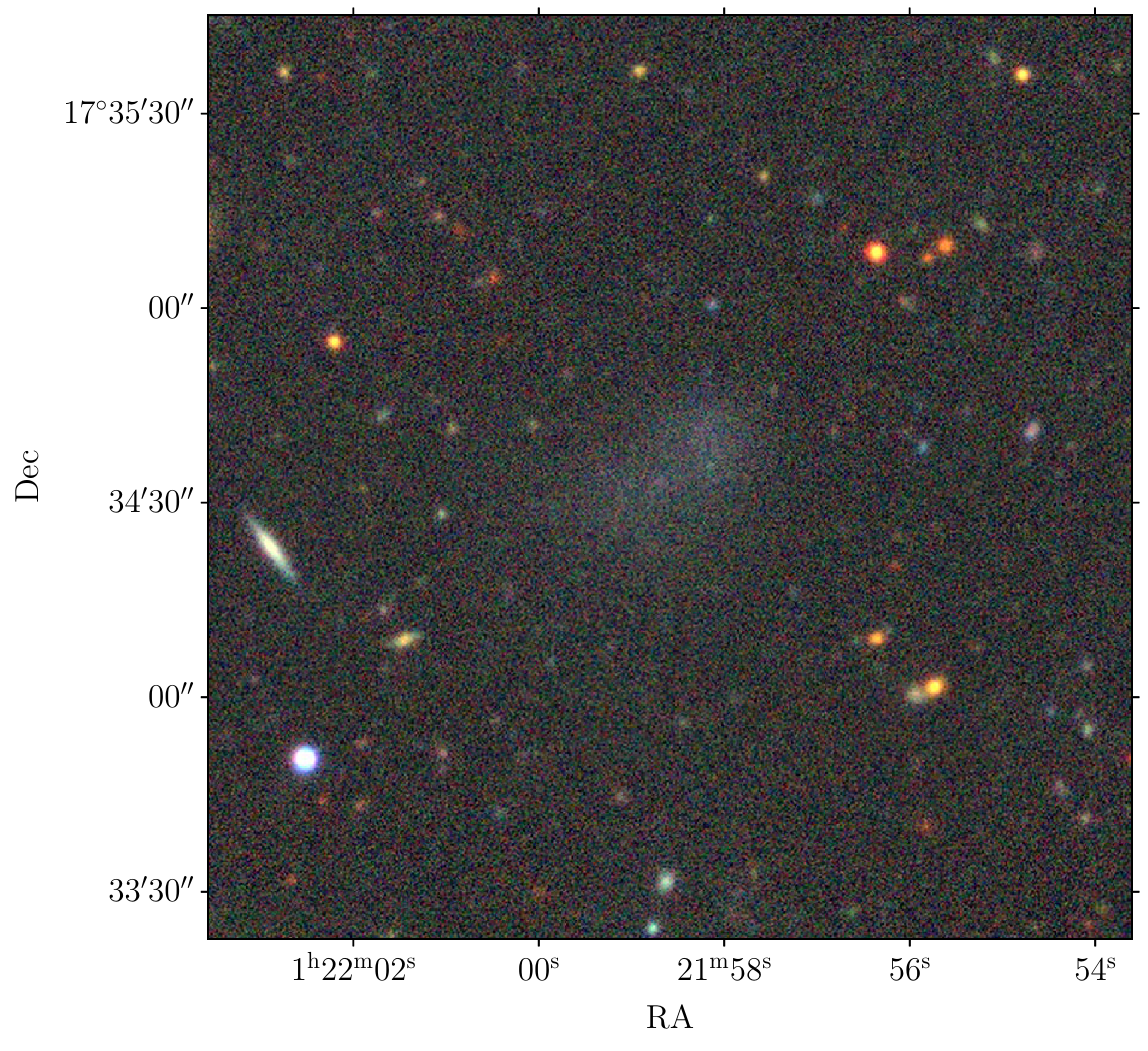}
    \includegraphics[width=\columnwidth]{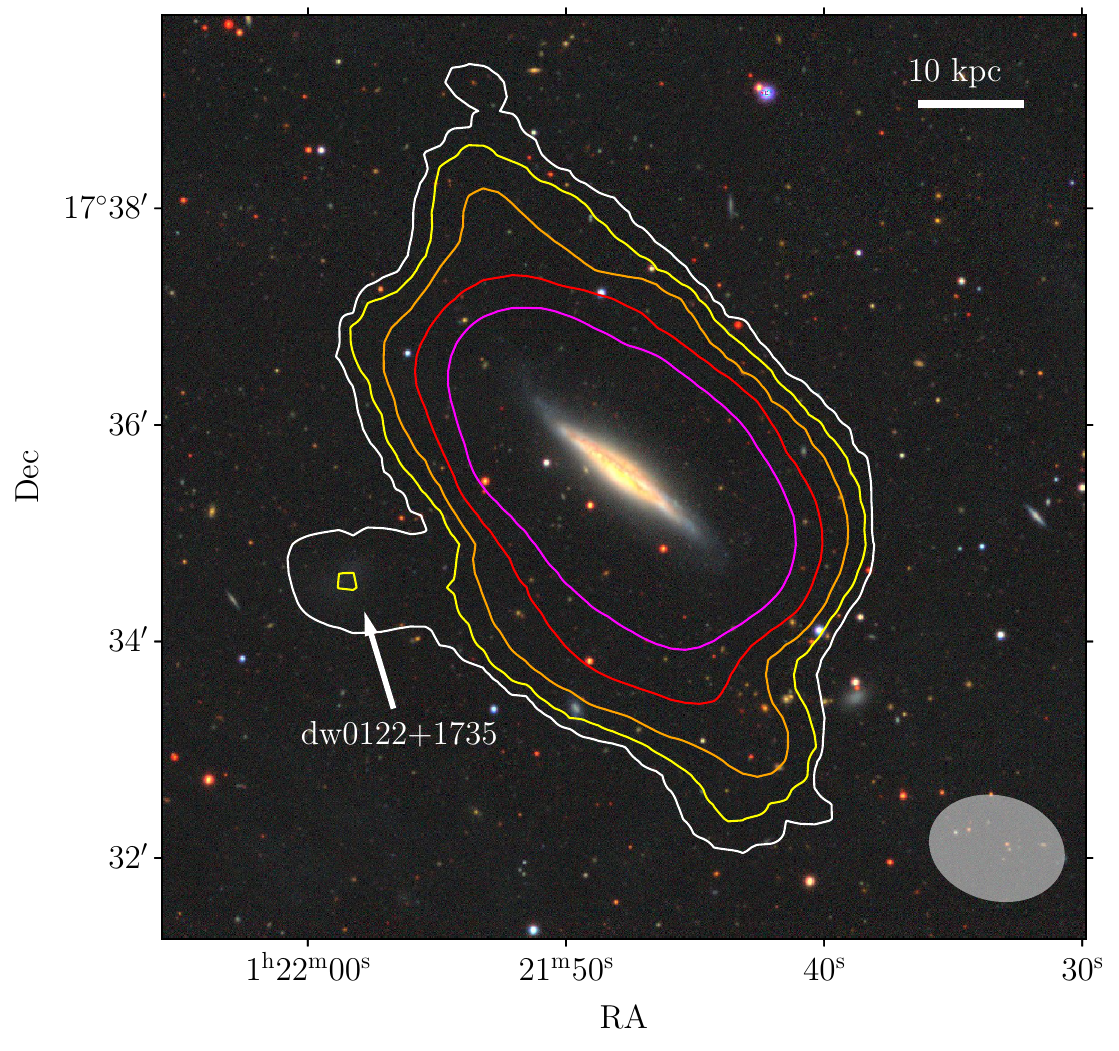}
    \caption{\textit{Left}: DECaLS $grz$ image of dw0122+1735. \textit{Right}: DECaLS $grz$ image of UGC~903 and dw0122+1735 (left of center) with \hi \ moment zero (robust=2) contours overlaid. The contours begin at 3$\sigma$ and each subsequent contour is double the previous. The semi-transparent ellipse in the lower-right corner indicates the synthesized beam size. There is a clear concentration of \hi \ centered on the optical source enlargerd in the left panel.}
    \label{fig:new_sat}
\end{figure*}

In the UGC~903 system we detected an additional satellite that was not in the SAGA-II sample, which we will refer to as dw0122+1735 (Figure~\ref{fig:new_sat}). This is a LSB dwarf approximately 2.7\arcmin \ to the SE of UGC~903 that we identified via its \hi \ line emission, which is lightly blended with that of UGC~903, but clearly concentrated on the LSB optical counterpart (Figure~\ref{fig:new_sat}, right). 

We used \texttt{GALFIT} \citep{Peng+2002,Peng+2010} and DECaLS images to fit the optical properties of this dwarf galaxy. As it is faint and LSB we used a $g+r$ co-add image to fit the structural properties, obtaining a half-light radius $R_\mathrm{eff} = 10.4 \pm 0.3$\arcsec, a S\'{e}rsic index $n=0.36\pm0.03$, and an axial ratio $a/b=0.66 \pm 0.02$. We then refit in each band separately to obtain the photometry, holding the parameters above fixed. As the object is slightly irregular, and not perfectly modeled with a S\'{e}rsic profile, we also perform manual aperture photometry and use the difference between the two as an estimate of the uncertainty. This gives $g = 19.4 \pm 0.4$, $\mu_{0,g} = 25.9 \pm 0.4$, $r = 18.8 \pm 0.3$, and $\mu_{0,r} = 25.2 \pm 0.3$ \citep[extinction corrected following][]{Schlafly+2011}. For consistency with other objects in Table~\ref{tab:sats} we use the stellar mass estimator from \citet{Mao+2021}, which gives $\log M_\ast/\mathrm{M_\odot} = 7.56$. We separate the \hi \ emission from UGC~903 using \texttt{SlicerAstro} to measure an \hi \ mass of $\log M_\mathrm{HI}/\mathrm{M_\odot} = 7.53\pm0.09$. These correspond to a gas fraction of 0.93. We take the flux-weighted average velocity of the \hi \ profile to estimate the radial velocity of dw0122+1735 as 2439~\kms \ ($\Delta cz = -77$~\kms\ relative to UGC~903).

Finally, we note that based on the photometry above and the distance estimate to the UGC~903 system (Table~\ref{tab:hosts}), dw0122+1735 falls well within the standard criteria \citep{vanDokkum+2015} to be considered an ultra-diffuse galaxy (UDG), with an effective radius of 1.9~kpc and $M_V = -13.9$.

\section{Discussion}
\label{sec:discussion}

\subsection{Gas fractions}
\label{sec:discuss_gas_frac}

Figure~\ref{fig:gas_frac} and Table~\ref{tab:sats} both indicate that unlike satellites in the LG, most SAGA satellites have significant \hi \ gas reservoirs.
At least part of the explanation for this difference in gas fractions is the different distribution of stellar masses between SAGA and the LG \citep[cf.][]{Karunakaran+2020a}. SAGA is only complete for relatively massive satellites ($\log M_\ast/\mathrm{M_\odot} \gtrsim 7.5$), while the MW only contains a total of three satellites in that stellar mass range (and meeting the other SAGA selection criteria), while M~31 has six \citep[e.g.][]{McConnachie+2012}. This means that SAGA is only seeing the massive end of the satellite population. For example, if we take the SMC stellar mass as approximately $10^{8.5}$~\Msol\ \citep[e.g.][]{Besla2015} then there are 14 satellites more massive than the SMC in the nine target SAGA systems, which make up about half of the \hi\ detections (three are non-detections), while the remaining half almost all have stellar masses above $10^{7.5}$~\Msol. The gas fractions of both the LMC and SMC appear to be roughly consistent with the observed trend in Figure~\ref{fig:gas_frac}, and it is this class of satellites (or slightly lower mass) that we are generally detecting with our \hi \ observations. 

We also note that although there are no direct equivalents in SAGA to the most the stringent gas fraction limits for LG satellites plotted in Figure~\ref{fig:gas_frac}, this is simply a function of sensitivity and distance. Likely some of the \hi \ non-detections from the SAGA sample are just as gas-poor as many of the LG satellites (of similar stellar mass), but given the distance to the SAGA systems it is not possible to obtain comparable limits on their \hi \ content.

However, despite these differences, meaningful comparisons can be made, especially to simulations that have attempted to reproduce the situation in the LG. For example, \citet{Simpson+2018} use the Auriga simulation suite \citep{Grand+2017} to assess quenching in MW-like systems. 
They find that ram pressure stripping is the dominant mechanism removing gas from satellites and that this acts rapidly ($<$1~Gyr) to quench them.
At a stellar mass of $\log M_\ast/\mathrm{M_\odot} \approx 8$ they predict that half of satellites (within 300~kpc of host) should be gas-poor.
Looking at the satellites in our sample that are $7.75 < \log M_\ast/\mathrm{M_\odot} < 8.25$, we indeed see that five are gas-poor and five are gas-rich. We consider a gas fraction of unity to be the dividing line and err on the side of over-counting gas-poor satellites when there are only upper limits for the \hi \ masses. We note that this is slightly different to the definition used by \citet{Simpson+2018}, which was a fixed threshold of $M_\mathrm{HI} = 10^5$~\Msol \ at all stellar masses.

At slightly higher stellar masses we see that most satellites have significant gas reservoirs. 
This means that either ram pressure stripping is no longer effective at removing their gas, or they are on their first infall (but smaller satellites are not).
Comparing again to \citet{Simpson+2018}, we see that this finding is also consistent. As shown in \citet{Simpson+2018} figure~4, $\log M_\ast/\mathrm{M_\odot} = 8$ is roughly the threshold at which ram pressure stripping becomes inefficient, with almost no satellites at $\log M_\ast/\mathrm{M_\odot} = 9$ being quenched.

These simple comparisons suggest that, despite the first impression that most SAGA satellites are star-forming and most LG satellites are quenched, the fraction and masses of the satellites detected in \hi \ is roughly consistent with the expectations from the Auriga simulations, which were designed to reproduce MW analog systems. However, in Figure~\ref{fig:gas_frac} we also noted a clear trend between the observed gas fraction and project host--satellite separation. This trend is presumably a signature of the gas removal mechanism. Given that there is expected to be considerable host-to-host scatter in CGM density \citep[e.g.][]{Ramesh+2023,Saeedzadeh+2023} it is unclear if ram pressure stripping should produce such a uniform trend when data are compiled from several different systems. We will revisit this point in more detail \S\ref{sec:sim_comp}.

\subsection{Star formation rates}
\label{sec:discuss_SFR}

\begin{figure*}
    \centering
    \includegraphics[width=0.67\textwidth]{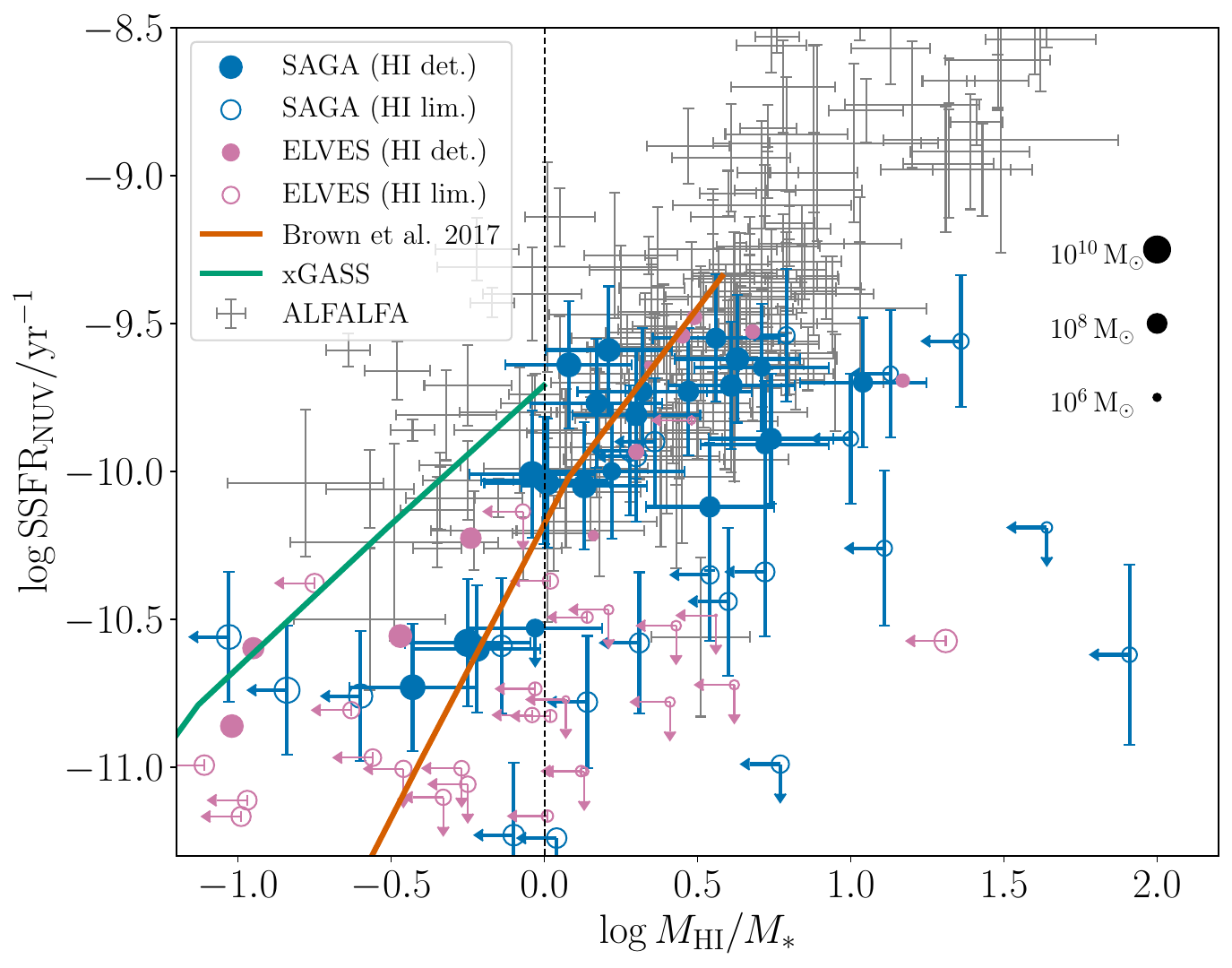}
    \caption{Specific star formation rate of SAGA satellites plotted as a function of their gas fractions. \hi \ detections are shown as filled points and non-detections as unfilled points. The size of the points is scaled by $\log M_\ast$, with examples shown on the right side of the plot. ELVES satellites with \hi \ and NUV SFR measurements \citep{Carlsten+2022,Karunakaran+2022} are included for comparison.  
    Grey error bars in the background show a stellar mass-matched comparison field sample from ALFALFA. 
    The orange line shows the \citet{Brown+2017} relation from satellites in groups of $12 < \log M_\mathrm{halo}/\mathrm{M_\odot} < 13$, and the green line shows the xGASS relation for field galaxies in the stellar mass range $9 < \log M_\ast/\mathrm{M_\odot} < 10$ \citep{Catinella+2018}.
    The vertical black dashed line indicates a gas fraction of unity.}
    \label{fig:gas_frac_SSFR}
\end{figure*}

The fit between H$\alpha$ and NUV-derived SFR estimates in Figure~\ref{fig:SFR_comp} shows that although the two metrics mostly agree, there is a slight tendency for H$\alpha$ to underpredict the SFR relative to NUV at low SFRs and the reverse at high SFRs. A similar trend was first noted for galaxies with low SF activity ($\mathrm{SFR} \lesssim 0.1~\mathrm{M_\odot\,yr^{-1}}$) in \citet{Lee+2009}. Even though that work used FUV (which is more directly comparable to H$\alpha$) rather than NUV, the slope that we fitted is consistent within the uncertainties. \citet{Lee+2009} also performed a thorough accounting of all conceivable sources of this discrepancy, but concluded that no one factor could explain it, with the possible exception of a varying stellar initial mass function. We have few detections in the low SFR regime and further investigation of this offset is beyond the scope of this paper. 

When the SSFRs of the SAGA satellites were investigated in Figure~\ref{fig:SSFR}, there were no objects with clearly elevated SF. In particular, we note that none of the four satellites that appear to be currently undergoing RPS have elevated SSFRs, nor do they show any obvious signs of fronts of SF. This suggests that ram pressure experienced by these satellites is insufficient to produce the compression-induced waves of SF sometimes seen in higher mass galaxies undergoing RPS \citep[e.g.][]{Gavazzi+1995,Vulcani+2018,Boselli+2021}. This finding is consistent with other work that has found that this type of SFR-enhancement is ineffective in galaxy groups \citep[e.g.][]{Vulcani+2021,Roberts+2021}, presumably because of the lower density of the intergalactic medium and the lower velocities involved.

In Figure~\ref{fig:SSFR} we also saw that there is a downturn in the SSFR of the \hi-detected satellites at small (projected) host--satellite separation ($D_\mathrm{proj} \lesssim 100$~kpc). In Figure~\ref{fig:gas_frac_SSFR} we show that this is likely a result of the suppressed gas fractions at small projected separations. At approximately a gas fraction of unity there is a sharp drop in SSFR of around 0.5~dex. Despite \hi \ not being the direct fuel for SF, this transition is so marked that there is only one SAGA object with a gas fraction below unity and $\log \mathrm{SSFR/yr^{-1}} > -10.5$. 

To verify that this transition is a feature specific to satellites, we drew a random selection of ALFALFA galaxies from the \citet{Durbala+2020} catalog that is 10 times the size of our SAGA sample, but matched in terms of its distribution of stellar masses.\footnote{Here we used the \citet{Taylor+2011} stellar mass estimates and the GALEX NUV SFR estimates from \citet{Durbala+2020}.} Unsurprisingly we see that the ALFALFA sample extends to higher SFR and gas fraction values. This is a predominantly a field galaxy population \citep[e.g.][]{Guo+2017,Jones+2020} and its intrinsic \hi-selection favors gas-rich galaxies. However, crucially we see a smooth extension of the main relation between gas fraction and SSFR below a $\log M_\mathrm{HI}/M_\ast = 1$ where there are no SAGA satellite galaxies, suggesting that this disconnect is not merely a feature of the relationship, but rather that it is specific to satellites. The sharpness of this transition also suggests that it is the underlying cause of the trend seen in Figure~\ref{fig:SSFR}, rather than there being a direct correspondence between host--satellite separation and SSFR, e.g. as a result of strong tidal interactions. 

In Figure~\ref{fig:gas_frac_SSFR} we also overlay two literature measurements of the relationship between gas fraction and SSFR. In green we show the relation from the extended GALEX Arecibo SDSS sample \citep[xGASS,][]{Catinella+2018}, which is another field sample but is not \hi-selected (unlike ALFALFA). This relation roughly overlaps the ALFALFA points at low gas fractions, but if extrapolated higher would be slightly to the left of the ALFALFA population (a result of the \hi-selection of ALFALFA). The orange relation is from the spectral line stacking work of \citet{Brown+2017} and shows the average relation for satellites in groups with halo masses in the range $12 < \log M_\mathrm{halo}/\mathrm{M_\odot} < 13$. This relation has a slightly lower SSFR at a gas fraction of unity than the xGASS relation, and then follows a much steeper decline in SSFR with decreasing gas fraction. This relation is a remarkable match to the trend in our \hi \ detections and again suggests that we are seeing behavior that is specific to satellites.

\citet{Brown+2017} argued that their results indicate that the timescale for gas removal from satellites was considerably shorter than that for the subsequent shutdown of SF. Several other works have also made similar suggestions \citep[e.g.][]{Wetzel+2013,Oman+2021}. However, if this were the case then we would expect to see a population of \hi-poor satellites with elevated SSFRs (for their gas fractions). Instead we see the opposite, a population of satellites with low SSFRs for their gas fractions, relative to field galaxies (i.e. ALFALFA and xGASS). This suggests that the shutdown of SF does operate on a similar timescale to the gas removal mechanism and that satellites are not largely stripped of their gas before SF starts to shutdown, at least not in the mass regime applicable to SAGA.

Finally, in Figure~\ref{fig:gas_frac_SSFR} we include the \hi \ values for the ELVES sample from \citet{Karunakaran+2022}. ELVES \citep{Carlsten+2022} has fewer MW-like hosts than SAGA, but they are significantly closer ($D<12$~Mpc instead of $25<D/\mathrm{Mpc}<40$ for SAGA) and thus ELVES is complete to much lower satellite masses. This sample has fewer detections than our SAGA sample and it is difficult to draw conclusions from so few values. However, curiously they do not appear to follow the same trend, with the low gas fraction points mostly lying between the two comparison relations. It may be that the current small samples from both SAGA and ELVES paint a somewhat misleading picture that may not be borne out by future observations. Ultimately, more data are needed to draw robust conclusions from this analysis. 

In Appendix~\ref{sec:gas_frac_SSFR_sims} we also explore the trend between gas fraction and SSFR in simulations. However, the comparison to SAGA is complicated by their offset gas fractions (see \S\ref{sec:sim_comp}).

\subsection{An overlooked, un-quenched satellite}

The newly identified satellite of UGC~903 (dw0122+1735) presented in \S\ref{sec:new_sat} was not included in the SAGA sample. In this section we briefly discuss the significance and the likely reason for this omission.

The (projected) separation between dw0122+1735 and UGC~903 is 2.7\arcmin \ (30~kpc). This means that it was not eliminated as a result of the 10~kpc exclusion radius that SAGA enforced in the vicinity of the hosts. At $r =18.8$  it is also significantly brighter than the nominal magnitude limit of SAGA \citep[$r = 20.75$,][]{Mao+2021}. Thus, the LSB of this object is the most likely cause of it being missed during the SAGA search for satellite candidates. There is also considerable Galactic cirrus in the vicinity of UGC~903, which may have contributed to it being overlooked.

There has been some debate in the literature \citep{Karunakaran+2021,Karunakaran+2023,Font+2022,Carlsten+2022} about whether or not SAGA is systematically missing LSB satellites that would otherwise meet their nominal selection criteria. This satellite, dw0122+1735, is likely an example of exactly that, but one lone example does not settle this debate, which is a statistical consideration. However, it does raise an important complication. \citet{Font+2022} and \citet{Carlsten+2022} both argued that SAGA under-counts the number of quenched satellites because (they argue) quenched dwarfs tend to have lower surface brightness. But, the possibility of missing un-quenched galaxies (due to the same hypothesized observational bias) was not considered. Although dw0122+1735 is not presently forming stars (it is undetected in UV and H$\alpha$), it does contain an appreciable neutral gas reservoir, which could fuel SF in the future, and therefore cannot be considered quenched. If other equivalent objects have been missed then it is even possible that the quenched fraction of satellites calculated from SAGA is underestimated and could disagree even more severely with simulations than indicated in \citet{Karunakaran+2021}.

Finally, as noted in \S\ref{sec:new_sat}, dw0122+1735 meets the usual surface brightness and physical size criteria to be considered a UDG. Gas-bearing UDGs are not uncommon \citep[e.g.][]{Leisman+2017,Jones+2018,Janowiecki+2019,Karunakaran+2020b}, however, this particular UDG is noteworthy as it is a satellite, gas-rich, and relatively red ($g-r = 0.6$), the combination of which makes it somewhat unusual. There is one similar object in the sample of \citet{Karunakaran+2020b}, but it is also an order of magnitude more massive.

\subsection{Ram pressure stripping toy model} \label{sec:toy_model}

\begin{figure}
    \centering
    \includegraphics[width=\columnwidth]{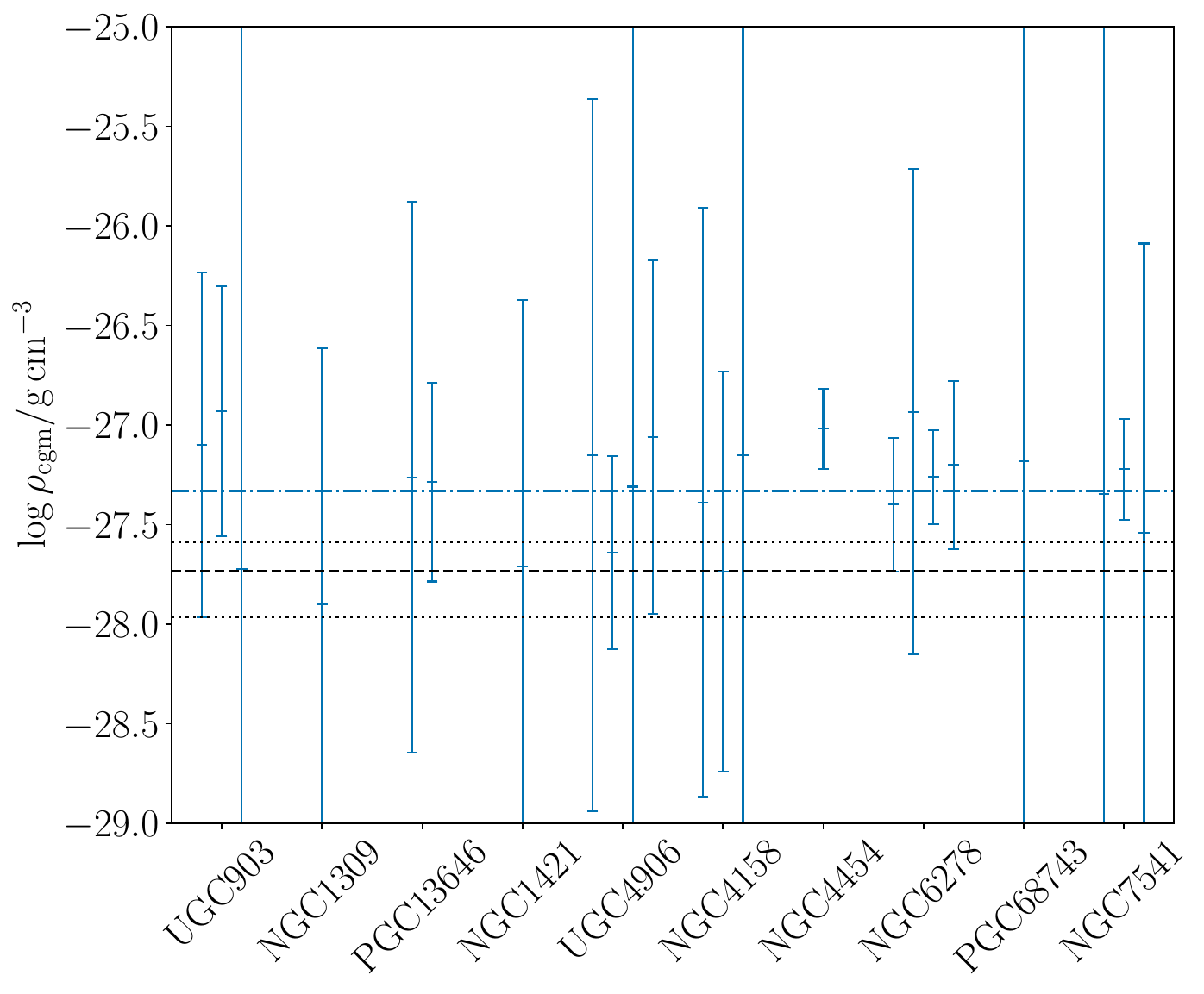}
    \caption{Estimates of CGM density from the satellites in each SAGA system that are detected in \hi \ (blue error bars). The blue horizontal dot-dash line indicates the weighted average of all the points ($\log \rho_\mathrm{cgm}/\mathrm{g\,cm^{-3}} \approx -27.3$) and the black dashed line (and dotted lines) indicate the CGM density estimate for the MW from \citet{Salem+2015}.}
    \label{fig:CGM_den}
\end{figure}

The \hi \ observations of the four satellites apparently in the process of being ram pressure stripped (Figure~\ref{fig:rps_examples}) are not at sufficiently high resolution to allow us to accurately measure the truncation of the leading edge of their \hi \ disks. The truncation of the \hi \ disk is a direct result of the impact of ram pressure and in the case of the LMC has been used to estimate the CGM density of the MW \citep{Salem+2015}. 

Although we cannot use the detailed \hi \ properties of the satellites to estimate the CGM densities of the SAGA hosts, we can use their combined global \hi \ properties if we make some simplifying assumptions and approximations. To do this we construct a toy model inspired by that of \citet{Wang+2020}. First, we assume that all satellites were previously field dwarfs, and when they originally fell into their current system, had \hi \ content typical of field dwarfs of similar stellar mass. We also assume that all subsequent \hi \ mass loss is the result of RPS from passing through the CGM of their present day hosts. Starting with these assumptions we can construct a toy model to estimate the CGM density as described below.

In what follows we will only consider the satellites that were detected in our \hi \ observations. This is because for these satellites it is reasonable to assume that they recently fell into the group and/or are still undergoing RPS. For the satellites devoid of gas, they could potentially have been satellites for many gigayears and their present day gas-poor state might be the result (at least in part) of effects other than RPS. For example, they may have been partially ram pressured stripped and then consumed the remainder of their gas through SF.

To approximate the ram pressure stripping threshold we will use the common Gunn \& Gott expression:
\begin{equation}
    \rho_\mathrm{cgm} v^2 = 2 \pi G \left( \Sigma_\ast(R) + \Sigma_\mathrm{gas}(R)\right) \Sigma_\mathrm{gas}(R)
    \label{eqn:GunnGottRPS}
\end{equation}
where $\rho_\mathrm{cgm}$ is the CGM mass density, $v$ is the velocity of the satellite through the CGM, $\Sigma_\ast$ is the stellar surface density, $\Sigma_\mathrm{gas}$ is the gas surface density, $R$ is the galactocentric radius, and $G$ is the gravitational constant. We note that this form of the expression implicitly assumes a 100\% self-gravitational coupling for the gas in the disk.

For the stellar surface density we simply assume an exponential disk with total mass estimated by \citet{Mao+2021} for each satellite. The scale radius is estimated via the stellar size--baryonic mass relation of \citet{Bradford+2015}. Although we could estimate this more accurately by re-fitting the light profiles of each satellite, in practice $\Sigma_\ast$ is not particularly important as the \hi \ distribution is typically much more extended than the stellar distribution and the $\Sigma_\mathrm{gas}^2(R)$ term dominates over the $\Sigma_\ast(R)\Sigma_\mathrm{gas}(R)$ term. Thus, this simple approximation is sufficient for our toy model.

To estimate the original (at infall) \hi \ mass of each satellite we use the stellar mass--gas mass (where $M_\mathrm{gas}=1.4M_\mathrm{HI}$) scaling relation of \citet{Bradford+2015} as this focuses on field galaxies in the appropriate stellar mass range. 
In addition to knowing the total original gas mass, we must also know how it was originally distributed in order to evaluate Equation~\ref{eqn:GunnGottRPS}. Fortunately, the radial distribution of \hi \ in galaxies is extraordinarily similar (after normalization of the scale radius and amplitude) over a wide range of masses \citep{Wang+2014,Wang+2020}. We use the averaged profile from \citet{Wang+2016} as the template profile for the original radial distribution of \hi \ in all of the SAGA satellites. The \hi \ radii are calculated using the \citet{Wang+2016} \hi \ size--mass relation.

The final piece of information that we need is to know how fast the satellites are traveling through the CGM. In practice the relevant quantity is the maximum value of $\rho_\mathrm{cgm} v^2$ (which presumably occurs at pericenter) as this corresponds to the smallest value of $R$ where the gas distribution will be stripped. The only constraint that we have on the value of $v$ at the orbital pericenter is a lower limit from the present day value of $\Delta cz$, the difference in the line-of-sight velocities of the host and satellite (Table~\ref{tab:sats}). For the upper limit we choose 275~\kms \ as this is the limit where the SAGA selection criteria would no longer consider a satellite candidate to be bound to the host.

With a means to estimate all of the relevant quantities we follow a Monte Carlo approach to estimate $\rho_\mathrm{cgm}$. The following steps are repeated 1000 times for each satellite in our sample.
\begin{enumerate}
    \item The initial \hi \ mass $M_\mathrm{HI,init}$ is drawn from the stellar mass--gas mass relation (with scatter), with the requirement that $M_\mathrm{HI,init} \geq M_\mathrm{HI,0}$, where the 0 subscript denotes the value today.
    \item The \hi \ scale radius is drawn from the \hi \ size--mass relation (with scatter) based on $M_\mathrm{HI,init}$.
    \item The stellar distribution scale radius is drawn from the stellar size--baryonic mass radius (with scatter) based on $M_{\ast}+1.4M_\mathrm{HI,init}$.
    \item The satellite maximum velocity relative to the CGM is drawn from a uniform distribution over the range $|\Delta cz| \leq 275$~\kms.
    \item Based on the amount of \hi \ mass that is missing, $M_\mathrm{HI,init}-M_\mathrm{HI,0}$, the RPS truncation radius of the gas distribution is calculated, along with the corresponding CGM density. 
\end{enumerate}

The median values of $\rho_\mathrm{cgm}$, with 1$\sigma$ error bars, are plotted for each of the satellites detected in \hi \ in Figure~\ref{fig:CGM_den}. The horizontal blue dash-dot line is the average value for all satellites and hosts ($\log \rho_\mathrm{cgm}/\mathrm{g\,cm^{-3}} \approx -27.3$). For comparison we consider an estimate of the MW CGM density (Figure~\ref{fig:CGM_den}, horizontal black dashed line). \citet{Salem+2015} calculated the peak ram pressure experienced by the LMC based on the truncation of its \hi \ disk along its leading edge, which in combination with its orbital parameters provided an estimate of the MW CGM density at the pericenter of the LMC's orbit ($\sim$50~kpc). The average CGM density value that we calculate is slightly higher than that of \citet{Salem+2015}, which might indicate that we are overestimating somewhat as it is unlikely that most of the SAGA satellites have passed closer to their host than 50~kpc (although most of their orbital parameters are unknown). However, considering the long list of significant assumptions that we needed to make above our estimate is still relatively close to that of the MW. 

Although there are many assumptions, the step that results in the majority of the uncertainty seems to be step 4, randomly drawing the maximum velocity of the satellite through the CGM. Both the mass surface density and the velocity are squared quantities in Equation~\ref{eqn:GunnGottRPS}, however, owing to the uniformity of most \hi \ disks, the latter is is much more uncertain. The satellites with the smallest error bars are those with the largest values of $|\Delta cz|$, i.e. the tightest constraints on the velocity. Similarly, the average value of $\rho_\mathrm{cgm}$ is highly sensitive to the upper limit for the velocity that we set above. For example, increasing this value to $\sim$350~\kms \ would result in agreement with the range of values for $\rho_\mathrm{cgm}$ for the MW. Therefore, we cannot draw strong conclusions from this toy model, but we note that the similarity to the MW values is remarkable given the number of assumptions. Very tight constraints via this approach would require detailed knowledge of the orbital parameters of the satellites, which unfortunately will not be feasible for most SAGA systems. However, a considerably larger sample than the one presented here would statistically sample the distribution of orbital parameters, and could be matched with theoretical priors for those parameters in order to construct a tighter constraint. In addition, a larger sample is likely to contain a significant number of satellites with large values of $|\Delta cz|$, which would further strengthen the constraint.

In the future we plan to obtain higher-resolution \hi \ imaging of the four satellites apparently undergoing RPS such that we can explicitly measure their truncation radii, thereby eliminating the need for most of the steps above and placing a direct constraint on the maximum ram pressure that they have experienced. The corresponding maximum velocity will still present a challenge for estimating the corresponding CGM density, however, with only four satellites we will attempt to construct tighter constraints on these velocities based on the directions of their ram pressure tails, their current host-satellite separations, and the values of $\Delta cz$.

\subsection{Gas fractions and projected separations in simulations} \label{sec:sim_comp}

Over the past few years there have been several investigations of the quenching of satellites in MW-like systems in cosmological and zoom-in simulations \citep[e.g.][]{Simpson+2018,Digby+2019,Akins+2021,Wright+2021,Engler+2023}. Although there is not a consensus on exactly how satellites quench in simulations, RPS is noted as a physical process that is likely important.

\citet{Engler+2023} analyzed the satellite populations in MW and M~31-like systems in the TNG50 simulation finding broad agreement between the properties of these simulated satellites and genuine LG satellites. They also measured a decline in \hi \ mass as a function of host--satellite separation for relatively massive ($M_\ast > 10^{8.5}$~\Msol) satellites in TNG50, analogous to that shown in Figure~\ref{fig:gas_frac}. They also find clear examples of ongoing RPS, with similar morphologies to the observations in Figure~\ref{fig:rps_examples}. Similar RPS tails have also been noted in the DC Justice League simulations \citep{Akins+2021}. \citet{Engler+2023} find that only a few percent of satellites should host visible RPS tails at $z=0$. As we have identified four strong candidates out of 45 satellites, it seems that the occurrence rate might be a factor of a few times higher. However, they consider satellites down to $M_\ast = 5\times10^6$~\Msol, and the timescale for gas loss and quenching is expected to scale inversely with satellite mass.

As noted by \citet{Crain+2017} for the Evolution and Assembly of GaLaxies and their Environments (EAGLE) simulations, and as we will discuss further below, it is common for current simulations to systematically under-predict the \hi \ masses of low-mass galaxies. However, as pointed out by \citet{Wright+2021} in their study of satellite quenching in EAGLE, when dealing with relative gas masses (e.g. between quenched and star-forming galaxies), or changes in gas mass, this bias is of minimal importance as the physical processes that are removing the gas are likely quite distinct from those driving this bias. In order words, it is the transition that is of most interest here and the simulations can still provide insight even if the absolute \hi \ content is offset.

\begin{figure}
    \centering
    \includegraphics[width=\columnwidth]{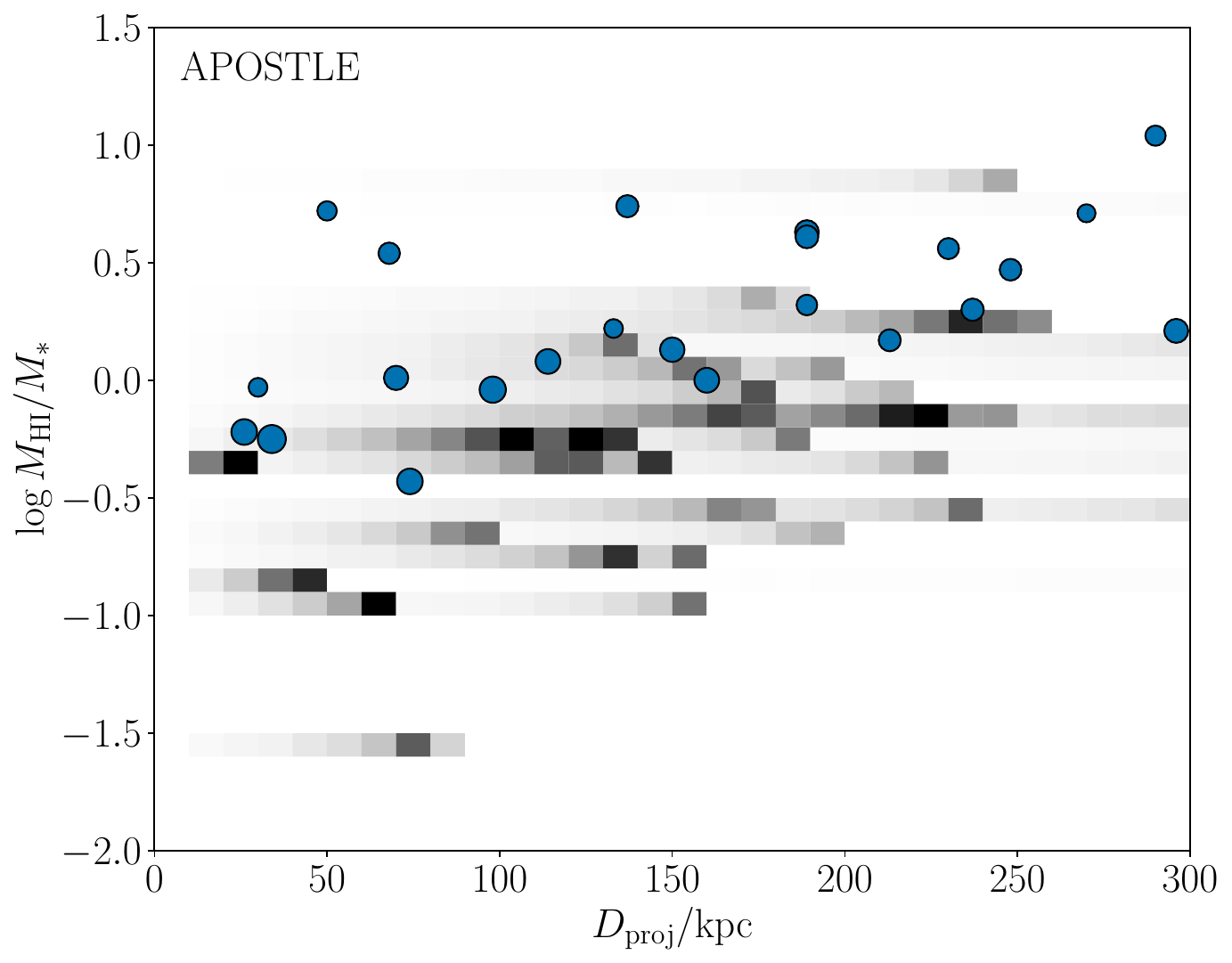}
    \includegraphics[width=\columnwidth]{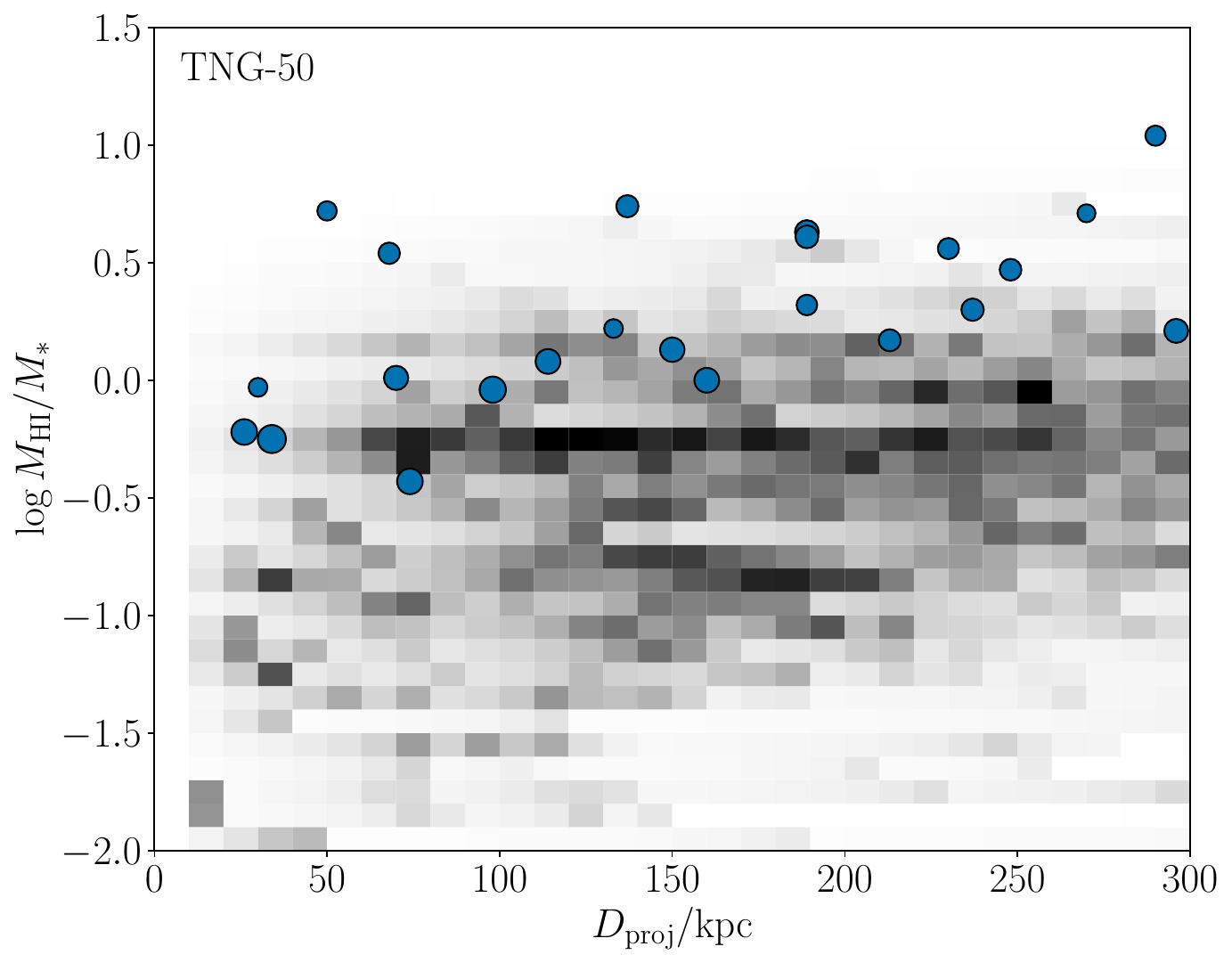}
    \caption{Comparison of observed gas fractions of SAGA satellites detected in \hi \ (blue circles) as a function of projected host--satellite separation and the greyscale maps showing the corresponding values averaged over 10,000 random orientations of MW-like satellite systems in the APOSTLE (top) and TNG50 (bottom) simulations.}
    \label{fig:sim_comp}
\end{figure}

To make a more direct comparison between the observed properties of satellites and theoretical expectations, we make use of the APOSTLE \citep{Fattahi+2016,Sawala+2016} and TNG50 \citep{Pillepich+2019,Nelson+2019a,Nelson+2019b} simulations. APOSTLE is dedicated to simulating systems that are comparable to the MW and LG, and is thus well-matched to our sample. It consists of 12 simulation volumes analogous to the MW and M~31 system, or a total of 24 MW-like satellite systems. For simplicity we use only the intermediate resolution L2 run of APOSTLE and require satellites to have a stellar mass of $\log M_\ast/\mathrm{M_\odot} > 7.1$ to roughly mimic the SAGA selection criteria and to ensure that each consists of at least 100 star particles in the simulation (which have masses of $1.1\times10^5$~\Msol). This gives a sample of 114 satellites. 

As TNG50 is not focused on MW-like systems we had to define our own set of hosts, taking all those in the stellar mass range $10.5 < \log M_\ast/\mathrm{M_\odot} < 11.5$ (cf. Table~\ref{tab:hosts}). Potential hosts with a more massive neighbor within 300~kpc were removed, as well as those that were themselves satellites within a massive halo ($\log M_\mathrm{total}/\mathrm{M_\odot} > 13$). Finally, as TNG50 contains some objects that have significant gas mass, but almost no dark matter, we place a lower limit of $10^8$~\Msol \ on the sub-halo mass of satellites to avoid erroneously including any such objects as satellites. This gives a total of 246 TNG50 hosts\footnote{Note that centrals that meet our criteria, but host no satellites meeting our criteria, are not included in the host count.} and 2030 satellites (with $\log M_\ast/\mathrm{M_\odot} > 7.1$). With this mass cut each satellite stellar component is made up of at least 100 particles (of mass $8.5\times10^4$~\Msol).

Satellite stellar masses are taken as the sum of all stellar mass particles that are gravitationally bound to the sub-halo. We estimate \hi \ masses for gas particles in APOSTLE following the convention for simulations using the EAGLE galaxy formation model established by \citet[][see also \citealp{Crain+2017} for further discussion]{Bahe+2016}, combining the partitioning scheme for neutral-to-ionized gas of \citet{Rahmati+2013} and the atomic-to-molecular partitioning scheme of \citet{BlitzRosolowsky2006}. For TNG50, we instead adopt the \hi \ profiles of \citet[][see also \citealp{Diemer+2018}]{Diemer+2019}. We use their ``projected'' (or ``map'', i.e. not ``volumetric'') profiles -- see \citet{Stevens+2019} for a discussion motivating this choice -- and the \citet{GnedinDraine2014} atomic-to-molecular partitioning scheme.

For both simulations we extracted the total \hi \ mass within 1, 5, 10, 30, and 50~kpc of the center of mass of each satellite. The \hi \ mass of each galaxy was then taken as the value within the first step that exceeds four times the stellar half mass radius. This was compared to instead selecting the step where the \hi \ surface density dropped below 1~\Msol/pc$^2$, and was found to be mostly consistent. However, the surface density approach gave spurious results for some galaxies (likely those with central holes in their \hi \ distributions), and thus the simpler approach was favored for being more resilient to outlying cases.

Figure~\ref{fig:sim_comp} shows the comparison of the gas fractions of the \hi-detected SAGA satellites (blue points) and the APOSTLE (top) and TNG50 (bottom) simulations as the background intensity maps. The maps for the simulations were generated by projecting each system on to 10,000 different, randomly chosen lines of sight. Additional selection criteria were enforced to match those from SAGA \citep{Mao+2021}. With each different projection only satellites with a projected separation from their host of $10<D_\mathrm{proj}/\mathrm{kpc}<300$ and a velocity separation $|\Delta cz| < 275$~\kms \ were retained. We also required the \hi \ masses of satellites to be above $10^{7.5}$~\Msol \ in order to roughly approximate our \hi \ detection limit (which in practice varies somewhat between different satellites and systems).\footnote{This value is intentionally about 0.5~dex lower than the typical detection limit in Table~\ref{tab:sats}. As will be discussed later, galaxies in both simulations have systematically lower \hi \ masses then real galaxies, precluding an exact like-for-like comparison.}

Both simulations show a qualitatively similar behavior to the observations of the SAGA sample, with gas fractions declining with projected separation. In the case of TNG50 the distribution of gas fractions appears constant to $D_\mathrm{proj} \approx 100$, after which a significant downturn begins. For APOSTLE there is a gradual change in gas fraction between $D_\mathrm{proj}=300$~kpc and 0~kpc (but the interpretation suffers from the small sample size compared to TNG50). The SAGA sample with \hi \ measurements is also currently quite small and the data could reasonably be fit with either of the functional forms discussed above, a plateau with a downturn around 100~kpc, or a roughly linear decline across the entire range of $D_\mathrm{proj}$ values.

A clear difference between the observations of the SAGA satellites and those in both simulations is that on average the observations lie at higher gas fractions (by approximately 0.5~dex). Even at $D_\mathrm{proj} = 300$~kpc the simulations do not reach the values seen in the observations, which implies that this discrepancy must extend into the field. As noted above, it is a known issue \citep[see also e.g.][figure 1]{Oman+2019} that simulations currently tend to under-predict the \hi \ content of low-mass galaxies. Thus, the satellites in the simulations likely never had as much \hi \ as the SAGA satellites did when they first fell into their present day systems.

Despite this offset, the existence of analogous trends in gas fraction in our observations and the APOSTLE and TNG50 simulations, as well as the reproduction of many other satellites properties \citep[e.g.][]{Sawala+2016,Engler+2023}, indicates that we can use the simulations to investigate the processes that are likely driving the observed trends. For many of the intermediate mass ($6 \lesssim \log M_\ast/\mathrm{M_\odot} \lesssim 8$) satellites in SAGA the dominant factor is likely RPS \citep[cf.][]{Digby+2019}. However, at higher masses there is likely a transition \citep[e.g.][]{Fillingham+2019} where the satellite's potential well becomes deep enough that RPS is no longer an effective means of quenching. Even if these high mass satellites are observed to have RPS tails, RPS likely cannot remove enough gas to quench their SF. Instead starvation could be the dominant mechanism that slowly shuts off SF, leading to much longer quenching timescales. 

In future work we hope to expand the sample to permit more detailed comparisons with simulations, such that we can begin to understand where this transition occurs. In particular, with a larger sample (both from simulations and observations) we could attempt to identify the radius at which the suppression of gas fraction and SFR begins for satellites of a fixed stellar mass. This would indicate the point where ram pressure begins to have a meaningful impact on the satellites. If ram pressure is the dominant mechanism then this characteristic radius should get progressively smaller for higher mass satellites, with the quenching mechanism eventually transitioning to starvation (as outlined above). Such findings would be strong evidence for ram pressure as the driving force for satellite quenching, with a transition to starvation at higher masses.

\section{Conclusions}
\label{sec:conclusions}

We have fully imaged eight SAGA systems (42 satellites) in both \hi \ and H$\alpha$, providing tracers of both the ongoing star formation and the gas that fuels it. A ninth system (3 satellites) has been partially observed. We observe a clear decline in gas fraction ($M_\mathrm{HI}/M_\ast$) as a function of projected host--satellite separation. Over the full range of separations considered ($\sim$25-300~kpc) the typical gas fraction changes by around an order of magnitude. A similar trend is seen in our comparisons to the APOSTLE and TNG50 simulations, however, there remains a $\sim$0.5~dex offset in gas fraction for all satellites, preventing a full like-for-like comparison. We also find that the fractions of gas-rich and gas-poor satellites are consistent with those from the Auriga simulations, although a larger sample is needed for detailed comparison. These simulations predict that RPS is the dominant mechanism driving gas removal, and it is likely that we are indirectly seeing its effects through this trend in gas fraction.

Four satellites also exhibit signs of likely ongoing RPS, in the form of clear one-sided \hi \ gas tails (Figure~\ref{fig:rps_examples}). All four are within 75~kpc (projected) of their host galaxy and are likely close to the pericenter of their orbits, where the CGM density and ram pressure are expected to be highest. In one case the tail appears to extend back around to the opposite side of the host (Figure~\ref{fig:rps_example_wide}). Only a handful of examples of satellites undergoing RPS in MW-mass systems are known, and thus these four represent a significant expansion of that sample. Furthermore, identifying four examples in only eight systems suggests this is a remarkably common phenomenon and that further observations of MW-like systems have a strong chance of uncovering additional examples. We plan to obtain higher-resolution \hi \ imaging in order to confirm and further explore the morphology of these features.

We constructed a RPS toy model to attempt to estimate the 3-dimensional mass density of the CGM based on the current and assumed initial gas content of the satellites. This approach (\S\ref{sec:toy_model}) contains some considerable assumptions and uncertainties, but permits an order of magnitude estimate of the CGM density. We find a value about 0.5~dex higher than estimated for the MW \citep{Salem+2015}. The constraining power of any individual galaxy depends strongly on how well aligned its orbital velocity vector is with the line-of-sight. We hope to place better constraints on the CGM density in these systems with an expanded sample as well as higher resolution observations that may permit the \hi \ disk truncation to be spatially resolved.

In the UGC~903 system our \hi \ observations revealed a previously unknown satellite that was not included in the SAGA sample, but does meet the nominal selection criteria. This satellite was likely overlooked because it has a very low surface brightness and, in fact, fits the standard definition of a UDG. As it is gas-bearing, it cannot be considered quenched as it may go on to form more stars in the future, but it is undetected in H$\alpha$ and UV and is therefore not presently forming stars. This finding complicates an active debate about whether SAGA is systematically missing LSB satellites and whether this would lead to a suppression of the quenched fraction observed \citep{Karunakaran+2021,Karunakaran+2023,Font+2022,Carlsten+2022}.

Our H$\alpha$ imaging produced SFR estimates that were largely consistent with the NUV estimates from \citet{Karunakaran+2021}. 
Thus, in this case the main advantage of H$\alpha$ over UV as a SF tracer is the angular resolution and we will present an assessment of the H$\alpha$ morphologies of the satellites in a subsequent paper. 
In Figure~\ref{fig:SFR_comp} we tentatively identified a trend analogous to that of \citet{Lee+2009}, where at very low SFRs there is a tendency for H$\alpha$ to slightly underpredict the SFR relative to UV. However, we currently have few detections in this regime.

In line with the trend in gas fractions, there is also a suppression of SSFR at small projected separations. This appears to be driven by the lower gas fractions at small separations. There is a sharp decline in SSFR for satellites with gas fractions below unity. Although there is currently only a small number of satellites in this regime, they are consistent with the \citet{Brown+2017} scaling relation (from \hi \ stacking) between gas fraction and SSFR for satellite galaxies. This suggests that the interplay between gas content and SF may be fundamentally different for the SAGA galaxies in comparison to field galaxies, by virtue of the fact that they are satellites rather than centrals. This behavior does not seem to be reproduced in the simulations that we compared to, but their offset gas fractions make this comparison problematic (see Appendix~\ref{sec:gas_frac_SSFR_sims}). 

\citet{Karunakaran+2021} sparked a continuing debate about the high fraction of SAGA satellites that are star-forming versus quenched, and whether this is inconsistent with our understanding of the satellites in LG and models that are designed to match them. Based on our observations of eight SAGA systems we find no significant disagreement in our H$\alpha$-based SFR estimates and the UV-based estimates of \citet{Karunakaran+2021}. However, we also do not find significant deviations in the expected fraction of gas-rich and gas-poor satellites from the Auriga simulations \citep{Simpson+2018}, and the general behavior of satellite gas fractions is roughly consistent with the APOSTLE and TNG50 simulations. Our findings above indicate that \hi \ observations in particular offer a unique insight into the ongoing processes in these systems. Ultimately, we need to expand the available observations of gas in these systems in order to enable more detailed comparisons and to draw more robust conclusions.

\begin{acknowledgments}
We thank the anonymous referee for their thorough reading of the paper and constructive suggestions.
This work used images from the Dark Energy Camera Legacy Survey (DECaLS; Proposal ID 2014B-0404; PIs: David Schlegel and Arjun Dey). Full acknowledgment at \url{https://www.legacysurvey.org/acknowledgment/}.
This work used data from the Karl G. Jansky Very Large Array. The National Radio Astronomy Observatory is a facility of the National Science Foundation operated under cooperative agreement by Associated Universities, Inc. The data were observed as part of program 22A-023 (PI: M.~Jones). 
DJS acknowledges support from NSF grant AST-2205863.
KS acknowledges support from the Natural Sciences and Engineering Research Council of Canada (NSERC).
AK acknowledges financial support from the grant (SEV-2017-0709) funded by MCIN/AEI/ 10.13039/501100011033 and from the grant POSTDOC\_21\_00845 funded by the Economic Transformation, Industry, Knowledge and Universities Council of the Regional Government of Andalusia.
DC acknowledges support from NSF grant AST-1814208.
KAO is supported by a Royal Society Dorothy Hodgkin Fellowship, by STFC through grant ST/T000244/1, and by the European Research Council (ERC) through Advanced Investigator grant to C.S. Frenk, DMIDAS (GA 786910).
\end{acknowledgments}

%

\vspace{5mm}
\facilities{VLA, CFHT, Blanco, GALEX}


\software{\href{https://www.astropy.org/index.html}{\texttt{astropy}} \citep{astropy2013,astropy2018}, \href{https://reproject.readthedocs.io/en/stable/}{\texttt{reproject}} \citep{reproject}, \href{https://sites.google.com/cfa.harvard.edu/saoimageds9}{\texttt{DS9}} \citep{DS9}, \href{https://aladin.u-strasbg.fr/}{\texttt{Aladin}} \citep{Aladin2000,Aladin2014}, \href{https://casa.nrao.edu/}{\texttt{CASA}} \citep{CASA}, \href{https://www.aperturephotometry.org}{\texttt{APT}} \citep{Laher+2012}, \texttt{Elixir-LSB} \citep{Ferrarese+2012}, \texttt{MegaPipe} \citep{Gwyn2008}, \href{https://github.com/Punzo/SlicerAstro}{\texttt{SlicerAstro}} \citep{Punzo+2016,Punzo+2017}, \href{https://users.obs.carnegiescience.edu/peng/work/galfit/galfit.html}{\texttt{GALFIT}} \citep{Peng+2002,Peng+2010}, \href{https://dustmaps.readthedocs.io/en/latest/}{\texttt{dustmaps}} \citep{dustmaps}, \href{https://matplotlib.org/}{\texttt{matplotlib}} \citep{matplotlib}, \href{https://numpy.org/}{\texttt{numpy}} \citep{numpy}, \href{https://scipy.org/}{\texttt{scipy}} \citep{scipy1,scipy2}, \href{https://pandas.pydata.org/}{\texttt{pandas}} \citep{pandas1,pandas2}, \href{https://hyperfit.icrar.org/}{\texttt{HyperFit}} \citep{hyperfit}, \href{https://ipyvolume.readthedocs.io/en/latest/}{\texttt{IPyvolume}}, \href{https://spectral-cube.readthedocs.io/en/latest/}{\texttt{spectral-cube}}, \href{https://gitlab.com/SoFiA-Admin/SoFiA}{\texttt{SoFiA}} \citep{SoFiA,Serra+2015}.}

\begin{longrotatetable}
\begin{deluxetable}{lcccccccccc}
\centerwidetable
\tablecaption{Satellite \hi \ and star formation properties}
\tablehead{
\colhead{Satellite} & \colhead{Host} & \colhead{RA} & \colhead{Dec} & \colhead{$D_\mathrm{proj}$} & \colhead{$\Delta cz$} &  \colhead{$\log M_\ast$} & \colhead{$\log \mathrm{SFR_{NUV}}$} & \colhead{$\log \mathrm{SFR_{H\alpha}}$} & \colhead{$\log M_\mathrm{HI}$} & \colhead{$\log M_\mathrm{HI}/M_\ast$} \\  
\colhead{} & \colhead{} & \colhead{deg} & \colhead{deg} & \colhead{kpc} & \colhead{km/s} & \colhead{$[\mathrm{M_\odot}]$} & \colhead{$[\mathrm{M_\odot/yr}]$} & \colhead{$[\mathrm{M_\odot/yr}]$} & \colhead{$[\mathrm{M_\odot}]$} & \colhead{}
}
\startdata
LS-429811-3398 & UGC903 & 20.2850 & 17.6022 & 105 & -78 & 8.02 & $-2.57\pm0.11$ & $-2.35\pm0.20$ & $<7.88$ & $<-0.14$ \\
LS-431187-1672 & UGC903 & 20.3280 & 17.7539 & 133 & -70 & 7.54 & $-2.46\pm0.11$ & $-2.59\pm0.20$ & $7.76\pm0.11$ & 0.22 \\
dw0122+1735\tablenotemark{a} & UGC903 & 20.4941 & 17.5761  & 30 & -77 & 7.56 & $<-2.97$ & $<-2.59$ & $7.53\pm0.09$ & -0.03 \\
LS-429812-2469 & UGC903 & 20.5362 & 17.5279 & 70 & 50 & 7.08 & $-2.46\pm0.10$ & $-2.28\pm0.20$ & $<7.87$ & $<0.79$ \\
LS-432563-224 & UGC903 & 20.7772 & 17.8916 & 290 & -1 & 7.89 & $-1.81\pm0.09$ & $-1.99\pm0.20$ & $8.93\pm0.05$ & 1.04 \\
\hline
DES-350665706 & NGC1309 & 50.1913 & -15.5749 & 220 & -244 & 7.74 & $-2.16\pm0.08$ & $-2.31\pm0.20$ & $<8.10$ & $<0.36$ \\
DES-353757883 & NGC1309 & 50.4652 & -15.7104 & 189 & -106 & 8.80 & $-0.82\pm0.08$ & $-0.78\pm0.20$ & $9.43\pm0.04$ & 0.63 \\
DES-353742769 & NGC1309 & 50.9464 & -15.4004 & 242 & 16 & 7.79 & $-2.99\pm0.10$ & $<-3.21$ & $<7.93$ & $<0.14$ \\
\hline
DES-371747881 & PGC13646 & 55.3397 & -13.1446 & 248 & 55 & 8.24 & $-1.49\pm0.08$ & $-0.96\pm0.20$ & $8.71\pm0.05$ & 0.47 \\
DES-373383928 & PGC13646 & 55.5682 & -13.2170 & 189 & -127 & 7.97 & $-1.76\pm0.08$ & $-1.34\pm0.20$ & $8.29\pm0.06$ & 0.32 \\
{\bf DES-373393030}\tablenotemark{b} & NGC1421 & 55.5841 & -13.4218 & 50 & -103 & 7.69 & $-2.22\pm0.08$ & $-1.76\pm0.20$ & $8.41\pm0.05$ & 0.72 \\
\hline
NSA-648311 & UGC4906 & 139.1897 & 53.4429 & 296 & -45 & 8.84 & $-0.75\pm0.08$ & $-0.41\pm0.20$ & $9.05\pm0.05$ & 0.21 \\
LS-595052-1698 & UGC4906 & 139.2389 & 53.0101 & 68 & -209 & 8.17 & $-1.95\pm0.08$ & $-2.51\pm0.21$ & $8.71\pm0.05$ & 0.54 \\
LS-595921-1395 & UGC4906 & 139.3098 & 53.2298 & 154 & 259 & 7.34 & $-3.01\pm0.10$ & $<-2.94$ & $<7.88$ & $<0.54$ \\
LS-595052-1940 & UGC4906 & 139.4253 & 53.0271 & 21 & -96 & 7.75 & $-2.20\pm0.09$ & $-3.26\pm0.23$ & $<8.05$\tablenotemark{c}& $<0.30$ \\
NSA-78956 & UGC4906 & 139.4444 & 53.2935 & 189 & -22 & 8.54 & $-1.17\pm0.08$ & $-1.15\pm0.20$ & $9.15\pm0.05$ & 0.61 \\
NSA-78947 & UGC4906 & 139.4972 & 52.7426 & 160 & 65 & 9.06 & $-0.97\pm0.08$ & $-0.94\pm0.20$ & $9.06\pm0.05$ & 0.00 \\
\hline
1237668298203070473 & NGC4158 & 182.4280 & 20.0469 & 230 & 60 & 8.09 & $-1.46\pm0.08$ & $-1.83\pm0.20$ & $8.65\pm0.05$ & 0.56 \\
LS-446799-3923 & NGC4158 & 182.6898 & 20.5927 & 270 & 118 & 7.39 & $-2.26\pm0.08$ & $-2.31\pm0.20$ & $8.10\pm0.08$ & 0.71 \\
LS-444092-4124 & NGC4158 & 182.8481 & 20.0633 & 78 & -84 & 6.78 & $-2.89\pm0.08$ & $<-3.32$ & $<7.91$ & $<1.13$ \\
NSA-542307 & NGC4158 & 182.9907 & 20.0279 & 150 & -37 & 9.02 & $-1.03\pm0.08$ & $-0.91\pm0.20$ & $9.15\pm0.04$\tablenotemark{d} & 0.13 \\
LS-444093-4832 & NGC4158 & 183.1177 & 20.1081 & 197 & 16 & 6.79 & $-3.10\pm0.09$ & $<-3.43$ & $<7.79$ & $<1.00$ \\
\hline
{\bf NSA-19694} & NGC4454 & 187.2117 & -2.0609 & 74 & 160 & 9.31 & $-1.42\pm0.08$ &  & $8.88\pm0.05$ & -0.43 \\
NSA-628407 & NGC4454 & 187.2415 & -1.6030 & 208 & 204 & 7.63 & $-2.71\pm0.09$ &  & $<8.35$ & $<0.72$ \\
LS-321038-2612 & NGC4454 & 187.5592 & -1.7291 & 250 & -163 & 6.78 & $-3.84\pm0.23$ &  & $<8.69$ & $<1.91$ \\
LS-321038-4238\tablenotemark{e} & NGC4454 & 187.5895 & -1.6383 & 297 & -152 & 7.84 & $-1.70\pm0.08$ &  &  &  \\
\tablebreak
LS-459129-1060 & NGC6278 & 254.9390 & 22.6673 & 291 & 82 & 6.85 & $-3.41\pm0.17$ & $<-2.86$ & $<7.96$ & $<1.11$ \\
LS-461785-3431 & NGC6278 & 254.9681 & 23.2804 & 239 & 150 & 7.27 & $<-3.72$ & $<-2.68$ & $<8.04$ & $<0.77$ \\
NSA-687367 & NGC6278 & 255.0608 & 23.1063 & 114 & 189 & 9.06 & $-0.58\pm0.08$ & $-0.20\pm0.20$ & $9.14\pm0.05$ & 0.08 \\
NSA-633932 & NGC6278 & 255.1382 & 22.8660 & 109 & -167 & 7.70 & $-2.88\pm0.13$ & $<-2.60$ & $<8.01$ & $<0.31$ \\
{\bf NSA-343657} & NGC6278 & 255.1879 & 23.0440 & 26 & -48 & 9.33 & $-1.27\pm0.08$ & $-1.30\pm0.20$ & $9.11\pm0.05$ & -0.22 \\
LS-459131-5790 & NGC6278 & 255.3576 & 22.8730 & 133 & 32 & 7.26 & $-3.18\pm0.15$ & $<-2.73$ & $<7.86$ & $<0.60$ \\
1237662301379166288 & NGC6278 & 255.3735 & 22.7311 & 218 & 15 & 7.95 & $-1.98\pm0.09$ & $-1.91\pm0.20$ & $<8.23$ & $<0.28$ \\
NSA-343648 & NGC6278 & 255.5052 & 23.1634 & 213 & 218 & 8.32 & $-1.45\pm0.09$ & $-1.36\pm0.20$ & $8.49\pm0.06$ & 0.17 \\
NSA-343463 & NGC6278 & 255.5795 & 23.0765 & 237 & 128 & 8.41 & $-1.40\pm0.08$ & $-1.23\pm0.20$ & $8.71\pm0.05$ & 0.30 \\
\hline
NSA-636047 & PGC68743 & 335.8363 & -3.6598 & 164 & 195 & 9.12 & $-1.62\pm0.09$ & $-1.24\pm0.20$ & $<8.28$ & $<-0.84$ \\
LS-310115-2825 & PGC68743 & 335.9540 & -3.7010 & 185 & -30 & 6.87 & $-2.69\pm0.10$ & $-2.87\pm0.20$ & $<8.23$ & $<1.36$ \\
LS-311554-3218 & PGC68743 & 335.9729 & -3.4296 & 40 & -85 & 8.00 & $-3.23\pm0.14$ & $<-2.61$ & $<7.90$ & $<-0.10$ \\
LS-312992-1567 & PGC68743 & 335.9799 & -3.2706 & 118 & 14 & 8.72 & $-2.04\pm0.09$ & $-1.66\pm0.20$ & $<8.12$ & $<-0.60$ \\
NSA-636065 & PGC68743 & 336.0479 & -3.4834 & 98 & -17 & 9.51 & $-0.50\pm0.08$ & $-0.44\pm0.20$ & $9.47\pm0.04$ & -0.04 \\
\hline
LS-360540-737 & NGC7541 & 348.5546 & 4.9151 & 267 & 125 & 6.18 & $<-4.01$ & $-2.80\pm0.20$ & $<7.82$ & $<1.64$ \\
LS-357668-2728 & NGC7541 & 348.6214 & 4.5073 & 44 & -121 & 7.74 & $<-3.50$ & $<-2.50$ & $<7.78$ & $<0.04$ \\
NSA-637123 & NGC7541 & 348.6438 & 4.4984 & 34 & 11 & 10.07 & $-0.51\pm0.08$ & $0.02\pm0.20$ & $9.82\pm0.04$\tablenotemark{f} & -0.25 \\
{\bf LS-359104-131} & NGC7541 & 348.6965 & 4.6396 & 70 & 182 & 8.93 & $-1.11\pm0.09$ & $-0.80\pm0.20$ & $8.94\pm0.05$ & 0.01 \\
LS-356233-3798 & NGC7541 & 348.7769 & 4.3732 & 123 & -13 & 8.86 & $-1.70\pm0.09$ & $-1.37\pm0.20$ & $<7.83$ & $<-1.03$ \\
LS-357669-3767 & NGC7541 & 348.8745 & 4.6131 & 137 & 68 & 8.37 & $-1.52\pm0.09$ & $-1.48\pm0.20$ & $9.11\pm0.04$ & 0.74 \\
\enddata
\tablenotetext{}{Bold entries indicate satellites with signs of ongoing RPS (Figure~\ref{fig:rps_examples}).}
\tablenotetext{a}{Not included in the SAGA-II catalog. \hi \ emission lightly blended with host. Straightforward to manually separate.}
\tablenotetext{b}{Incorrect host in SAGA-II. Velocity and projected separation values have been updated.}
\tablenotetext{c}{Satellite overlaps host's \hi \ distribution, but does not appear to contain its own \hi.}
\tablenotetext{d}{RFI contamination. Manually separated. Uncertainty in \hi \ mass likely underestimated.}
\tablenotetext{e}{Not observed with the VLA.}
\tablenotetext{f}{Possible minor interaction with host. \hi \ manually separated.}
\label{tab:sats}
\end{deluxetable}
\end{longrotatetable}



\appendix

\section{Gas fraction and specific star formation rates in simulated satellites} \label{sec:gas_frac_SSFR_sims}

\begin{figure*}
    \centering
    \includegraphics[width=0.67\textwidth]{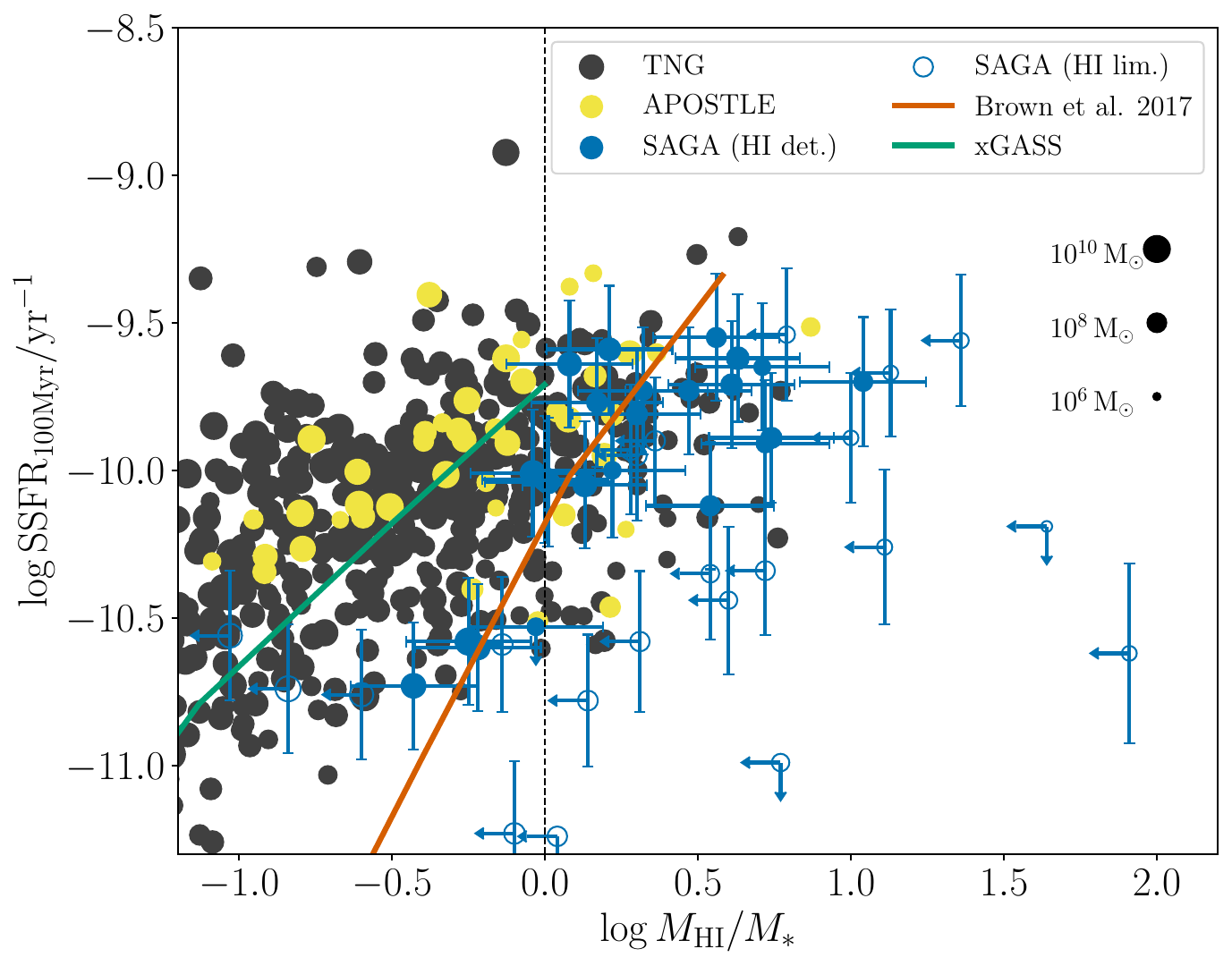}
    \caption{Specific star formation rates of satellites plotted as a function of their gas fractions. Equivalent to Figure~\ref{fig:gas_frac_SSFR} but showing the APSOTLE and TNG50 simulations in comparison to the SAGA sample observations.
    The size of the points is scaled $\log M_\ast$, with examples shown on the right side of the plot.  
    The orange line show the \citet{Brown+2017} relation from satellites in groups of $12 < \log M_\mathrm{halo}/\mathrm{M_\odot} < 13$, and the green line shows the xGASS relation for field galaxies in the stellar mass range $9 < \log M_\ast/\mathrm{M_\odot} < 10$ \citep{Catinella+2018}.
    The vertical black dashed line indicates a gas fraction of unity.}
    \label{fig:gas_frac_SSFR_sims}
\end{figure*}

Figure~\ref{fig:gas_frac_SSFR} showed the relationship between gas fraction and SSFR for the SAGA satellites and various comparison samples. The sharp downturn in SSFR around a gas fraction of unity suggests that this relationship may be different for field and satellite galaxies. Here we make the same plot (Figure~\ref{fig:gas_frac_SSFR_sims}), but for satellites from the TNG50 and APOSTLE simulations. Satellites were selected as those within 300~kpc (physical) and $\Delta v < |275|$~\kms \ of their host. Satellites within 10~kpc of their host were removed, in line the with SAGA selection. We also removed any satellites with $\log M_\ast/\mathrm{M_\odot} < 7.1$ to make the sample more directly comparable to SAGA and to ensure all the simulated satellites are well resolved.

Here we see that, unlike the SAGA satellites, the simulated satellites mostly follow the field galaxy (xGASS) relation between gas fraction and SSFR. As expected (cf. \S\ref{sec:sim_comp}) the simulated satellites are generally more gas-poor than those in the observations. While at first glance it might appear as if the SAGA observations are mostly only sampling the gas-rich tail of the population seen in the simulations, the stellar mass cut mentioned above means that both the observational and simulated samples correspond to a mass regime where SAGA is relatively complete. 
The \hi \ observations are naturally more likely to detect gas-rich satellites, but even taking the non-detections into account, a significant fraction of the SAGA sample fall at the very extreme of the gas-fraction--SSFR parameter space sampled by the simulations.
Thus, what we are mostly seeing here is that the simulations are under-predicting the \hi \ masses, by around 0.5~dex. However, if the simulations merely increased the gas masses of these satellites, then they would likely continue to follow the green relation on which they currently fall, and would therefore probably end up with SSFRs that are higher than the SAGA satellites. This finding indicates that the resolution to this issue is not as straightforward as simply making the simulated satellites more gas-rich. It may be that the relationship between gas and star formation in satellite galaxies behaves in a different manner to field galaxies, in a way that is not fully captured by current simulations.

\bibliography{refs}{}

\begin{thebibliography}{}
\expandafter\ifx\csname natexlab\endcsname\relax\def\natexlab#1{#1}\fi
\providecommand{\url}[1]{\href{#1}{#1}}
\providecommand{\dodoi}[1]{doi:~\href{http://doi.org/#1}{\nolinkurl{#1}}}
\providecommand{\doeprint}[1]{\href{http://ascl.net/#1}{\nolinkurl{http://ascl.net/#1}}}
\providecommand{\doarXiv}[1]{\href{https://arxiv.org/abs/#1}{\nolinkurl{https://arxiv.org/abs/#1}}}

\bibitem[{{Abbott} {et~al.}(2018){Abbott}, {Abdalla}, {Allam}, {Amara},
  {Annis}, {Asorey}, {Avila}, {Ballester}, {Banerji}, {Barkhouse}, {Baruah},
  {Baumer}, {Bechtol}, {Becker}, {Benoit-L{\'e}vy}, {Bernstein}, {Bertin},
  {Blazek}, {Bocquet}, {Brooks}, {Brout}, {Buckley-Geer}, {Burke}, {Busti},
  {Campisano}, {Cardiel-Sas}, {Carnero Rosell}, {Carrasco Kind}, {Carretero},
  {Castander}, {Cawthon}, {Chang}, {Chen}, {Conselice}, {Costa}, {Crocce},
  {Cunha}, {D'Andrea}, {da Costa}, {Das}, {Daues}, {Davis}, {Davis}, {De
  Vicente}, {DePoy}, {DeRose}, {Desai}, {Diehl}, {Dietrich}, {Dodelson},
  {Doel}, {Drlica-Wagner}, {Eifler}, {Elliott}, {Evrard}, {Farahi}, {Fausti
  Neto}, {Fernandez}, {Finley}, {Flaugher}, {Foley}, {Fosalba}, {Friedel},
  {Frieman}, {Garc{\'\i}a-Bellido}, {Gaztanaga}, {Gerdes}, {Giannantonio},
  {Gill}, {Glazebrook}, {Goldstein}, {Gower}, {Gruen}, {Gruendl}, {Gschwend},
  {Gupta}, {Gutierrez}, {Hamilton}, {Hartley}, {Hinton}, {Hislop}, {Hollowood},
  {Honscheid}, {Hoyle}, {Huterer}, {Jain}, {James}, {Jeltema}, {Johnson},
  {Johnson}, {Kacprzak}, {Kent}, {Khullar}, {Klein}, {Kovacs}, {Koziol},
  {Krause}, {Kremin}, {Kron}, {Kuehn}, {Kuhlmann}, {Kuropatkin}, {Lahav},
  {Lasker}, {Li}, {Li}, {Liddle}, {Lima}, {Lin}, {L{\'o}pez-Reyes}, {MacCrann},
  {Maia}, {Maloney}, {Manera}, {March}, {Marriner}, {Marshall}, {Martini},
  {McClintock}, {McKay}, {McMahon}, {Melchior}, {Menanteau}, {Miller},
  {Miquel}, {Mohr}, {Morganson}, {Mould}, {Neilsen}, {Nichol}, {Nogueira},
  {Nord}, {Nugent}, {Nunes}, {Ogando}, {Old}, {Pace}, {Palmese},
  {Paz-Chinch{\'o}n}, {Peiris}, {Percival}, {Petravick}, {Plazas}, {Poh},
  {Pond}, {Porredon}, {Pujol}, {Refregier}, {Reil}, {Ricker}, {Rollins},
  {Romer}, {Roodman}, {Rooney}, {Ross}, {Rykoff}, {Sako}, {Sanchez}, {Sanchez},
  {Santiago}, {Saro}, {Scarpine}, {Scolnic}, {Serrano}, {Sevilla-Noarbe},
  {Sheldon}, {Shipp}, {Silveira}, {Smith}, {Smith}, {Smith}, {Soares-Santos},
  {Sobreira}, {Song}, {Stebbins}, {Suchyta}, {Sullivan}, {Swanson}, {Tarle},
  {Thaler}, {Thomas}, {Thomas}, {Troxel}, {Tucker}, {Vikram}, {Vivas},
  {Walker}, {Wechsler}, {Weller}, {Wester}, {Wolf}, {Wu}, {Yanny}, {Zenteno},
  {Zhang}, {Zuntz}, {DES Collaboration}, {Juneau}, {Fitzpatrick}, {Nikutta},
  {Nidever}, {Olsen}, {Scott}, \& {NOAO Data Lab}}]{DESDR1}
{Abbott}, T.~M.~C., {Abdalla}, F.~B., {Allam}, S., {et~al.} 2018, \apjs, 239,
  18, \dodoi{10.3847/1538-4365/aae9f0}

\bibitem[{{Abolfathi} {et~al.}(2018){Abolfathi}, {Aguado}, {Aguilar}, {Allende
  Prieto}, {Almeida}, {Ananna}, {Anders}, {Anderson}, {Andrews}, {Anguiano},
  {Arag{\'o}n-Salamanca}, {Argudo-Fern{\'a}ndez}, {Armengaud}, {Ata},
  {Aubourg}, {Avila-Reese}, {Badenes}, {Bailey}, {Balland}, {Barger},
  {Barrera-Ballesteros}, {Bartosz}, {Bastien}, {Bates}, {Baumgarten},
  {Bautista}, {Beaton}, {Beers}, {Belfiore}, {Bender}, {Bernardi}, {Bershady},
  {Beutler}, {Bird}, {Bizyaev}, {Blanc}, {Blanton}, {Blomqvist}, {Bolton},
  {Boquien}, {Borissova}, {Bovy}, {Bradna Diaz}, {Brandt}, {Brinkmann},
  {Brownstein}, {Bundy}, {Burgasser}, {Burtin}, {Busca}, {Ca{\~n}as},
  {Cano-D{\'\i}az}, {Cappellari}, {Carrera}, {Casey}, {Cervantes Sodi}, {Chen},
  {Cherinka}, {Chiappini}, {Choi}, {Chojnowski}, {Chuang}, {Chung}, {Clerc},
  {Cohen}, {Comerford}, {Comparat}, {Correa do Nascimento}, {da Costa},
  {Cousinou}, {Covey}, {Crane}, {Cruz-Gonzalez}, {Cunha}, {da Silva Ilha},
  {Damke}, {Darling}, {Davidson}, {Dawson}, {de Icaza Lizaola}, {de la
  Macorra}, {de la Torre}, {De Lee}, {de Sainte Agathe}, {Deconto Machado},
  {Dell'Agli}, {Delubac}, {Diamond-Stanic}, {Donor}, {Downes}, {Drory}, {du Mas
  des Bourboux}, {Duckworth}, {Dwelly}, {Dyer}, {Ebelke}, {Davis Eigenbrot},
  {Eisenstein}, {Elsworth}, {Emsellem}, {Eracleous}, {Erfanianfar},
  {Escoffier}, {Fan}, {Fern{\'a}ndez Alvar}, {Fernandez-Trincado}, {Fernando
  Cirolini}, {Feuillet}, {Finoguenov}, {Fleming}, {Font-Ribera}, {Freischlad},
  {Frinchaboy}, {Fu}, {G{\'o}mez Maqueo Chew}, {Galbany}, {Garc{\'\i}a
  P{\'e}rez}, {Garcia-Dias}, {Garc{\'\i}a-Hern{\'a}ndez}, {Garma Oehmichen},
  {Gaulme}, {Gelfand}, {Gil-Mar{\'\i}n}, {Gillespie}, {Goddard}, {Gonz{\'a}lez
  Hern{\'a}ndez}, {Gonzalez-Perez}, {Grabowski}, {Green}, {Grier}, {Gueguen},
  {Guo}, {Guy}, {Hagen}, {Hall}, {Harding}, {Hasselquist}, {Hawley}, {Hayes},
  {Hearty}, {Hekker}, {Hernandez}, {Hernandez Toledo}, {Hogg},
  {Holley-Bockelmann}, {Holtzman}, {Hou}, {Hsieh}, {Hunt}, {Hutchinson},
  {Hwang}, {Jimenez Angel}, {Johnson}, {Jones}, {J{\"o}nsson}, {Jullo}, {Khan},
  {Kinemuchi}, {Kirkby}, {Kirkpatrick}, {Kitaura}, {Knapp}, {Kneib},
  {Kollmeier}, {Lacerna}, {Lane}, {Lang}, {Law}, {Le Goff}, {Lee}, {Li}, {Li},
  {Lian}, {Liang}, {Lima}, {Lin}, {Long}, {Lucatello}, {Lundgren}, {Mackereth},
  {MacLeod}, {Mahadevan}, {Maia}, {Majewski}, {Manchado}, {Maraston},
  {Mariappan}, {Marques-Chaves}, {Masseron}, {Masters}, {McDermid}, {McGreer},
  {Melendez}, {Meneses-Goytia}, {Merloni}, {Merrifield}, {Meszaros}, {Meza},
  {Minchev}, {Minniti}, {Mueller}, {Muller-Sanchez}, {Muna}, {Mu{\~n}oz},
  {Myers}, {Nair}, {Nandra}, {Ness}, {Newman}, {Nichol}, {Nidever},
  {Nitschelm}, {Noterdaeme}, {O'Connell}, {Oelkers}, {Oravetz}, {Oravetz},
  {Ort{\'\i}z}, {Osorio}, {Pace}, {Padilla}, {Palanque-Delabrouille},
  {Palicio}, {Pan}, {Pan}, {Parikh}, {P{\^a}ris}, {Park}, {Peirani},
  {Pellejero-Ibanez}, {Penny}, {Percival}, {Perez-Fournon}, {Petitjean},
  {Pieri}, {Pinsonneault}, {Pisani}, {Prada}, {Prakash}, {Queiroz}, {Raddick},
  {Raichoor}, {Barboza Rembold}, {Richstein}, {Riffel}, {Riffel}, {Rix},
  {Robin}, {Rodr{\'\i}guez Torres}, {Rom{\'a}n-Z{\'u}{\~n}iga}, {Ross},
  {Rossi}, {Ruan}, {Ruggeri}, {Ruiz}, {Salvato}, {S{\'a}nchez}, {S{\'a}nchez},
  {Sanchez Almeida}, {S{\'a}nchez-Gallego}, {Santana Rojas}, {Santiago},
  {Schiavon}, {Schimoia}, {Schlafly}, {Schlegel}, {Schneider}, {Schuster},
  {Schwope}, {Seo}, {Serenelli}, {Shen}, {Shen}, {Shetrone}, {Shull}, {Silva
  Aguirre}, {Simon}, {Skrutskie}, {Slosar}, {Smethurst}, {Smith}, {Sobeck},
  {Somers}, {Souter}, {Souto}, {Spindler}, {Stark}, {Stassun}, {Steinmetz},
  {Stello}, {Storchi-Bergmann}, {Streblyanska}, {Stringfellow}, {Su{\'a}rez},
  {Sun}, {Szigeti}, {Taghizadeh-Popp}, {Talbot}, {Tang}, {Tao}, {Tayar},
  {Tembe}, {Teske}, {Thakar}, {Thomas}, {Tissera}, {Tojeiro}, {Tremonti},
  {Troup}, {Urry}, {Valenzuela}, {van den Bosch}, {Vargas-Gonz{\'a}lez},
  {Vargas-Maga{\~n}a}, {Vazquez}, {Villanova}, {Vogt}, {Wake}, {Wang},
  {Weaver}, {Weijmans}, {Weinberg}, {Westfall}, {Whelan}, {Wilcots}, {Wild},
  {Williams}, {Wilson}, {Wood-Vasey}, {Wylezalek}, {Xiao}, {Yan}, {Yang},
  {Ybarra}, {Y{\`e}che}, {Zakamska}, {Zamora}, {Zarrouk}, {Zasowski}, {Zhang},
  {Zhao}, {Zhao}, {Zheng}, {Zheng}, {Zhou}, {Zhu}, {Zinn}, \& {Zou}}]{SDSSDR14}
{Abolfathi}, B., {Aguado}, D.~S., {Aguilar}, G., {et~al.} 2018, \apjs, 235, 42,
  \dodoi{10.3847/1538-4365/aa9e8a}

\bibitem[{{Akins} {et~al.}(2021){Akins}, {Christensen}, {Brooks}, {Munshi},
  {Applebaum}, {Engelhardt}, \& {Chamberland}}]{Akins+2021}
{Akins}, H.~B., {Christensen}, C.~R., {Brooks}, A.~M., {et~al.} 2021, \apj,
  909, 139, \dodoi{10.3847/1538-4357/abe2ab}

\bibitem[{{Astropy Collaboration} {et~al.}(2013){Astropy Collaboration},
  {Robitaille}, {Tollerud}, {Greenfield}, {Droettboom}, {Bray}, {Aldcroft},
  {Davis}, {Ginsburg}, {Price-Whelan}, {Kerzendorf}, {Conley}, {Crighton},
  {Barbary}, {Muna}, {Ferguson}, {Grollier}, {Parikh}, {Nair}, {Unther},
  {Deil}, {Woillez}, {Conseil}, {Kramer}, {Turner}, {Singer}, {Fox}, {Weaver},
  {Zabalza}, {Edwards}, {Azalee Bostroem}, {Burke}, {Casey}, {Crawford},
  {Dencheva}, {Ely}, {Jenness}, {Labrie}, {Lim}, {Pierfederici}, {Pontzen},
  {Ptak}, {Refsdal}, {Servillat}, \& {Streicher}}]{astropy2013}
{Astropy Collaboration}, {Robitaille}, T.~P., {Tollerud}, E.~J., {et~al.} 2013,
  \aap, 558, A33, \dodoi{10.1051/0004-6361/201322068}

\bibitem[{{Astropy Collaboration} {et~al.}(2018){Astropy Collaboration},
  {Price-Whelan}, {Sip{\H{o}}cz}, {G{\"u}nther}, {Lim}, {Crawford}, {Conseil},
  {Shupe}, {Craig}, {Dencheva}, {Ginsburg}, {VanderPlas}, {Bradley},
  {P{\'e}rez-Su{\'a}rez}, {de Val-Borro}, {Aldcroft}, {Cruz}, {Robitaille},
  {Tollerud}, {Ardelean}, {Babej}, {Bach}, {Bachetti}, {Bakanov}, {Bamford},
  {Barentsen}, {Barmby}, {Baumbach}, {Berry}, {Biscani}, {Boquien}, {Bostroem},
  {Bouma}, {Brammer}, {Bray}, {Breytenbach}, {Buddelmeijer}, {Burke},
  {Calderone}, {Cano Rodr{\'\i}guez}, {Cara}, {Cardoso}, {Cheedella}, {Copin},
  {Corrales}, {Crichton}, {D'Avella}, {Deil}, {Depagne}, {Dietrich}, {Donath},
  {Droettboom}, {Earl}, {Erben}, {Fabbro}, {Ferreira}, {Finethy}, {Fox},
  {Garrison}, {Gibbons}, {Goldstein}, {Gommers}, {Greco}, {Greenfield},
  {Groener}, {Grollier}, {Hagen}, {Hirst}, {Homeier}, {Horton}, {Hosseinzadeh},
  {Hu}, {Hunkeler}, {Ivezi{\'c}}, {Jain}, {Jenness}, {Kanarek}, {Kendrew},
  {Kern}, {Kerzendorf}, {Khvalko}, {King}, {Kirkby}, {Kulkarni}, {Kumar},
  {Lee}, {Lenz}, {Littlefair}, {Ma}, {Macleod}, {Mastropietro}, {McCully},
  {Montagnac}, {Morris}, {Mueller}, {Mumford}, {Muna}, {Murphy}, {Nelson},
  {Nguyen}, {Ninan}, {N{\"o}the}, {Ogaz}, {Oh}, {Parejko}, {Parley}, {Pascual},
  {Patil}, {Patil}, {Plunkett}, {Prochaska}, {Rastogi}, {Reddy Janga},
  {Sabater}, {Sakurikar}, {Seifert}, {Sherbert}, {Sherwood-Taylor}, {Shih},
  {Sick}, {Silbiger}, {Singanamalla}, {Singer}, {Sladen}, {Sooley},
  {Sornarajah}, {Streicher}, {Teuben}, {Thomas}, {Tremblay}, {Turner},
  {Terr{\'o}n}, {van Kerkwijk}, {de la Vega}, {Watkins}, {Weaver}, {Whitmore},
  {Woillez}, {Zabalza}, \& {Astropy Contributors}}]{astropy2018}
{Astropy Collaboration}, {Price-Whelan}, A.~M., {Sip{\H{o}}cz}, B.~M., {et~al.}
  2018, \aj, 156, 123, \dodoi{10.3847/1538-3881/aabc4f}

\bibitem[{{Bah{\'e}} {et~al.}(2016){Bah{\'e}}, {Crain}, {Kauffmann}, {Bower},
  {Schaye}, {Furlong}, {Lagos}, {Schaller}, {Trayford}, {Dalla Vecchia}, \&
  {Theuns}}]{Bahe+2016}
{Bah{\'e}}, Y.~M., {Crain}, R.~A., {Kauffmann}, G., {et~al.} 2016, \mnras, 456,
  1115, \dodoi{10.1093/mnras/stv2674}

\bibitem[{{Beale} {et~al.}(2020){Beale}, {Donovan Meyer}, {Tollerud}, {Putman},
  \& {Peek}}]{Beale+2020}
{Beale}, L., {Donovan Meyer}, J., {Tollerud}, E.~J., {Putman}, M.~E., \&
  {Peek}, J.~E.~G. 2020, \apj, 903, 59, \dodoi{10.3847/1538-4357/abb81a}

\bibitem[{{Ben{\'\i}tez-Llambay} {et~al.}(2013){Ben{\'\i}tez-Llambay},
  {Navarro}, {Abadi}, {Gottl{\"o}ber}, {Yepes}, {Hoffman}, \&
  {Steinmetz}}]{Benitez-Llambay+2013}
{Ben{\'\i}tez-Llambay}, A., {Navarro}, J.~F., {Abadi}, M.~G., {et~al.} 2013,
  \apjl, 763, L41, \dodoi{10.1088/2041-8205/763/2/L41}

\bibitem[{{Ben{\'\i}tez-Llambay} {et~al.}(2017){Ben{\'\i}tez-Llambay},
  {Navarro}, {Frenk}, {Sawala}, {Oman}, {Fattahi}, {Schaller}, {Schaye},
  {Crain}, \& {Theuns}}]{Benitez-Llambay+2017}
{Ben{\'\i}tez-Llambay}, A., {Navarro}, J.~F., {Frenk}, C.~S., {et~al.} 2017,
  \mnras, 465, 3913, \dodoi{10.1093/mnras/stw2982}

\bibitem[{{Bennet} {et~al.}(2019){Bennet}, {Sand}, {Crnojevi{\'c}}, {Spekkens},
  {Karunakaran}, {Zaritsky}, \& {Mutlu-Pakdil}}]{Bennet+2019}
{Bennet}, P., {Sand}, D.~J., {Crnojevi{\'c}}, D., {et~al.} 2019, \apj, 885,
  153, \dodoi{10.3847/1538-4357/ab46ab}

\bibitem[{{Bernard} {et~al.}(2012){Bernard}, {Ferguson}, {Barker}, {Irwin},
  {Jablonka}, \& {Arimoto}}]{Bernard+2012}
{Bernard}, E.~J., {Ferguson}, A. M.~N., {Barker}, M.~K., {et~al.} 2012, \mnras,
  426, 3490, \dodoi{10.1111/j.1365-2966.2012.22025.x}

\bibitem[{{Besla}(2015)}]{Besla2015}
{Besla}, G. 2015, arXiv e-prints, arXiv:1511.03346,
  \dodoi{10.48550/arXiv.1511.03346}

\bibitem[{{Blanton} {et~al.}(2011){Blanton}, {Kazin}, {Muna}, {Weaver}, \&
  {Price-Whelan}}]{Blanton+2011}
{Blanton}, M.~R., {Kazin}, E., {Muna}, D., {Weaver}, B.~A., \& {Price-Whelan},
  A. 2011, \aj, 142, 31, \dodoi{10.1088/0004-6256/142/1/31}

\bibitem[{{Blitz} \& {Rosolowsky}(2006)}]{BlitzRosolowsky2006}
{Blitz}, L., \& {Rosolowsky}, E. 2006, \apj, 650, 933, \dodoi{10.1086/505417}

\bibitem[{{Boch} \& {Fernique}(2014)}]{Aladin2014}
{Boch}, T., \& {Fernique}, P. 2014, in Astronomical Society of the Pacific
  Conference Series, Vol. 485, Astronomical Data Analysis Software and Systems
  XXIII, ed. N.~{Manset} \& P.~{Forshay}, 277

\bibitem[{{Bonnarel} {et~al.}(2000){Bonnarel}, {Fernique}, {Bienaym{\'e}},
  {Egret}, {Genova}, {Louys}, {Ochsenbein}, {Wenger}, \&
  {Bartlett}}]{Aladin2000}
{Bonnarel}, F., {Fernique}, P., {Bienaym{\'e}}, O., {et~al.} 2000, \aaps, 143,
  33, \dodoi{10.1051/aas:2000331}

\bibitem[{{Boselli} {et~al.}(2022){Boselli}, {Fossati}, \&
  {Sun}}]{Boselli+2022}
{Boselli}, A., {Fossati}, M., \& {Sun}, M. 2022, \aapr, 30, 3,
  \dodoi{10.1007/s00159-022-00140-3}

\bibitem[{{Boselli} {et~al.}(2018){Boselli}, {Fossati}, {Cuillandre},
  {Boissier}, {Boquien}, {Buat}, {Burgarella}, {Consolandi}, {Cortese},
  {C{\^o}t{\'e}}, {C{\^o}t{\'e}}, {Durrell}, {Ferrarese}, {Fumagalli},
  {Gavazzi}, {Gwyn}, {Hensler}, {Koribalski}, {Roediger}, {Roehlly}, {Russeil},
  {Sun}, {Toloba}, {Vollmer}, \& {Zavagno}}]{Boselli+2018}
{Boselli}, A., {Fossati}, M., {Cuillandre}, J.~C., {et~al.} 2018, \aap, 615,
  A114, \dodoi{10.1051/0004-6361/201732410}

\bibitem[{{Boselli} {et~al.}(2021){Boselli}, {Lupi}, {Epinat}, {Amram},
  {Fossati}, {Anderson}, {Boissier}, {Boquien}, {Consolandi}, {C{\^o}t{\'e}},
  {Cuillandre}, {Ferrarese}, {Galbany}, {Gavazzi}, {G{\'o}mez-L{\'o}pez},
  {Gwyn}, {Hensler}, {Hutchings}, {Kuncarayakti}, {Longobardi}, {Peng},
  {Plana}, {Postma}, {Roediger}, {Roehlly}, {Schimd}, {Trinchieri}, \&
  {Vollmer}}]{Boselli+2021}
{Boselli}, A., {Lupi}, A., {Epinat}, B., {et~al.} 2021, \aap, 646, A139,
  \dodoi{10.1051/0004-6361/202039046}

\bibitem[{{Boselli} {et~al.}(2023){Boselli}, {Serra}, {de Gasperin}, {Vollmer},
  {Amram}, {Edler}, {Fossati}, {Consolandi}, {C{\^o}t{\'e}}, {Cuillandre},
  {Ferrarese}, {Gwyn}, {Postma}, {Boquien}, {Braine}, {Combes}, {Gavazzi},
  {Hensler}, {Miville-Deschenes}, {Murgia}, {Roediger}, {Roehlly}, {Smith},
  {Zhang}, \& {Zabel}}]{Boselli+2023}
{Boselli}, A., {Serra}, P., {de Gasperin}, F., {et~al.} 2023, \aap, 676, A92,
  \dodoi{10.1051/0004-6361/202346812}

\bibitem[{{Bradford} {et~al.}(2015){Bradford}, {Geha}, \&
  {Blanton}}]{Bradford+2015}
{Bradford}, J.~D., {Geha}, M.~C., \& {Blanton}, M.~R. 2015, \apj, 809, 146,
  \dodoi{10.1088/0004-637X/809/2/146}

\bibitem[{{Brown} {et~al.}(2017){Brown}, {Catinella}, {Cortese}, {Lagos},
  {Dav{\'e}}, {Kilborn}, {Haynes}, {Giovanelli}, \&
  {Rafieferantsoa}}]{Brown+2017}
{Brown}, T., {Catinella}, B., {Cortese}, L., {et~al.} 2017, \mnras, 466, 1275,
  \dodoi{10.1093/mnras/stw2991}

\bibitem[{{Bureau} \& {Carignan}(2002)}]{Bureau+2002}
{Bureau}, M., \& {Carignan}, C. 2002, \aj, 123, 1316, \dodoi{10.1086/338899}

\bibitem[{{Carlsten} {et~al.}(2020){Carlsten}, {Greco}, {Beaton}, \&
  {Greene}}]{Carlsten+2020}
{Carlsten}, S.~G., {Greco}, J.~P., {Beaton}, R.~L., \& {Greene}, J.~E. 2020,
  \apj, 891, 144, \dodoi{10.3847/1538-4357/ab7758}

\bibitem[{{Carlsten} {et~al.}(2022){Carlsten}, {Greene}, {Beaton}, {Danieli},
  \& {Greco}}]{Carlsten+2022}
{Carlsten}, S.~G., {Greene}, J.~E., {Beaton}, R.~L., {Danieli}, S., \& {Greco},
  J.~P. 2022, \apj, 933, 47, \dodoi{10.3847/1538-4357/ac6fd7}

\bibitem[{{Catinella} {et~al.}(2018){Catinella}, {Saintonge}, {Janowiecki},
  {Cortese}, {Dav{\'e}}, {Lemonias}, {Cooper}, {Schiminovich}, {Hummels},
  {Fabello}, {Ger{\'e}b}, {Kilborn}, \& {Wang}}]{Catinella+2018}
{Catinella}, B., {Saintonge}, A., {Janowiecki}, S., {et~al.} 2018, \mnras, 476,
  875, \dodoi{10.1093/mnras/sty089}

\bibitem[{{Chung} {et~al.}(2009){Chung}, {van Gorkom}, {Kenney}, {Crowl}, \&
  {Vollmer}}]{Chung+2009}
{Chung}, A., {van Gorkom}, J.~H., {Kenney}, J. D.~P., {Crowl}, H., \&
  {Vollmer}, B. 2009, \aj, 138, 1741, \dodoi{10.1088/0004-6256/138/6/1741}

\bibitem[{{Cortese} {et~al.}(2021){Cortese}, {Catinella}, \&
  {Smith}}]{Cortese+2021}
{Cortese}, L., {Catinella}, B., \& {Smith}, R. 2021, \pasa, 38, e035,
  \dodoi{10.1017/pasa.2021.18}

\bibitem[{{Crain} {et~al.}(2017){Crain}, {Bah{\'e}}, {Lagos}, {Rahmati},
  {Schaye}, {McCarthy}, {Marasco}, {Bower}, {Schaller}, {Theuns}, \& {van der
  Hulst}}]{Crain+2017}
{Crain}, R.~A., {Bah{\'e}}, Y.~M., {Lagos}, C. d.~P., {et~al.} 2017, \mnras,
  464, 4204, \dodoi{10.1093/mnras/stw2586}

\bibitem[{{Crnojevi{\'c}} {et~al.}(2016){Crnojevi{\'c}}, {Sand}, {Spekkens},
  {Caldwell}, {Guhathakurta}, {McLeod}, {Seth}, {Simon}, {Strader}, \&
  {Toloba}}]{Crnojevic+2016}
{Crnojevi{\'c}}, D., {Sand}, D.~J., {Spekkens}, K., {et~al.} 2016, \apj, 823,
  19, \dodoi{10.3847/0004-637X/823/1/19}

\bibitem[{{de Boer} {et~al.}(1998){de Boer}, {Braun}, {Vallenari}, \&
  {Mebold}}]{deBoer+1998}
{de Boer}, K.~S., {Braun}, J.~M., {Vallenari}, A., \& {Mebold}, U. 1998, \aap,
  329, L49, \dodoi{10.48550/arXiv.astro-ph/9711052}

\bibitem[{{Dey} {et~al.}(2019){Dey}, {Schlegel}, {Lang}, {Blum}, {Burleigh},
  {Fan}, {Findlay}, {Finkbeiner}, {Herrera}, {Juneau}, {Landriau}, {Levi},
  {McGreer}, {Meisner}, {Myers}, {Moustakas}, {Nugent}, {Patej}, {Schlafly},
  {Walker}, {Valdes}, {Weaver}, {Y{\`e}che}, {Zou}, {Zhou}, {Abareshi},
  {Abbott}, {Abolfathi}, {Aguilera}, {Alam}, {Allen}, {Alvarez}, {Annis},
  {Ansarinejad}, {Aubert}, {Beechert}, {Bell}, {BenZvi}, {Beutler}, {Bielby},
  {Bolton}, {Brice{\~n}o}, {Buckley-Geer}, {Butler}, {Calamida}, {Carlberg},
  {Carter}, {Casas}, {Castander}, {Choi}, {Comparat}, {Cukanovaite}, {Delubac},
  {DeVries}, {Dey}, {Dhungana}, {Dickinson}, {Ding}, {Donaldson}, {Duan},
  {Duckworth}, {Eftekharzadeh}, {Eisenstein}, {Etourneau}, {Fagrelius},
  {Farihi}, {Fitzpatrick}, {Font-Ribera}, {Fulmer}, {G{\"a}nsicke},
  {Gaztanaga}, {George}, {Gerdes}, {Gontcho}, {Gorgoni}, {Green}, {Guy},
  {Harmer}, {Hernandez}, {Honscheid}, {Huang}, {James}, {Jannuzi}, {Jiang},
  {Joyce}, {Karcher}, {Karkar}, {Kehoe}, {Kneib}, {Kueter-Young}, {Lan},
  {Lauer}, {Le Guillou}, {Le Van Suu}, {Lee}, {Lesser}, {Perreault Levasseur},
  {Li}, {Mann}, {Marshall}, {Mart{\'\i}nez-V{\'a}zquez}, {Martini}, {du Mas des
  Bourboux}, {McManus}, {Meier}, {M{\'e}nard}, {Metcalfe},
  {Mu{\~n}oz-Guti{\'e}rrez}, {Najita}, {Napier}, {Narayan}, {Newman}, {Nie},
  {Nord}, {Norman}, {Olsen}, {Paat}, {Palanque-Delabrouille}, {Peng},
  {Poppett}, {Poremba}, {Prakash}, {Rabinowitz}, {Raichoor}, {Rezaie},
  {Robertson}, {Roe}, {Ross}, {Ross}, {Rudnick}, {Safonova}, {Saha},
  {S{\'a}nchez}, {Savary}, {Schweiker}, {Scott}, {Seo}, {Shan}, {Silva},
  {Slepian}, {Soto}, {Sprayberry}, {Staten}, {Stillman}, {Stupak}, {Summers},
  {Sien Tie}, {Tirado}, {Vargas-Maga{\~n}a}, {Vivas}, {Wechsler}, {Williams},
  {Yang}, {Yang}, {Yapici}, {Zaritsky}, {Zenteno}, {Zhang}, {Zhang}, {Zhou}, \&
  {Zhou}}]{Dey+2019}
{Dey}, A., {Schlegel}, D.~J., {Lang}, D., {et~al.} 2019, \aj, 157, 168,
  \dodoi{10.3847/1538-3881/ab089d}

\bibitem[{{Diemer} {et~al.}(2018){Diemer}, {Stevens}, {Forbes}, {Marinacci},
  {Hernquist}, {Lagos}, {Sternberg}, {Pillepich}, {Nelson}, {Popping},
  {Villaescusa-Navarro}, {Torrey}, \& {Vogelsberger}}]{Diemer+2018}
{Diemer}, B., {Stevens}, A. R.~H., {Forbes}, J.~C., {et~al.} 2018, \apjs, 238,
  33, \dodoi{10.3847/1538-4365/aae387}

\bibitem[{{Diemer} {et~al.}(2019){Diemer}, {Stevens}, {Lagos}, {Calette},
  {Tacchella}, {Hernquist}, {Marinacci}, {Nelson}, {Pillepich},
  {Rodriguez-Gomez}, {Villaescusa-Navarro}, \& {Vogelsberger}}]{Diemer+2019}
{Diemer}, B., {Stevens}, A. R.~H., {Lagos}, C. d.~P., {et~al.} 2019, \mnras,
  487, 1529, \dodoi{10.1093/mnras/stz1323}

\bibitem[{{Digby} {et~al.}(2019){Digby}, {Navarro}, {Fattahi}, {Simpson},
  {Oman}, {Gomez}, {Frenk}, {Grand}, \& {Pakmor}}]{Digby+2019}
{Digby}, R., {Navarro}, J.~F., {Fattahi}, A., {et~al.} 2019, \mnras, 485, 5423,
  \dodoi{10.1093/mnras/stz745}

\bibitem[{{Dressler}(1980)}]{Dressler1980}
{Dressler}, A. 1980, \apj, 236, 351, \dodoi{10.1086/157753}

\bibitem[{{Durbala} {et~al.}(2020){Durbala}, {Finn}, {Crone Odekon}, {Haynes},
  {Koopmann}, \& {O'Donoghue}}]{Durbala+2020}
{Durbala}, A., {Finn}, R.~A., {Crone Odekon}, M., {et~al.} 2020, \aj, 160, 271,
  \dodoi{10.3847/1538-3881/abc018}

\bibitem[{{Engler} {et~al.}(2023){Engler}, {Pillepich}, {Joshi}, {Pasquali},
  {Nelson}, \& {Grebel}}]{Engler+2023}
{Engler}, C., {Pillepich}, A., {Joshi}, G.~D., {et~al.} 2023, \mnras, 522,
  5946, \dodoi{10.1093/mnras/stad1357}

\bibitem[{{Espada} {et~al.}(2011){Espada}, {Verdes-Montenegro}, {Huchtmeier},
  {Sulentic}, {Verley}, {Leon}, \& {Sabater}}]{Espada+2011}
{Espada}, D., {Verdes-Montenegro}, L., {Huchtmeier}, W.~K., {et~al.} 2011,
  \aap, 532, A117, \dodoi{10.1051/0004-6361/201016117}

\bibitem[{{Fattahi} {et~al.}(2016){Fattahi}, {Navarro}, {Sawala}, {Frenk},
  {Oman}, {Crain}, {Furlong}, {Schaller}, {Schaye}, {Theuns}, \&
  {Jenkins}}]{Fattahi+2016}
{Fattahi}, A., {Navarro}, J.~F., {Sawala}, T., {et~al.} 2016, \mnras, 457, 844,
  \dodoi{10.1093/mnras/stv2970}

\bibitem[{{Ferrarese} {et~al.}(2012){Ferrarese}, {C{\^o}t{\'e}}, {Cuillandre},
  {Gwyn}, {Peng}, {MacArthur}, {Duc}, {Boselli}, {Mei}, {Erben}, {McConnachie},
  {Durrell}, {Mihos}, {Jord{\'a}n}, {Lan{\c{c}}on}, {Puzia}, {Emsellem},
  {Balogh}, {Blakeslee}, {van Waerbeke}, {Gavazzi}, {Vollmer}, {Kavelaars},
  {Woods}, {Ball}, {Boissier}, {Courteau}, {Ferriere}, {Gavazzi},
  {Hildebrandt}, {Hudelot}, {Huertas-Company}, {Liu}, {McLaughlin}, {Mellier},
  {Milkeraitis}, {Schade}, {Balkowski}, {Bournaud}, {Carlberg}, {Chapman},
  {Hoekstra}, {Peng}, {Sawicki}, {Simard}, {Taylor}, {Tully}, {van Driel},
  {Wilson}, {Burdullis}, {Mahoney}, \& {Manset}}]{Ferrarese+2012}
{Ferrarese}, L., {C{\^o}t{\'e}}, P., {Cuillandre}, J.-C., {et~al.} 2012, \apjs,
  200, 4, \dodoi{10.1088/0067-0049/200/1/4}

\bibitem[{{Fillingham} {et~al.}(2019){Fillingham}, {Cooper}, {Kelley},
  {Rodriguez Wimberly}, {Boylan-Kolchin}, {Bullock}, {Garrison-Kimmel},
  {Pawlowski}, \& {Wheeler}}]{Fillingham+2019}
{Fillingham}, S.~P., {Cooper}, M.~C., {Kelley}, T., {et~al.} 2019, arXiv
  e-prints, arXiv:1906.04180, \dodoi{10.48550/arXiv.1906.04180}

\bibitem[{{Fitzpatrick}(1999)}]{Fitzpatrick+1999}
{Fitzpatrick}, E.~L. 1999, \pasp, 111, 63, \dodoi{10.1086/316293}

\bibitem[{{Font} {et~al.}(2022){Font}, {McCarthy}, {Belokurov}, {Brown}, \&
  {Stafford}}]{Font+2022}
{Font}, A.~S., {McCarthy}, I.~G., {Belokurov}, V., {Brown}, S.~T., \&
  {Stafford}, S.~G. 2022, \mnras, 511, 1544, \dodoi{10.1093/mnras/stac183}

\bibitem[{{Fritz} {et~al.}(1997){Fritz}, {Mebold}, {C{\^o}t{\'e}}, \&
  {Dickey}}]{Fritz+1997}
{Fritz}, T., {Mebold}, U., {C{\^o}t{\'e}}, S., \& {Dickey}, J. 1997, Acta
  Cosmologica, 23, 197

\bibitem[{{Gavazzi} {et~al.}(1995){Gavazzi}, {Contursi}, {Carrasco}, {Boselli},
  {Kennicutt}, {Scodeggio}, \& {Jaffe}}]{Gavazzi+1995}
{Gavazzi}, G., {Contursi}, A., {Carrasco}, L., {et~al.} 1995, \aap, 304, 325

\bibitem[{{Geha} {et~al.}(2017){Geha}, {Wechsler}, {Mao}, {Tollerud}, {Weiner},
  {Bernstein}, {Hoyle}, {Marchi}, {Marshall}, {Mu{\~n}oz}, \& {Lu}}]{Geha+2017}
{Geha}, M., {Wechsler}, R.~H., {Mao}, Y.-Y., {et~al.} 2017, \apj, 847, 4,
  \dodoi{10.3847/1538-4357/aa8626}

\bibitem[{{Giovanelli} {et~al.}(2005){Giovanelli}, {Haynes}, {Kent},
  {Perillat}, {Saintonge}, {Brosch}, {Catinella}, {Hoffman}, {Stierwalt},
  {Spekkens}, {Lerner}, {Masters}, {Momjian}, {Rosenberg}, {Springob},
  {Boselli}, {Charmandaris}, {Darling}, {Davies}, {Garcia Lambas}, {Gavazzi},
  {Giovanardi}, {Hardy}, {Hunt}, {Iovino}, {Karachentsev}, {Karachentseva},
  {Koopmann}, {Marinoni}, {Minchin}, {Muller}, {Putman}, {Pantoja}, {Salzer},
  {Scodeggio}, {Skillman}, {Solanes}, {Valotto}, {van Driel}, \& {van
  Zee}}]{Giovanelli+2005}
{Giovanelli}, R., {Haynes}, M.~P., {Kent}, B.~R., {et~al.} 2005, \aj, 130,
  2598, \dodoi{10.1086/497431}

\bibitem[{{Gnedin} \& {Draine}(2014)}]{GnedinDraine2014}
{Gnedin}, N.~Y., \& {Draine}, B.~T. 2014, \apj, 795, 37,
  \dodoi{10.1088/0004-637X/795/1/37}

\bibitem[{{Grand} {et~al.}(2017){Grand}, {G{\'o}mez}, {Marinacci}, {Pakmor},
  {Springel}, {Campbell}, {Frenk}, {Jenkins}, \& {White}}]{Grand+2017}
{Grand}, R. J.~J., {G{\'o}mez}, F.~A., {Marinacci}, F., {et~al.} 2017, \mnras,
  467, 179, \dodoi{10.1093/mnras/stx071}

\bibitem[{{Grebel} {et~al.}(2003){Grebel}, {Gallagher}, \&
  {Harbeck}}]{Grebel+2003}
{Grebel}, E.~K., {Gallagher}, John~S., I., \& {Harbeck}, D. 2003, \aj, 125,
  1926, \dodoi{10.1086/368363}

\bibitem[{Green(2018)}]{dustmaps}
Green, G.~M. 2018, Journal of Open Source Software, 3, 695,
  \dodoi{10.21105/joss.00695}

\bibitem[{{Greene} {et~al.}(2023){Greene}, {Danieli}, {Carlsten}, {Beaton},
  {Jiang}, \& {Li}}]{Greene+2023}
{Greene}, J.~E., {Danieli}, S., {Carlsten}, S., {et~al.} 2023, \apj, 949, 94,
  \dodoi{10.3847/1538-4357/acc58c}

\bibitem[{{Gunn} \& {Gott}(1972)}]{Gunn+Gott1972}
{Gunn}, J.~E., \& {Gott}, J.~Richard, I. 1972, \apj, 176, 1,
  \dodoi{10.1086/151605}

\bibitem[{{Guo} {et~al.}(2017){Guo}, {Li}, {Zheng}, {Mo}, {Jing}, {Zu}, {Lim},
  \& {Xu}}]{Guo+2017}
{Guo}, H., {Li}, C., {Zheng}, Z., {et~al.} 2017, \apj, 846, 61,
  \dodoi{10.3847/1538-4357/aa85e7}

\bibitem[{{Gwyn}(2008)}]{Gwyn2008}
{Gwyn}, S. D.~J. 2008, \pasp, 120, 212, \dodoi{10.1086/526794}

\bibitem[{{Haynes} {et~al.}(2018){Haynes}, {Giovanelli}, {Kent}, {Adams},
  {Balonek}, {Craig}, {Fertig}, {Finn}, {Giovanardi}, {Hallenbeck}, {Hess},
  {Hoffman}, {Huang}, {Jones}, {Koopmann}, {Kornreich}, {Leisman}, {Miller},
  {Moorman}, {O'Connor}, {O'Donoghue}, {Papastergis}, {Troischt}, {Stark}, \&
  {Xiao}}]{Haynes+2018}
{Haynes}, M.~P., {Giovanelli}, R., {Kent}, B.~R., {et~al.} 2018, \apj, 861, 49,
  \dodoi{10.3847/1538-4357/aac956}

\bibitem[{{Hess} \& {Wilcots}(2013)}]{Hess+2013}
{Hess}, K.~M., \& {Wilcots}, E.~M. 2013, \aj, 146, 124,
  \dodoi{10.1088/0004-6256/146/5/124}

\bibitem[{{Huchra} {et~al.}(2012){Huchra}, {Macri}, {Masters}, {Jarrett},
  {Berlind}, {Calkins}, {Crook}, {Cutri}, {Erdo{\v{g}}du}, {Falco}, {George},
  {Hutcheson}, {Lahav}, {Mader}, {Mink}, {Martimbeau}, {Schneider},
  {Skrutskie}, {Tokarz}, \& {Westover}}]{Huchra+2012}
{Huchra}, J.~P., {Macri}, L.~M., {Masters}, K.~L., {et~al.} 2012, \apjs, 199,
  26, \dodoi{10.1088/0067-0049/199/2/26}

\bibitem[{{Hunter}(2007)}]{matplotlib}
{Hunter}, J.~D. 2007, Computing in Science and Engineering, 9, 90,
  \dodoi{10.1109/MCSE.2007.55}

\bibitem[{{Iglesias-P{\'a}ramo} {et~al.}(2006){Iglesias-P{\'a}ramo}, {Buat},
  {Takeuchi}, {Xu}, {Boissier}, {Boselli}, {Burgarella}, {Madore}, {Gil de
  Paz}, {Bianchi}, {Barlow}, {Byun}, {Donas}, {Forster}, {Friedman}, {Heckman},
  {Jelinski}, {Lee}, {Malina}, {Martin}, {Milliard}, {Morrissey}, {Neff},
  {Rich}, {Schiminovich}, {Seibert}, {Siegmund}, {Small}, {Szalay}, {Welsh}, \&
  {Wyder}}]{Iglesias-Paramo+2006}
{Iglesias-P{\'a}ramo}, J., {Buat}, V., {Takeuchi}, T.~T., {et~al.} 2006, \apjs,
  164, 38, \dodoi{10.1086/502628}

\bibitem[{{J{\'a}chym} {et~al.}(2014){J{\'a}chym}, {Combes}, {Cortese}, {Sun},
  \& {Kenney}}]{Jachym+2014}
{J{\'a}chym}, P., {Combes}, F., {Cortese}, L., {Sun}, M., \& {Kenney}, J. D.~P.
  2014, \apj, 792, 11, \dodoi{10.1088/0004-637X/792/1/11}

\bibitem[{{Janowiecki} {et~al.}(2019){Janowiecki}, {Jones}, {Leisman}, \&
  {Webb}}]{Janowiecki+2019}
{Janowiecki}, S., {Jones}, M.~G., {Leisman}, L., \& {Webb}, A. 2019, \mnras,
  490, 566, \dodoi{10.1093/mnras/stz1868}

\bibitem[{{Johnson} {et~al.}(2015){Johnson}, {Kamphuis}, {Koribalski}, {Wang},
  {Oh}, {Hill}, {O'Sullivan}, {Haan}, \& {Serra}}]{Johnson+2015}
{Johnson}, M.~C., {Kamphuis}, P., {Koribalski}, B.~S., {et~al.} 2015, \mnras,
  451, 3192, \dodoi{10.1093/mnras/stv1180}

\bibitem[{{Jones} {et~al.}(2020){Jones}, {Hess}, {Adams}, \&
  {Verdes-Montenegro}}]{Jones+2020}
{Jones}, M.~G., {Hess}, K.~M., {Adams}, E. A.~K., \& {Verdes-Montenegro}, L.
  2020, \mnras, 494, 2090, \dodoi{10.1093/mnras/staa810}

\bibitem[{{Jones} {et~al.}(2018){Jones}, {Papastergis}, {Pandya}, {Leisman},
  {Romanowsky}, {Yung}, {Somerville}, \& {Adams}}]{Jones+2018}
{Jones}, M.~G., {Papastergis}, E., {Pandya}, V., {et~al.} 2018, \aap, 614, A21,
  \dodoi{10.1051/0004-6361/201732409}

\bibitem[{{Jones} {et~al.}(2022){Jones}, {Sand}, {Bellazzini}, {Spekkens},
  {Cannon}, {Mutlu-Pakdil}, {Karunakaran}, {Beccari}, {Magrini}, {Cresci},
  {Inoue}, {Fuson}, {Adams}, {Battaglia}, {Bennet}, {Crnojevi{\'c}},
  {Caldwell}, {Guhathakurta}, {Haynes}, {Mu{\~n}oz}, {Seth}, {Strader},
  {Toloba}, \& {Zaritsky}}]{Jones+2022}
{Jones}, M.~G., {Sand}, D.~J., {Bellazzini}, M., {et~al.} 2022, \apjl, 926,
  L15, \dodoi{10.3847/2041-8213/ac51dc}

\bibitem[{{Jones} {et~al.}(2023){Jones}, {Verdes-Montenegro}, {Moldon}, {Damas
  Segovia}, {Borthakur}, {Luna}, {Yun}, {del Olmo}, {Perea}, {Cannon}, {Lopez
  Gutierrez}, {Cluver}, {Garrido}, \& {Sanchez}}]{Jones+2023}
{Jones}, M.~G., {Verdes-Montenegro}, L., {Moldon}, J., {et~al.} 2023, \aap,
  670, A21, \dodoi{10.1051/0004-6361/202244622}

\bibitem[{{Joye} \& {Mandel}(2003)}]{DS9}
{Joye}, W.~A., \& {Mandel}, E. 2003, in Astronomical Society of the Pacific
  Conference Series, Vol. 295, Astronomical Data Analysis Software and Systems
  XII, ed. H.~E. {Payne}, R.~I. {Jedrzejewski}, \& R.~N. {Hook}, 489

\bibitem[{{Karunakaran} {et~al.}(2023){Karunakaran}, {Sand}, {Jones},
  {Spekkens}, {Bennet}, {Crnojevi{\'c}},
  {Mutlu-Pakd{\i}{\ensuremath{\dot{}}}l}, \& {Zaritsky}}]{Karunakaran+2023}
{Karunakaran}, A., {Sand}, D.~J., {Jones}, M.~G., {et~al.} 2023, \mnras, 524,
  5314, \dodoi{10.1093/mnras/stad2208}

\bibitem[{{Karunakaran} {et~al.}(2020{\natexlab{a}}){Karunakaran}, {Spekkens},
  {Bennet}, {Sand}, {Crnojevi{\'c}}, \& {Zaritsky}}]{Karunakaran+2020a}
{Karunakaran}, A., {Spekkens}, K., {Bennet}, P., {et~al.} 2020{\natexlab{a}},
  \aj, 159, 37, \dodoi{10.3847/1538-3881/ab5af1}

\bibitem[{{Karunakaran} {et~al.}(2022){Karunakaran}, {Spekkens}, {Carroll},
  {Sand}, {Bennet}, {Crnojevi{\'c}}, {Jones}, \&
  {Mutlu-Pakd{\i}l}}]{Karunakaran+2022}
{Karunakaran}, A., {Spekkens}, K., {Carroll}, R., {et~al.} 2022, \mnras, 516,
  1741, \dodoi{10.1093/mnras/stac2329}

\bibitem[{{Karunakaran} {et~al.}(2020{\natexlab{b}}){Karunakaran}, {Spekkens},
  {Zaritsky}, {Donnerstein}, {Kadowaki}, \& {Dey}}]{Karunakaran+2020b}
{Karunakaran}, A., {Spekkens}, K., {Zaritsky}, D., {et~al.} 2020{\natexlab{b}},
  \apj, 902, 39, \dodoi{10.3847/1538-4357/abb464}

\bibitem[{{Karunakaran} {et~al.}(2021){Karunakaran}, {Spekkens}, {Oman},
  {Simpson}, {Fattahi}, {Sand}, {Bennet}, {Crnojevi{\'c}}, {Frenk},
  {G{\'o}mez}, {Grand}, {Jones}, {Marinacci}, {Mutlu-Pakdil}, {Navarro}, \&
  {Zaritsky}}]{Karunakaran+2021}
{Karunakaran}, A., {Spekkens}, K., {Oman}, K.~A., {et~al.} 2021, \apjl, 916,
  L19, \dodoi{10.3847/2041-8213/ac0e3a}

\bibitem[{{Kenney} \& {Koopmann}(1999)}]{Kenney+1999}
{Kenney}, J. D.~P., \& {Koopmann}, R.~A. 1999, \aj, 117, 181,
  \dodoi{10.1086/300683}

\bibitem[{{Kenney} {et~al.}(2004){Kenney}, {van Gorkom}, \&
  {Vollmer}}]{Kenney+2004}
{Kenney}, J. D.~P., {van Gorkom}, J.~H., \& {Vollmer}, B. 2004, \aj, 127, 3361,
  \dodoi{10.1086/420805}

\bibitem[{{Kennicutt}(1998)}]{Kennicutt1998}
{Kennicutt}, Robert~C., J. 1998, \araa, 36, 189,
  \dodoi{10.1146/annurev.astro.36.1.189}

\bibitem[{{Koopmann} \& {Kenney}(2004)}]{Koopmann+2004}
{Koopmann}, R.~A., \& {Kenney}, J. D.~P. 2004, \apj, 613, 866,
  \dodoi{10.1086/423191}

\bibitem[{{Laher} {et~al.}(2012){Laher}, {Gorjian}, {Rebull}, {Masci},
  {Fowler}, {Helou}, {Kulkarni}, \& {Law}}]{Laher+2012}
{Laher}, R.~R., {Gorjian}, V., {Rebull}, L.~M., {et~al.} 2012, \pasp, 124, 737,
  \dodoi{10.1086/666883}

\bibitem[{{Lee} {et~al.}(2009){Lee}, {Gil de Paz}, {Tremonti}, {Kennicutt},
  {Salim}, {Bothwell}, {Calzetti}, {Dalcanton}, {Dale}, {Engelbracht}, {Funes},
  {Johnson}, {Sakai}, {Skillman}, {van Zee}, {Walter}, \& {Weisz}}]{Lee+2009}
{Lee}, J.~C., {Gil de Paz}, A., {Tremonti}, C., {et~al.} 2009, \apj, 706, 599,
  \dodoi{10.1088/0004-637X/706/1/599}

\bibitem[{{Lee} {et~al.}(2011){Lee}, {Gil de Paz}, {Kennicutt}, {Bothwell},
  {Dalcanton}, {Jos{\'e} G. Funes S.}, {Johnson}, {Sakai}, {Skillman},
  {Tremonti}, \& {van Zee}}]{Lee+2011}
{Lee}, J.~C., {Gil de Paz}, A., {Kennicutt}, Robert~C., J., {et~al.} 2011,
  \apjs, 192, 6, \dodoi{10.1088/0067-0049/192/1/6}

\bibitem[{{Leisman} {et~al.}(2017){Leisman}, {Haynes}, {Janowiecki},
  {Hallenbeck}, {J{\'o}zsa}, {Giovanelli}, {Adams}, {Bernal Neira}, {Cannon},
  {Janesh}, {Rhode}, \& {Salzer}}]{Leisman+2017}
{Leisman}, L., {Haynes}, M.~P., {Janowiecki}, S., {et~al.} 2017, \apj, 842,
  133, \dodoi{10.3847/1538-4357/aa7575}

\bibitem[{{Longhetti} \& {Saracco}(2009)}]{Longhetti+2009}
{Longhetti}, M., \& {Saracco}, P. 2009, \mnras, 394, 774,
  \dodoi{10.1111/j.1365-2966.2008.14375.x}

\bibitem[{{Luks} \& {Rohlfs}(1992)}]{Luks+Rholfs1992}
{Luks}, T., \& {Rohlfs}, K. 1992, \aap, 263, 41

\bibitem[{{Makarov} {et~al.}(2014){Makarov}, {Prugniel}, {Terekhova},
  {Courtois}, \& {Vauglin}}]{Makarov+2014}
{Makarov}, D., {Prugniel}, P., {Terekhova}, N., {Courtois}, H., \& {Vauglin},
  I. 2014, \aap, 570, A13, \dodoi{10.1051/0004-6361/201423496}

\bibitem[{{Mao} {et~al.}(2021){Mao}, {Geha}, {Wechsler}, {Weiner}, {Tollerud},
  {Nadler}, \& {Kallivayalil}}]{Mao+2021}
{Mao}, Y.-Y., {Geha}, M., {Wechsler}, R.~H., {et~al.} 2021, \apj, 907, 85,
  \dodoi{10.3847/1538-4357/abce58}

\bibitem[{{McConnachie}(2012)}]{McConnachie+2012}
{McConnachie}, A.~W. 2012, \aj, 144, 4, \dodoi{10.1088/0004-6256/144/1/4}

\bibitem[{{McConnachie} {et~al.}(2007){McConnachie}, {Venn}, {Irwin}, {Young},
  \& {Geehan}}]{McConnachie+2007}
{McConnachie}, A.~W., {Venn}, K.~A., {Irwin}, M.~J., {Young}, L.~M., \&
  {Geehan}, J.~J. 2007, \apjl, 671, L33, \dodoi{10.1086/524887}

\bibitem[{{McMullin} {et~al.}(2007){McMullin}, {Waters}, {Schiebel}, {Young},
  \& {Golap}}]{CASA}
{McMullin}, J.~P., {Waters}, B., {Schiebel}, D., {Young}, W., \& {Golap}, K.
  2007, in Astronomical Society of the Pacific Conference Series, Vol. 376,
  Astronomical Data Analysis Software and Systems XVI, ed. R.~A. {Shaw},
  F.~{Hill}, \& D.~J. {Bell}, 127

\bibitem[{{Millman} \& {Aivazis}(2011)}]{scipy2}
{Millman}, K.~J., \& {Aivazis}, M. 2011, Computing in Science and Engineering,
  13, 9, \dodoi{10.1109/MCSE.2011.36}

\bibitem[{{Nelson} {et~al.}(2019{\natexlab{a}}){Nelson}, {Springel},
  {Pillepich}, {Rodriguez-Gomez}, {Torrey}, {Genel}, {Vogelsberger}, {Pakmor},
  {Marinacci}, {Weinberger}, {Kelley}, {Lovell}, {Diemer}, \&
  {Hernquist}}]{Nelson+2019a}
{Nelson}, D., {Springel}, V., {Pillepich}, A., {et~al.} 2019{\natexlab{a}},
  Computational Astrophysics and Cosmology, 6, 2,
  \dodoi{10.1186/s40668-019-0028-x}

\bibitem[{{Nelson} {et~al.}(2019{\natexlab{b}}){Nelson}, {Pillepich},
  {Springel}, {Pakmor}, {Weinberger}, {Genel}, {Torrey}, {Vogelsberger},
  {Marinacci}, \& {Hernquist}}]{Nelson+2019b}
{Nelson}, D., {Pillepich}, A., {Springel}, V., {et~al.} 2019{\natexlab{b}},
  \mnras, 490, 3234, \dodoi{10.1093/mnras/stz2306}

\bibitem[{{Odekon} {et~al.}(2016){Odekon}, {Koopmann}, {Haynes}, {Finn},
  {McGowan}, {Micula}, {Reed}, {Giovanelli}, \& {Hallenbeck}}]{Odekon+2016}
{Odekon}, M.~C., {Koopmann}, R.~A., {Haynes}, M.~P., {et~al.} 2016, \apj, 824,
  110, \dodoi{10.3847/0004-637X/824/2/110}

\bibitem[{{Oliphant}(2007)}]{scipy1}
{Oliphant}, T.~E. 2007, Computing in Science and Engineering, 9, 10,
  \dodoi{10.1109/MCSE.2007.58}

\bibitem[{{Oman} {et~al.}(2021){Oman}, {Bah{\'e}}, {Healy}, {Hess}, {Hudson},
  \& {Verheijen}}]{Oman+2021}
{Oman}, K.~A., {Bah{\'e}}, Y.~M., {Healy}, J., {et~al.} 2021, \mnras, 501,
  5073, \dodoi{10.1093/mnras/staa3845}

\bibitem[{{Oman} {et~al.}(2019){Oman}, {Marasco}, {Navarro}, {Frenk}, {Schaye},
  \& {Ben{\'\i}tez-Llambay}}]{Oman+2019}
{Oman}, K.~A., {Marasco}, A., {Navarro}, J.~F., {et~al.} 2019, \mnras, 482,
  821, \dodoi{10.1093/mnras/sty2687}

\bibitem[{pandas~development team(2020)}]{pandas2}
pandas~development team, T. 2020, pandas-dev/pandas: Pandas, latest,  Zenodo,
  \dodoi{10.5281/zenodo.3509134}

\bibitem[{{Paturel} {et~al.}(2003){Paturel}, {Petit}, {Prugniel}, {Theureau},
  {Rousseau}, {Brouty}, {Dubois}, \& {Cambr{\'e}sy}}]{Paturel+2003}
{Paturel}, G., {Petit}, C., {Prugniel}, P., {et~al.} 2003, \aap, 412, 45,
  \dodoi{10.1051/0004-6361:20031411}

\bibitem[{{Peng} {et~al.}(2002){Peng}, {Ho}, {Impey}, \& {Rix}}]{Peng+2002}
{Peng}, C.~Y., {Ho}, L.~C., {Impey}, C.~D., \& {Rix}, H.-W. 2002, \aj, 124,
  266, \dodoi{10.1086/340952}

\bibitem[{{Peng} {et~al.}(2010){Peng}, {Ho}, {Impey}, \& {Rix}}]{Peng+2010}
---. 2010, \aj, 139, 2097, \dodoi{10.1088/0004-6256/139/6/2097}

\bibitem[{{Pillepich} {et~al.}(2019){Pillepich}, {Nelson}, {Springel},
  {Pakmor}, {Torrey}, {Weinberger}, {Vogelsberger}, {Marinacci}, {Genel}, {van
  der Wel}, \& {Hernquist}}]{Pillepich+2019}
{Pillepich}, A., {Nelson}, D., {Springel}, V., {et~al.} 2019, \mnras, 490,
  3196, \dodoi{10.1093/mnras/stz2338}

\bibitem[{{Poggianti} {et~al.}(2017){Poggianti}, {Moretti}, {Gullieuszik},
  {Fritz}, {Jaff{\'e}}, {Bettoni}, {Fasano}, {Bellhouse}, {Hau}, {Vulcani},
  {Biviano}, {Omizzolo}, {Paccagnella}, {D'Onofrio}, {Cava}, {Sheen}, {Couch},
  \& {Owers}}]{Poggianti+2017}
{Poggianti}, B.~M., {Moretti}, A., {Gullieuszik}, M., {et~al.} 2017, \apj, 844,
  48, \dodoi{10.3847/1538-4357/aa78ed}

\bibitem[{{Punzo} {et~al.}(2016){Punzo}, {van der Hulst}, {Roerdink}, \&
  {Fillion-Robin}}]{Punzo+2016}
{Punzo}, D., {van der Hulst}, T., {Roerdink}, J., \& {Fillion-Robin}, J.-C.
  2016, {SlicerAstro: Astronomy (HI) extension for 3D Slicer}, Astrophysics
  Source Code Library, record ascl:1611.021.
\newblock \doeprint{1611.021}

\bibitem[{Punzo {et~al.}(2017)Punzo, van der Hulst, Roerdink, Fillion-Robin,
  \& Yu}]{Punzo+2017}
Punzo, D., van der Hulst, J., Roerdink, J., Fillion-Robin, J., \& Yu, L.
  2017, Astronomy and Computing, 19, 45,
  \dodoi{https://doi.org/10.1016/j.ascom.2017.03.004}

\bibitem[{{Putman} {et~al.}(2021){Putman}, {Zheng}, {Price-Whelan}, {Grcevich},
  {Johnson}, {Tollerud}, \& {Peek}}]{Putman+2021}
{Putman}, M.~E., {Zheng}, Y., {Price-Whelan}, A.~M., {et~al.} 2021, \apj, 913,
  53, \dodoi{10.3847/1538-4357/abe391}

\bibitem[{{Rahmati} {et~al.}(2013){Rahmati}, {Pawlik}, {Rai{\v{c}}evi{\'c}}, \&
  {Schaye}}]{Rahmati+2013}
{Rahmati}, A., {Pawlik}, A.~H., {Rai{\v{c}}evi{\'c}}, M., \& {Schaye}, J. 2013,
  \mnras, 430, 2427, \dodoi{10.1093/mnras/stt066}

\bibitem[{{Ramatsoku} {et~al.}(2019){Ramatsoku}, {Serra}, {Poggianti},
  {Moretti}, {Gullieuszik}, {Bettoni}, {Deb}, {Fritz}, {van Gorkom},
  {Jaff{\'e}}, {Tonnesen}, {Verheijen}, {Vulcani}, {Hugo}, {J{\'o}zsa},
  {Maccagni}, {Makhathini}, {Ramaila}, {Smirnov}, \& {Thorat}}]{Ramatsoku+2019}
{Ramatsoku}, M., {Serra}, P., {Poggianti}, B.~M., {et~al.} 2019, \mnras, 487,
  4580, \dodoi{10.1093/mnras/stz1609}

\bibitem[{{Ramesh} {et~al.}(2023){Ramesh}, {Nelson}, \&
  {Pillepich}}]{Ramesh+2023}
{Ramesh}, R., {Nelson}, D., \& {Pillepich}, A. 2023, \mnras, 518, 5754,
  \dodoi{10.1093/mnras/stac3524}

\bibitem[{{Roberts} {et~al.}(2021){Roberts}, {van Weeren}, {McGee}, {Botteon},
  {Ignesti}, \& {Rottgering}}]{Roberts+2021}
{Roberts}, I.~D., {van Weeren}, R.~J., {McGee}, S.~L., {et~al.} 2021, \aap,
  652, A153, \dodoi{10.1051/0004-6361/202141118}

\bibitem[{{Robitaille} {et~al.}(2020){Robitaille}, {Deil}, \&
  {Ginsburg}}]{reproject}
{Robitaille}, T., {Deil}, C., \& {Ginsburg}, A. 2020, {reproject: Python-based
  astronomical image reprojection}.
\newblock \doeprint{2011.023}

\bibitem[{{Robotham} \& {Obreschkow}(2015)}]{hyperfit}
{Robotham}, A.~S.~G., \& {Obreschkow}, D. 2015, \pasa, 32, e033,
  \dodoi{10.1017/pasa.2015.33}

\bibitem[{{Saeedzadeh} {et~al.}(2023){Saeedzadeh}, {Jung}, {Rennehan}, {Babul},
  {Tremmel}, {Quinn}, {Shao}, {Sharma}, {Mayer}, {OSullivan}, \& {Ilani
  Loubser}}]{Saeedzadeh+2023}
{Saeedzadeh}, V., {Jung}, S.~L., {Rennehan}, D., {et~al.} 2023, arXiv e-prints,
  arXiv:2304.03798, \dodoi{10.48550/arXiv.2304.03798}

\bibitem[{{Salem} {et~al.}(2015){Salem}, {Besla}, {Bryan}, {Putman}, {van der
  Marel}, \& {Tonnesen}}]{Salem+2015}
{Salem}, M., {Besla}, G., {Bryan}, G., {et~al.} 2015, \apj, 815, 77,
  \dodoi{10.1088/0004-637X/815/1/77}

\bibitem[{{Sawala} {et~al.}(2016){Sawala}, {Frenk}, {Fattahi}, {Navarro},
  {Bower}, {Crain}, {Dalla Vecchia}, {Furlong}, {Helly}, {Jenkins}, {Oman},
  {Schaller}, {Schaye}, {Theuns}, {Trayford}, \& {White}}]{Sawala+2016}
{Sawala}, T., {Frenk}, C.~S., {Fattahi}, A., {et~al.} 2016, \mnras, 457, 1931,
  \dodoi{10.1093/mnras/stw145}

\bibitem[{{Schlafly} \& {Finkbeiner}(2011)}]{Schlafly+2011}
{Schlafly}, E.~F., \& {Finkbeiner}, D.~P. 2011, \apj, 737, 103,
  \dodoi{10.1088/0004-637X/737/2/103}

\bibitem[{{Serra} {et~al.}(2014){Serra}, {Westmeier}, {Giese}, {Jurek},
  {Fl{\"o}er}, {Popping}, {Winkel}, {van der Hulst}, {Meyer}, {Koribalski},
  {Staveley-Smith}, \& {Courtois}}]{SoFiA}
{Serra}, P., {Westmeier}, T., {Giese}, N., {et~al.} 2014, {SoFiA: Source
  Finding Application}, Astrophysics Source Code Library.
\newblock \doeprint{1412.001}

\bibitem[{{Serra} {et~al.}(2015){Serra}, {Westmeier}, {Giese}, {Jurek},
  {Fl{\"o}er}, {Popping}, {Winkel}, {van der Hulst}, {Meyer}, {Koribalski},
  {Staveley-Smith}, \& {Courtois}}]{Serra+2015}
---. 2015, \mnras, 448, 1922, \dodoi{10.1093/mnras/stv079}

\bibitem[{{Simpson} {et~al.}(2018){Simpson}, {Grand}, {G{\'o}mez}, {Marinacci},
  {Pakmor}, {Springel}, {Campbell}, \& {Frenk}}]{Simpson+2018}
{Simpson}, C.~M., {Grand}, R. J.~J., {G{\'o}mez}, F.~A., {et~al.} 2018, \mnras,
  478, 548, \dodoi{10.1093/mnras/sty774}

\bibitem[{{Spekkens} {et~al.}(2014){Spekkens}, {Urbancic}, {Mason}, {Willman},
  \& {Aguirre}}]{Spekkens+2014}
{Spekkens}, K., {Urbancic}, N., {Mason}, B.~S., {Willman}, B., \& {Aguirre},
  J.~E. 2014, \apjl, 795, L5, \dodoi{10.1088/2041-8205/795/1/L5}

\bibitem[{{Stevens} {et~al.}(2019){Stevens}, {Diemer}, {Lagos}, {Nelson},
  {Pillepich}, {Brown}, {Catinella}, {Hernquist}, {Weinberger}, {Vogelsberger},
  \& {Marinacci}}]{Stevens+2019}
{Stevens}, A. R.~H., {Diemer}, B., {Lagos}, C. d.~P., {et~al.} 2019, \mnras,
  483, 5334, \dodoi{10.1093/mnras/sty3451}

\bibitem[{{Taylor} {et~al.}(2011){Taylor}, {Hopkins}, {Baldry}, {Brown},
  {Driver}, {Kelvin}, {Hill}, {Robotham}, {Bland-Hawthorn}, {Jones}, {Sharp},
  {Thomas}, {Liske}, {Loveday}, {Norberg}, {Peacock}, {Bamford}, {Brough},
  {Colless}, {Cameron}, {Conselice}, {Croom}, {Frenk}, {Gunawardhana},
  {Kuijken}, {Nichol}, {Parkinson}, {Phillipps}, {Pimbblet}, {Popescu},
  {Prescott}, {Sutherland}, {Tuffs}, {van Kampen}, \&
  {Wijesinghe}}]{Taylor+2011}
{Taylor}, E.~N., {Hopkins}, A.~M., {Baldry}, I.~K., {et~al.} 2011, \mnras, 418,
  1587, \dodoi{10.1111/j.1365-2966.2011.19536.x}

\bibitem[{{Tonnesen} \& {Bryan}(2009)}]{Tonnesen+2009}
{Tonnesen}, S., \& {Bryan}, G.~L. 2009, \apj, 694, 789,
  \dodoi{10.1088/0004-637X/694/2/789}

\bibitem[{{van der Walt} {et~al.}(2011){van der Walt}, {Colbert}, \&
  {Varoquaux}}]{numpy}
{van der Walt}, S., {Colbert}, S.~C., \& {Varoquaux}, G. 2011, Computing in
  Science and Engineering, 13, 22, \dodoi{10.1109/MCSE.2011.37}

\bibitem[{{van Dokkum} {et~al.}(2015){van Dokkum}, {Abraham}, {Merritt},
  {Zhang}, {Geha}, \& {Conroy}}]{vanDokkum+2015}
{van Dokkum}, P.~G., {Abraham}, R., {Merritt}, A., {et~al.} 2015, \apjl, 798,
  L45, \dodoi{10.1088/2041-8205/798/2/L45}

\bibitem[{{Vollmer} {et~al.}(2001){Vollmer}, {Cayatte}, {Balkowski}, \&
  {Duschl}}]{Vollmer+2001}
{Vollmer}, B., {Cayatte}, V., {Balkowski}, C., \& {Duschl}, W.~J. 2001, \apj,
  561, 708, \dodoi{10.1086/323368}

\bibitem[{{Vulcani} {et~al.}(2018){Vulcani}, {Poggianti}, {Gullieuszik},
  {Moretti}, {Tonnesen}, {Jaff{\'e}}, {Fritz}, {Fasano}, \&
  {Bettoni}}]{Vulcani+2018}
{Vulcani}, B., {Poggianti}, B.~M., {Gullieuszik}, M., {et~al.} 2018, \apjl,
  866, L25, \dodoi{10.3847/2041-8213/aae68b}

\bibitem[{{Vulcani} {et~al.}(2021){Vulcani}, {Poggianti}, {Moretti},
  {Franchetto}, {Bacchini}, {McGee}, {Jaff{\'e}}, {Mingozzi}, {Werle},
  {Tomi{\v{c}}i{\'c}}, {Fritz}, {Bettoni}, {Wolter}, \&
  {Gullieuszik}}]{Vulcani+2021}
{Vulcani}, B., {Poggianti}, B.~M., {Moretti}, A., {et~al.} 2021, \apj, 914, 27,
  \dodoi{10.3847/1538-4357/abf655}

\bibitem[{{Wang} {et~al.}(2020){Wang}, {Catinella}, {Saintonge}, {Pan},
  {Serra}, \& {Shao}}]{Wang+2020}
{Wang}, J., {Catinella}, B., {Saintonge}, A., {et~al.} 2020, \apj, 890, 63,
  \dodoi{10.3847/1538-4357/ab68dd}

\bibitem[{{Wang} {et~al.}(2016){Wang}, {Koribalski}, {Serra}, {van der Hulst},
  {Roychowdhury}, {Kamphuis}, \& {Chengalur}}]{Wang+2016}
{Wang}, J., {Koribalski}, B.~S., {Serra}, P., {et~al.} 2016, \mnras, 460, 2143,
  \dodoi{10.1093/mnras/stw1099}

\bibitem[{{Wang} {et~al.}(2014){Wang}, {Fu}, {Aumer}, {Kauffmann}, {J{\'o}zsa},
  {Serra}, {Huang}, {Brinchmann}, {van der Hulst}, \& {Bigiel}}]{Wang+2014}
{Wang}, J., {Fu}, J., {Aumer}, M., {et~al.} 2014, \mnras, 441, 2159,
  \dodoi{10.1093/mnras/stu649}

\bibitem[{{W}es {M}c{K}inney(2010)}]{pandas1}
{W}es {M}c{K}inney. 2010, in {P}roceedings of the 9th {P}ython in {S}cience
  {C}onference, ed. {S}t\'efan van~der {W}alt \& {J}arrod {M}illman, 56 -- 61,
  \dodoi{10.25080/Majora-92bf1922-00a}

\bibitem[{{Wetzel} {et~al.}(2013){Wetzel}, {Tinker}, {Conroy}, \& {van den
  Bosch}}]{Wetzel+2013}
{Wetzel}, A.~R., {Tinker}, J.~L., {Conroy}, C., \& {van den Bosch}, F.~C. 2013,
  \mnras, 432, 336, \dodoi{10.1093/mnras/stt469}

\bibitem[{{Wright} {et~al.}(2021){Wright}, {Tremmel}, {Brooks}, {Munshi},
  {Nagai}, {Sharma}, \& {Quinn}}]{Wright+2021}
{Wright}, A.~C., {Tremmel}, M., {Brooks}, A.~M., {et~al.} 2021, \mnras, 502,
  5370, \dodoi{10.1093/mnras/stab081}

\bibitem[{{Wright} {et~al.}(2022){Wright}, {Lagos}, {Power}, {Stevens},
  {Cortese}, \& {Poulton}}]{Wright+2022}
{Wright}, R.~J., {Lagos}, C. d.~P., {Power}, C., {et~al.} 2022, \mnras, 516,
  2891, \dodoi{10.1093/mnras/stac2042}

\bibitem[{{Yang} {et~al.}(2022){Yang}, {Ianjamasimanana}, {Hammer}, {Higgs},
  {Namumba}, {Carignan}, {J{\'o}zsa}, \& {McConnachie}}]{Yang+2022}
{Yang}, Y., {Ianjamasimanana}, R., {Hammer}, F., {et~al.} 2022, \aap, 660, L11,
  \dodoi{10.1051/0004-6361/202243307}

\bibitem[{{Young} {et~al.}(2007){Young}, {Skillman}, {Weisz}, \&
  {Dolphin}}]{Young+2007}
{Young}, L.~M., {Skillman}, E.~D., {Weisz}, D.~R., \& {Dolphin}, A.~E. 2007,
  \apj, 659, 331, \dodoi{10.1086/512153}

\bibitem[{{Zaritsky} {et~al.}(1993){Zaritsky}, {Smith}, {Frenk}, \&
  {White}}]{Zaritsky+1993}
{Zaritsky}, D., {Smith}, R., {Frenk}, C., \& {White}, S. D.~M. 1993, \apj, 405,
  464, \dodoi{10.1086/172379}

\bibitem[{{Zaritsky} {et~al.}(1997){Zaritsky}, {Smith}, {Frenk}, \&
  {White}}]{Zaritsky+1997}
---. 1997, \apj, 478, 39, \dodoi{10.1086/303784}

\bibitem[{{Zhu} \& {Putman}(2023)}]{Zhu+2023}
{Zhu}, J., \& {Putman}, M.~E. 2023, \mnras, 521, 3765,
  \dodoi{10.1093/mnras/stad695}

\end{thebibliography}
\bibliographystyle{aasjournal}



\end{document}